\documentclass[twocolumn,superscriptaddress,longbibliography,aps,pra,preprintnumbers]{revtex4-1}
\usepackage{hyperref}
\usepackage{epsfig}
\usepackage{graphicx}
\usepackage{epstopdf}
\usepackage{tikz}
\usepackage{float}
\usepackage{amsmath}
\usepackage{amssymb}
\usepackage{bm}
\usepackage[normalem]{ulem}

\newcommand{\mb}[1]{ { \mbox{\boldmath{$#1$}}}  }

\begin{document}

\title{Kondo and Majorana signatures near the singlet-doublet quantum phase transition}

\author{G. G\'orski}
\affiliation{Faculty of Mathematics and Natural Sciences, University of Rzesz\'ow, 
       35-310 Rzesz\'ow, Poland}

\author{J. Bara\'nski}
\affiliation{Institute of Physics, Polish Academy of Sciences, 02-668 Warsaw, Poland}

\author{I. Weymann}
\affiliation{Faculty of Physics, A.\ Mickiewicz University, 61-614 Pozna\'n, Poland}

\author{T. Doma\'{n}ski}
\email[e-mail: ]{doman@kft.umcs.lublin.pl}
\affiliation{Institute of Physics, M.\ Curie-Sk\l{}odowska University, 20-031 Lublin, Poland}

\date{\today}

\begin{abstract}
We study the low energy spectrum of a correlated quantum dot embedded between 
the normal conducting and superconducting reservoirs and hybridized with 
the topological superconducting nanowire, hosting the Majorana end-modes. 
We investigate the leaking Majorana quasiparticle and inspect its interplay 
with the proximity induced on-dot pairing and correlations. In particular, we 
focus on the subgap Kondo effect near the quantum phase transition/crossover 
from the spinfull (doublet) to the spinless (BCS-type singlet) configurations. 
Treating the correlations perturbatively and within the NRG approach we study 
its signatures observable  in the Andreev (particle-to-hole conversion) 
tunneling  spectroscopy. We find, that the leaking Majorana mode has a
spin-selective influence on the subgap Kondo effect.
\end{abstract}

\maketitle

\section{Introduction}
\label{sec:intro}

Intensive studies have been recently devoted to  quasiparticles, 
resembling the Majorana fermions 
\cite{Alicea-12,Flensberg-12,Stanescu-13,Beenakker-13,Franz-15,
Aguado-2017,Lutchyn-2017} that are
identical with their own antiparticles. These exotic objects have been
predicted at defects \cite{Volovik-1999} or boundaries of 
topological superconductors \cite{Read-2000,Kitaev-2001} and 
their non-Abelian character make them appealing for quantum computing 
or novel spintronic devices \cite{DasSarma-2016}.  
Majorana quasiparticles have been predicted in various setups
\cite{Tewari-2007,Fu-2008,Nilsson-2008,Sato-2009,Tworzydlo-2010,
Sau-2010,Oreg-2010,Lutchyn-2010,Choy-2011,ultracold,Aguado-2012}, 
but their experimental realization has been so far reported only
in the ballistic tunneling \cite{Mourik-12,Kouwenhoven-2016} and 
STM measurements \cite{Yazdani-14,Kisiel-15,Franke-15,Yazdani-2017}
through the nanowires proximity-coupled to the bulk $s$-wave superconductors. 

\begin{figure}
\centering
\includegraphics[width=0.8\linewidth]{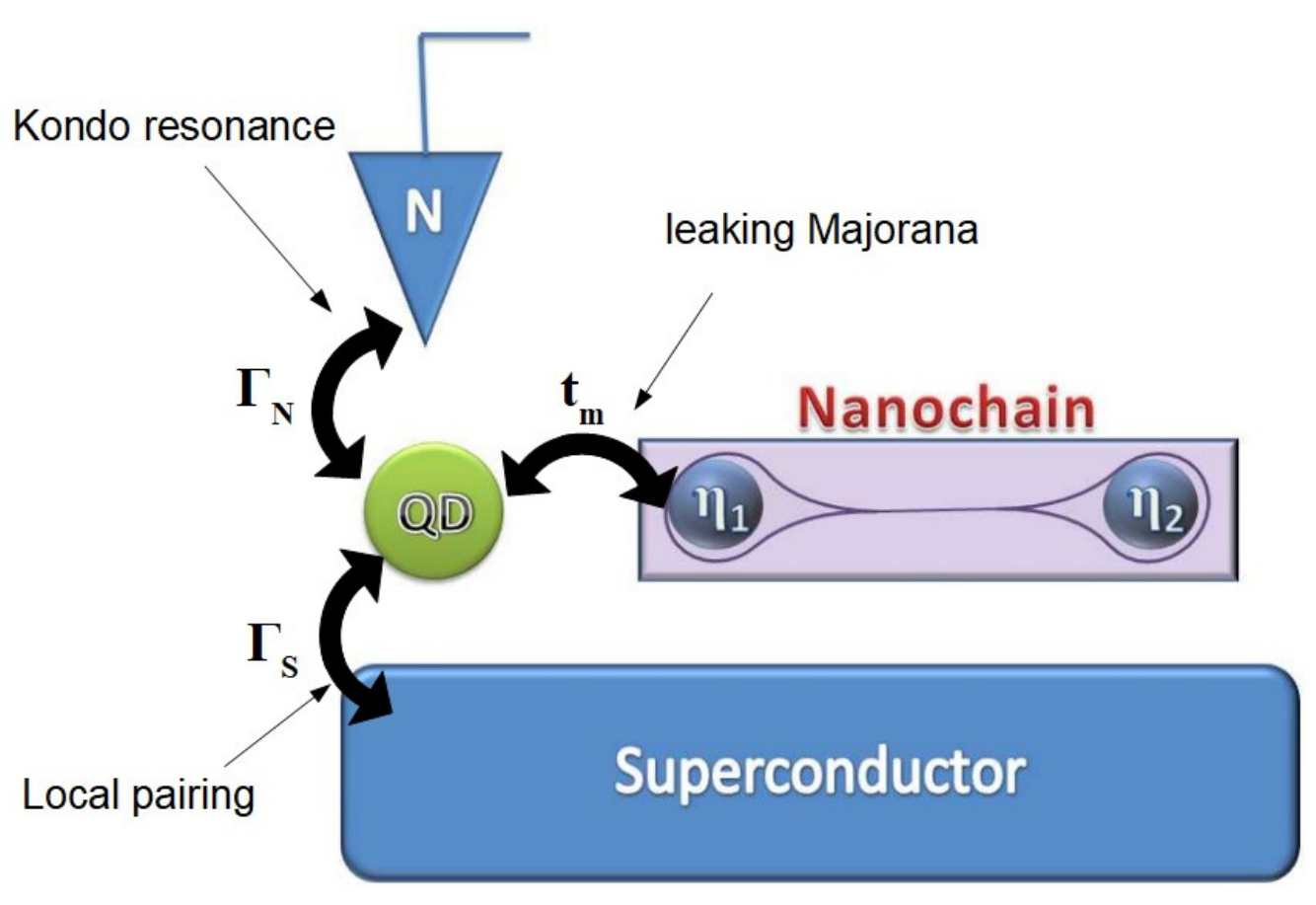}

\vspace{-0.2cm}
\caption{Scheme of the quantum dot (QD) deposited on 
super\-conducting substrate (S) and hybridized with the Rashba nanowire 
[hosting the Majorana end-modes $\eta_{1}$ and $\eta_{2}$] which is 
probed by the metallic tip (N) via the Andreev tunneling.}
\label{schematics}
\end{figure}

Inspired by the work by M.T. Deng {\em et al.} \cite{Deng-2017}, 
who provided experimental evidence for the Majorana mode leaking 
into the quantum dot side-attached to InAs nanowire, 
we propose here a slightly different setup (Fig. \ref{schematics}) 
for studying interplay between: electron pairing and the Kondo effect 
in presence of the Majorana quasiparticle. Leakage of such mode has 
been initially predicted by E. Vernek {\em et al.} \cite{Vernek-2014}. 
Coalescence of the Andreev states into the zero-energy Majorana state 
has been thoroughly discussed by various groups
\cite{DasSarma-2017,Klinovaja-2017,Ptok-2017}, addressing also 
correlation effects on the Hartree-Fock level \cite{Prada-2017},  
within the equation of motion approach \cite{Baranski-2017}, and using 
NRG \cite{Chirla-2016} (but for very weak coupling $\Gamma_{N}$ to 
the conducting lead). Our present study is complementary to 
the former analysis, focusing on the subgap Kondo effect driven 
by an effective exchange interaction of the correlated quantum dot 
with the normal lead \cite{Zitko-2015a,Domanski-2016}. Such situation 
could be realized in STM-type geometry, similar to what has been 
used by the Princeton \cite{Yazdani-14} and Basel \cite{Kisiel-15} 
groups. For instance, one can use a nanoscopic Fe chain with one 
(or a few) side-coupled nonmagnetic atoms (like Ag or Au) deposited 
on the superconducting substrate (e.g. Pb or Al) and probe it 
either by the normal \cite{Yazdani-14,Kisiel-15} or ferromagnetic 
\cite{Yazdani-2017} STM tip. 

The quantum dot (QD) embedded between a superconducting (S) and 
metallic (N) leads and side-coupled to the Majorana nanowire  
(Fig.\ \ref{schematics}) is formally affected by three reservoirs. 
We assume, however, that charge tunneling occurs only via metallic 
tip -- quantum dot -- superconductor circuit. In other 
words, the nanowire plays here a role of `floating' lead. 
Without the Majorana mode a relationship between the subgap 
Kondo effect and the  on-dot pairing has been analyzed for N-QD-S 
setup by several authors \cite{Zitko-2015a,Domanski-2016,Lee-2017}. 
Here we extend this analysis, considering influence of the Majorana 
mode on the low-energy (subgap) electronic states. We study the 
resulting spectroscopic signatures, focusing on the quantum 
phase transition/crossover from the spinfull (singly occupied)  
to the spinless (BCS-type) configurations 
\cite{Zitko-2015a,Domanski-2016}.

Correlations and the Majorana mode have been 
already studied for QD embedded between the metallic 
\cite{Baranger-2011,Lopez-2014,Lee-2013,Cao-2012, 
Vernek-2014,Gong-2014,Lutchyn-2015,Stefanski-2015,Li-2015} or ferromagnetic 
\cite{Weymann-2017,Wojcik-2017} electrodes. It has been shown, that 
the leaking Majorana quasiparticle affects the Kondo resonance of such
`normal' QD and its signatures are observable in the linear conductance. 
In particular, for the long nanowires (with negligible overlap between 
Majorana modes) the linear conductance reaches $3e^2/2h$, whereas 
for the short ones (with the overlapping Majoranas) the conductance 
reaches $2e^2/h$ \cite{Lopez-2014,Lee-2013,Weymann-2017,Wojcik-2017}.
Analysis of the thermoelectric properties of such N-QD-N setup 
has revealed that for small overlap the thermopower would 
reverse its sign \cite{Lopez-2014, Weymann-2017, Wojcik-2017}.

For junctions, comprising the normal and superconducting electrodes
a relationship between the Majorana quasiparticle with the subgap
Kondo effect is much less explored \cite{Chirla-2016, Baranski-2017, 
Gong-2014,Wang-2016}. Through the proximity effect, the QD acquires  
pairing \cite{Balatsky-2006,Domanski-2010} thereby all processes engaging 
a given spin simultaneously affect its opposite counter-partner 
\cite{Golub-2015}. In such situation, the Majorana quasiparticle 
hybridized with, let's say $\uparrow$ electron, would affect the 
spectrum of $\downarrow$ electrons. Our present analysis shows, 
that the leaking Majorana quasiparticle has a {\em spin-selective 
influence on the subgap Kondo effect}. Since both spin components 
are  important for the Andreev scattering we discuss in some detail 
the resulting subgap transport properties.  

The paper is organized as follows. In Sec.\ \ref{sec.model} we formulate
the microscopic model. In Sec.\ \ref{sec.free} we study the subgap 
QD spectrum and the tunneling conductance, neglecting the correlations.
Next, in Sec.\ \ref{sec.Kondo}, we consider the correlated QD in
the subgap Kondo regime. In Sec.\ \ref{sec.sum} we summarize the 
main results. Appendices provide details, concerning 
influence of the magnetic field, finite polarization of Majorana 
modes, their overlap, etc.

\section{Low energy model}
\label{sec.model}

Empirical realizations of the topological superconductivity in 
the semiconducting wires \cite{Mourik-12,Kouwenhoven-2016} or
magnetic atoms' chain \cite{Yazdani-14,Kisiel-15,Franke-15,Yazdani-2017}
rely on the $p$-wave pairing (of identical spins) between the nearest 
neighbor sites, reminiscent of the Kitaev scenario \cite{Kitaev-2001}. 
In this paper we assume that such pairing is induced for $\uparrow$ electrons, 
so only this particular spin component of QD  is {\em directly} coupled 
to the Majorana quasiparticle \cite{Vernek-2014,Vernek-2015}. 
Due to the proximity induced on-dot pairing,  the other
($\downarrow$) spin is {\em indirectly} influenced by the Majorana
quasiparticle as well. For this reason, any process engaging $\uparrow$ 
electrons would simultaneously (although with different efficiency) 
affect the opposite spin \cite{Golub-2015}. This would be very important 
for the Andreev (particle to hole conversion) scattering,
which is the only subgap transport channel at low temperatures.

Our setup (Fig.\ \ref{schematics}) can be  described by the following 
Anderson-type Hamiltonian
\begin{eqnarray}
H= \sum_{\beta=S,N} \left( H_{\beta} + H_{\beta - QD} \right) 
+H_{QD} +H_{MQD} ,
\label{HAnd}
\end{eqnarray}
where $H_{N}=\sum_{k, \sigma} \xi_{k N}c^{\dagger}_{k \sigma N}c_{k\sigma N}$ 
describes the metallic electrode,  $H_{S}=\sum_{k, \sigma} \xi_{kS}
c^{\dagger}_{k \sigma S}c_{k \sigma S} - \sum_k ( \Delta c^{\dagger}
_{k\uparrow S}c^{\dagger}_{-k \downarrow S}+h.c.)$ refers to $s$-wave 
superconducting substrate and electron energies $\xi_{k\beta}$ are measured 
with respect to the chemical potentials $\mu_{\beta}$.  
The correlated QD is described  by $H_{QD}=\sum_{\sigma} 
\epsilon d^{\dagger}_{\sigma} d_{\sigma}+Un_{\downarrow} n_{\uparrow}$, where 
$\epsilon$ denotes the energy level and $U$ stands for the repulsive interaction 
between opposite spin electrons.  The QD is coupled to both external reservoirs 
via $H_{\beta - QD}=\sum_{k,\sigma}
(V_{k\beta}d^{\dagger}_{ \sigma} c_{k \sigma \beta}+h.c.)$, where 
$V_{k\beta}$ denote the matrix elements. 
In a wide bandwidth limit, it is convenient to introduce the auxiliary
couplings $\Gamma_{\beta}=2\pi\sum_{k} |V_{k\beta}|^{2} \delta (
\omega - \xi_{k\beta})$, which we assume to be constant. 
It can be shown  \cite{Bauer-2007,Yamada-2011,Rodero-2011,Baranski-2013}, 
that for $|\omega| \ll \Delta$ the superconducting electrode  induces 
the static pairing 
$H_{S}+H_{S-QD} \approx 
- \frac{\Gamma_S}{2}(d_{\uparrow} 
d_{\downarrow}+d^{\dagger}_{\downarrow} d^{\dagger}_{\uparrow})$. 
We  make use of this low energy model, whose extension to  
arbitrary values of $\Delta$ has been discussed for instance in Ref.\ 
\cite{DasSarma-2017}.

The zero-energy end modes of the topological nanowire can be modeled 
by the following term \cite{Lutchyn-2015}
\begin{eqnarray}
H_{MQD} = i\epsilon_m \eta_{1} \eta_{2} + \lambda \left(  
d^{\dagger}_{\uparrow} \eta_{1} + \eta_{1} d_{\uparrow}  \right)  
\label{Majorana_part}
\end{eqnarray}
with the hermitian operators $\eta_{i}=\eta_{i}^{\dagger}$, 
where $\epsilon_{m}$ corresponds to their overlap. We recast these
Majorana operators by the standard fermionic ones \cite{Franz-15}
$\eta_{1}=\frac{1}{\sqrt{2}}(f+f^{\dagger})$ and
$\eta_{2}=\frac{-i}{\sqrt{2}}(f-f^{\dagger})$
so that  (\ref{Majorana_part}) can be expressed as
\begin{eqnarray}
H_{MQD} = t_m (d^{\dagger}_{\uparrow} - d_{\uparrow}) 
( f + f^{\dagger} ) + \epsilon_m f^{\dagger} f 
- \frac{\epsilon_m}{2}  ,
\label{majorana_dot}
\end{eqnarray}
where $t_{m}=\lambda/\sqrt{2}$.

\section{Majorana vs electron pairing} 
\label{sec.free}

We first consider the case of uncorrelated QD ($U=0$).
Let us calculate  the Green's function  
${\cal{G}}(\omega)=\langle\langle \Psi ; \Psi^{\dagger} \rangle \rangle$ 
in the following matrix notation $\Psi= (d_{\uparrow}, d_{\downarrow}^{\dagger}, f, f^{\dagger})$
\begin{widetext}
\begin{eqnarray} 
\lim_{U=0} 
{\cal{G}}^{-1}(\omega) =
\left( \begin{array}{cccc}  
\omega-\epsilon+i\Gamma_N/2 & \Gamma_S/2 & -t_{m} & -t_{m}\\
\Gamma_S/2 & \omega+\epsilon+i\Gamma_N/2 & 0 & 0 \\
-t_{m} & 0 & \omega-\epsilon_{m}-t_{m}^2/b & -t_{m}^2/b\\
-t_{m} & 0 & -t_{m}^2/b       & \omega+\epsilon_{m}-t_{m}^2/b 
\end{array}\right),
\label{Gr44}
\end{eqnarray} 
\end{widetext} 
where $b=\omega+\epsilon+i\Gamma_N/2-(\Gamma_S/2)^2/(\omega-\epsilon+i\Gamma_N/2)$.
For $\epsilon_m=0$ (i.e.\ without any overlap between the Majorana modes) 
the Green's function (\ref{Gr44}) simplifies to
\begin{eqnarray}
{\cal{G}}_{11}(\omega) & = & \frac{\omega+\epsilon+i\frac{\Gamma_N}{2}}
{D_1(\omega)}+\frac{2t_m^2(\omega+\epsilon+i\frac{\Gamma_N}{2})^2}{D(\omega)},
\label{G11} \\
{\cal{G}}_{22}(\omega) & = & \frac{\omega-\epsilon+i\frac{\Gamma_N}{2}}
{D_1(\omega)}+\frac{2t_m^2 \left( \frac{\Gamma_S}{2}\right)^2}{D(\omega)},
\label{G22} \\
{\cal{G}}_{12}(\omega) & = & \frac{-\frac{\Gamma_S}{2}}{D_1(\omega)}
-\frac{2t_m^2(\omega+\epsilon+i\frac{\Gamma_N}{2})\frac{\Gamma_S}{2}}{D(\omega)},
\label{G12}
\end{eqnarray}
where  $D(\omega)\equiv D_1(\omega)[\omega D_1(\omega)-4t_m^2(\omega+i\Gamma_N/2)]$ 
and $D_1(\omega)\equiv (\omega+i\Gamma_N/2)^2-\epsilon^2-(\Gamma_S/2)^2$. These 
Green's functions \eqref{G11}-\eqref{G12} are composed of the part, representing
solution for the quantum dot coupled only to N and S electrodes ($t_m=0$) 
(first terms of r.h.s Eqs. \eqref{G11}-\eqref{G12}) and additional term 
dependent on coupling to the Majorana fermions.  
In the `superconducting atomic limit' ($\Gamma_N\rightarrow 0$) such Green's function
is characterized by five poles: two corresponding to the Andreev bound 
states ($\pm \sqrt{\epsilon^2+(\Gamma_S/2)^2}$) and three additional states 
resulting from the Majorana fermions 
(0, $\pm \sqrt{\epsilon^2+(\Gamma_S/2)^2+(2t_m)^2}$). 

\subsection{Free quasiparticle spectrum}

\begin{figure}
\centering
\includegraphics[width=\linewidth]{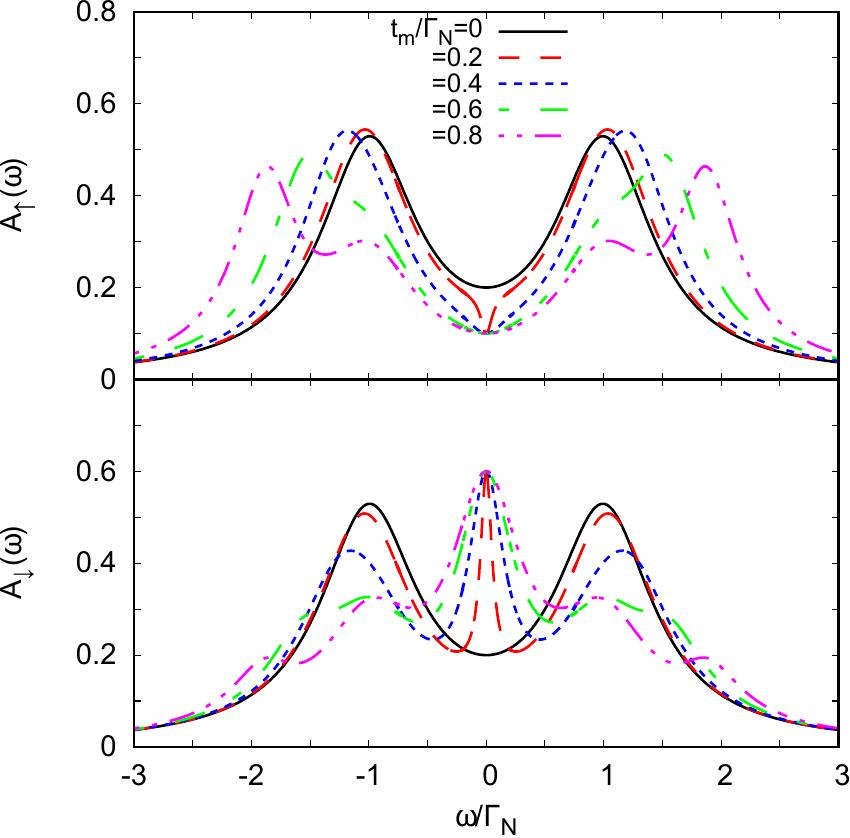}
\caption{The normalized spectral function $A_{\sigma}(\omega)=
\frac{\pi}{2}\Gamma_N\rho_{\sigma}(\omega)$ of the uncorrelated dot
$U=0$ obtained for $\Gamma_S=2\Gamma_N$, $\epsilon=0$ and various 
couplings $t_m$.}
\label{spectrum_free}
\end{figure}

Figure~\ref{spectrum_free} shows the spin-resolved normalized spectral function 
$A_{\sigma}(\omega)=\frac{\pi}{2}\Gamma_N\rho_{\sigma}(\omega)$ of 
the uncorrelated QD obtained at half-filling ($\epsilon=0$) for 
various couplings $t_m$. As a reference shape we also present the spectrum in 
absence of the Majorana quasiparticles ($t_m=0$), revealing the Andreev quasiparticle 
peaks at $\omega=\pm\sqrt{\epsilon^{2}+(\Gamma_{S}/2)^{2}}$ whose broadening is 
$\Gamma_{N}$. For $t_{m}\neq 0$ the spin-resolved spectra are no longer identical
due to the direct (indirect) coupling of  $\uparrow$ ($\downarrow$)
QD electrons with the side-attached Majorana state. The most significant 
differences show up near $\omega \sim 0$. In particular, direct hybridization of
$\uparrow$ electrons depletes their spectrum near the Majorana state. 
Exactly at $\omega=0$ their spectral function is reduced  
by half, $A_{\uparrow}(0)_{\left| t_{m}\neq 0 \right.}=0.5 A_{\uparrow}
(0)_{\left| t_m=0 \right.}$,
similarly to what has been reported for the same geometry with 
both nonsuperconducting leads  \cite{Lee-2013,Vernek-2015,Weymann-2017}. 
Contrary to this behavior, the spin $\downarrow$ electrons (indirectly 
coupled to the Majorana state via  on-dot pairing) clearly gain the electronic 
states. Again, at $\omega=0$ the spectral function $A_{\downarrow}(0)$ 
does not depend on $t_m$ (unless $t_{m}$ vanishes). This constructive
feedback of the side-attached Majorana state on $\downarrow$ electrons
has no analogy to any normal systems \cite{Lee-2013,Vernek-2015,Weymann-2017}.

Upon increasing the coupling $t_m$ we observe a gradual splitting of the Andreev 
quasiparticles, leading to emergence of the effective `molecular' structure. 
We can notice some differences appearing in the spectrum $A_{\sigma}(\omega)$ 
of $\uparrow$ and $\downarrow$ electrons, especially in the low energy region.

\subsection{Quasiparticle features in tunneling spectroscopy}

Low energy quasiparticles of the QD, which is side-attached to the Majorana 
mode, can be probed in our setup (Fig.\ \ref{schematics}) only indirectly,  
via the tunneling current. When voltage $V$ applied between the normal 
tip and superconducting substrate is smaller than the energy gap $\Delta$ 
the charge transport is provided at low temperatures solely by the Andreev 
reflections \cite{Andreev-1964}. For noninteracting systems such transport 
mechanism can be quantitatively determined from the Landauer-type formula
\begin{eqnarray} 
I_{A}(V) = \frac{e}{h} \int \!\!  d\omega \; T_{A}(\omega)
\left[ f(\omega\!-\!eV)\!-\!f(\omega\!+\!eV)\right] ,
\label{I_A}
\end{eqnarray} 
where $f(x)=\left[ 1 + \mbox{\rm exp}(x/k_{B}T) \right]^{-1}$ is the 
Fermi distribution. The energy-dependent transmittance  
\begin{eqnarray} 
 T_{A}(\omega)=\Gamma_{N}^{2} \; \left| {\cal{G}}_{12}(\omega) 
\right|^{2} + \Gamma_{N}^{2} \; \left|  {\cal{G}}_{21}(\omega) 
\right|^{2}
\label{T_A}
\end{eqnarray} 
describes a probability of electron (from STM tip) with spin $\sigma$ 
to be converted into a hole (reflected back to the STM tip) with an opposite 
spin $\bar{\sigma}$, injecting one Cooper pair into the superconducting
substrate. Similar expression (\ref{I_A}) is valid (subject to certain
approximations) also for the correlated quantum dots \cite{Krawiec-2004}.
It has been emphasized \cite{Baranski-2013}, that the differential 
conductance $G_{A}(V)=dI_{A}(V)/dV$ can detect the subgap 
quasiparticle states, at expense of mixing the particle and hole
degrees of freedom. In particular, at zero temperature the differential 
conductance simplifies to $G_{A}(V)=\frac{2e^{2}}{h}\left[ T_{A}(\omega
=+eV)+T_{A}(\omega=-eV)\right]$. 

\begin{figure}
\centering
\includegraphics[width=\linewidth]{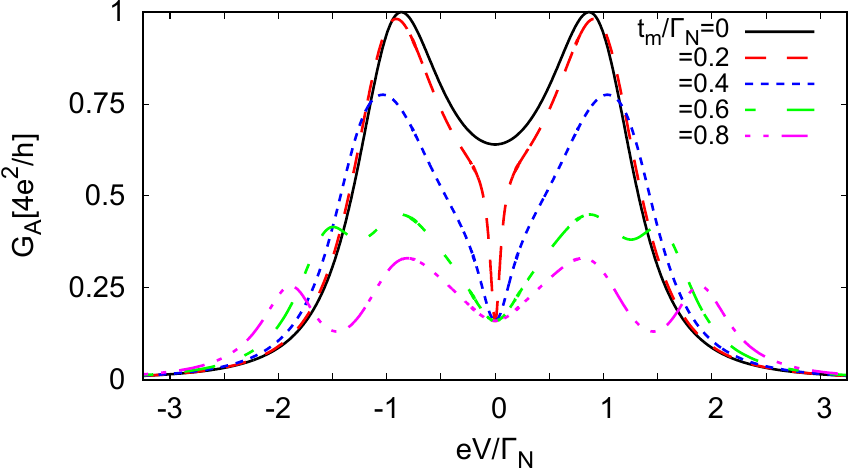}
\caption{The differential Andreev conductance obtained at zero 
temperature for the same model parameters as in Fig.\ \ref{spectrum_free}.}
\label{Tra1}
\end{figure}

Figure~\ref{Tra1} shows the differential Andreev conductance obtained 
at zero temperature for different values of $t_m$, assuming $\epsilon_m=0$. 
We observe, that  
for all finite couplings $t_{m}\neq 0$ the linear 
conductance $G_A(V\! =\! 0)$ drops to the value $\frac{1}{4} G_A(V\! =\! 0)_{t_m=0}$. 
This result is qualitatively different from what has been obtained 
for N-QD-N junctions, where 
$G(V\! =\! 0)_{t_m\neq 0}=\frac{3}{4} G(V\! =\! 0)_{t_m= 0}$ \cite{Lee-2013}.
Upon increasing the coupling $t_{m}$  
the nonlinear conductance 
$G_A(V \! \neq\! 0)$ develops four local maxima, two of them at
$\pm\sqrt{\epsilon^2+(\Gamma_S/2)^2}$ and additional pair at
$\pm\sqrt{\epsilon^2+(\Gamma_S/2)^2+(2t_m)^2}$. These local maxima
are no longer equal to the perfect Andreev conductance $4e^{2}/h$.
They originate from the Andreev
states mixed with the Majorana quasiparticle 
(see Fig.\ \ref{spectrum_free}).

In N-QD-N junctions with the side-attached Majorana nanowire the 
weak coupling $t_{m}$ leads to the Fano-type interference patterns 
\cite{Schuray-2017}. Unlike the usual $\pi$ shift one obtains in that 
case only a fraction of such phase. In consequence the density of states 
is reduced by half and the corresponding linear conductance drops to $3/4$ 
of its original value, namely to $e^2/h +\frac{1}{2}e^2/h$ as compared to 
the maximum $2e^2/h$ for $t_m=0$ case. In our N-QD-S setup (Fig.\ 
\ref{schematics}) both spins participate in forming the local pairs. 
The Andreev current depends on the squared anomalous Green's 
functions ${\cal{G}}_{12}(\omega)$ and ${\cal{G}}_{21}(\omega)$ therefore 
the linear conductance $G_A(V=0)$ is strongly reduced (down to $25\%$) 
for arbitrary coupling $t_{m}\neq 0$ (Fig.\ \ref{Tra1}).

\section{Majorana vs Kondo}
\label{sec.Kondo}

We now analyze the correlated quantum dot, focusing on the subgap 
Kondo effect originating from the Coulomb potential $U$ and the 
coupling $\Gamma_{N}$ to the normal STM tip. It has been emphasized 
\cite{Yamada-2011,Rodero-2011}, that (in absence of the Majorana 
quasiparticle) the induced on-dot pairing has unique 
relationship with the Coulomb repulsion. For instance, by increasing 
the ratio $\Gamma_{S}/U$ the subgap Kondo peak would broaden
\cite{Domanski-2016,Zitko-2015a} and this behavior shall occur 
upon approaching the quantum phase transition from 
the (spinfull) doublet side \cite{Bauer-2007}.

Our main purpose here is to examine how this subgap Kondo effect 
(appearing at zero energy) gets along with the leaking 
Majorana mode. Some earlier studies of the correlated quantum dot 
coupled to both normal (conducting) electrodes in presence of 
the side-attached Rashba chain indicated a competition between 
the Kondo and Majorana physics \cite{Vernek-2015,Lee-2013,Cheng-2014,
Beek-2016,Weymann-2017,Wojcik-2017}. For sufficiently long wire 
($\epsilon_m=0$) the Kondo effect is preserved only for the
spin-channel $\downarrow$ (which is not coupled to the Majorana 
zero-energy mode), whereas for the other spin-channel $\uparrow$ 
there appears a dip in the spectral density at $\omega=0$ 
(reminiscent to what we observed in the upper panel of Fig.\ 
\ref{spectrum_free}). In consequence, the total transmission
is partly blocked suppressing the linear conductance from 
$2e^2/h$ to the fractional value $3e^2/2h$ 
\cite{Vernek-2015,Lee-2013,Weymann-2017,Wojcik-2017,Lopez-2014}.
For the short Rashba wires ($\epsilon_m\neq0$) the Kondo physics 
may survive in both spin channels, with its width dependent on 
$\epsilon_m$. The initial Kondo features are fully recovered 
in both of the spin-channels only when $\epsilon_m \gg 
(|\epsilon|,U,\Gamma)$. 

When the correlated quantum dot is embedded between the metallic and 
superconducting leads (N-QD-S) the eventual subgap Kondo effect is
controlled by $U/\Gamma_{S}$ ratio and $\epsilon$ 
\cite{Domanski-2016,Zitko-2015a,Yamada-2011,Tanaka-2007,Baranski-2013}, 
which decide whether QD ground-state is the (spinfull) doublet $\left| 
\sigma \right>$ or the (spinless) BCS-type $u \left| 0 \right> - 
v\left| \uparrow \downarrow \right>$ configuration. In particular, 
for the half-filled QD ($\epsilon=-\frac{U}{2}$) the BCS singlet 
is realized for $U<\Gamma_{S}$, whereas the doublet is preferred 
for $U>\Gamma_{S}$ \cite{Bauer-2007}. Obviously, the Kondo physics 
might occur only for the latter one, owing to antiferromagnetic 
exchange interactions driven between the QD and normal lead 
\cite{Domanski-2016,Domanski-2017}.
In what follows, we confront this subgap Kondo effect 
with the leaking Majorana quasiparticle.

\subsection{Perturbative treatment of correlations}

To study the correlation effects we start by treating the Coulomb term
$Un_{\downarrow}n_{\uparrow}$ in a perturbative manner. We compute the 
self-energy matrix $\mb{\Sigma}(\omega)$ of the single particle Green's 
function from the Dyson equation ${\cal{G}}^{-1}(\omega)=\left[{\cal{G}}
^{U=0}(\omega)\right]^{-1}-\mb{\Sigma}(\omega)$ within the second-order
perturbation theory (SOPT) \cite{Vecino-2003}, which yields the following 
diagonal and off-diagonal selfenergies \cite{Domanski-2016,Yamada-2011}
\begin{eqnarray}
{\mb\Sigma}_{11}(\omega) & = & U\langle d_{\downarrow}^{\dagger} d_{\downarrow}\rangle\label{diag11_sigma}\\
 & + & U^{2}\int\limits _{-\infty}^{\infty}{\frac{\left(-\frac{1}{\pi}\right)\mbox{{\rm Im}}{{\mb\Sigma}_{11}^{(2)}(\omega')}}{{\omega-\omega'+i0^{+}}}d\omega'},\nonumber \\
{\mb\Sigma}_{22}(\omega) & = & U\langle d_{\uparrow} d_{\uparrow}^{\dagger}\rangle\label{diag22_sigma}\\
 & + & U^{2}\int\limits _{-\infty}^{\infty}{\frac{\left(-\frac{1}{\pi}\right)\mbox{{\rm Im}}{{\mb\Sigma}_{22}^{(2)}(\omega')}}{{\omega-\omega'+i0^{+}}}d\omega'},\nonumber \\
{\mb\Sigma}_{12}(\omega) & = & U\langle d_{\downarrow} d_{\uparrow}\rangle\label{offdiag_sigma}\\
 & - & U^{2}\int\limits _{-\infty}^{\infty}{\frac{\left(-\frac{1}{\pi}\right)\mbox{{\rm Im}}{{\mb\Sigma}_{12}^{(2)}(\omega')}}{{\omega-\omega'+i0^{+}}}d\omega'}.\nonumber 
\end{eqnarray}
The imaginary parts of the second-order contributions
${\mb\Sigma}_{ij}^{(2)}(\omega)$ are expressed by the  convolutions \cite{Domanski-2017}
\begin{eqnarray}
-\frac{1}{\pi}\mbox{{\rm Im}}{\mb\Sigma}_{11(22)}^{(2)}(\omega) & = & \int\limits _{-\infty}^{\infty}\left[{\mb\Pi}_{1}(\omega+\omega')\rho_{22(11)}^{+}(\omega')\right.\nonumber \\
 & + & \left.{\mb\Pi}_{2}(\omega+\omega')\rho_{22(11)}^{-}(\omega')\right]\;d\omega',\label{imag_diag}\\
-\frac{1}{\pi}\mbox{{\rm Im}}{\mb\Sigma}_{12}^{(2)}(\omega) & = & \int\limits _{-\infty}^{\infty}\left[{\mb\Pi}_{1}(\omega+\omega')\rho_{21}^{+}(\omega')\right.\nonumber \\
 & + & \left.{\mb\Pi}_{2}(\omega+\omega')\rho_{21}^{-}(\omega')\right]\;d\omega',\label{imag_offdiag}
\end{eqnarray}
where 
\begin{eqnarray}
{\mb\Pi}_{1(2)}(\omega) & = & \int\limits _{-\infty}^{\infty}\left[\rho_{11}^{-(+)}(\omega')\rho_{22}^{-(+)}(\omega-\omega')\right.\nonumber \\
 & - & \left.\rho_{12}^{-(+)}(\omega')\rho_{21}^{-(+)}(\omega-\omega')\right]\;d\omega',\label{Pi_prop}
\end{eqnarray}
and  $\rho_{ij}^{\pm}(\omega)\equiv \frac{-1}{\pi}
\mathrm{Im}\,{\cal{G}}^{0}_{ij}(\omega\!+\!i0^{+})f(\pm\omega)$ denote the occupancies 
obtained from the uncorrelated Green's functions \eqref{Gr44}, taking into account 
the effective dot level and influence of the superconducting electrode \cite{Yamada-2011}.

\begin{figure}
\centering
\includegraphics[width=0.9\linewidth]{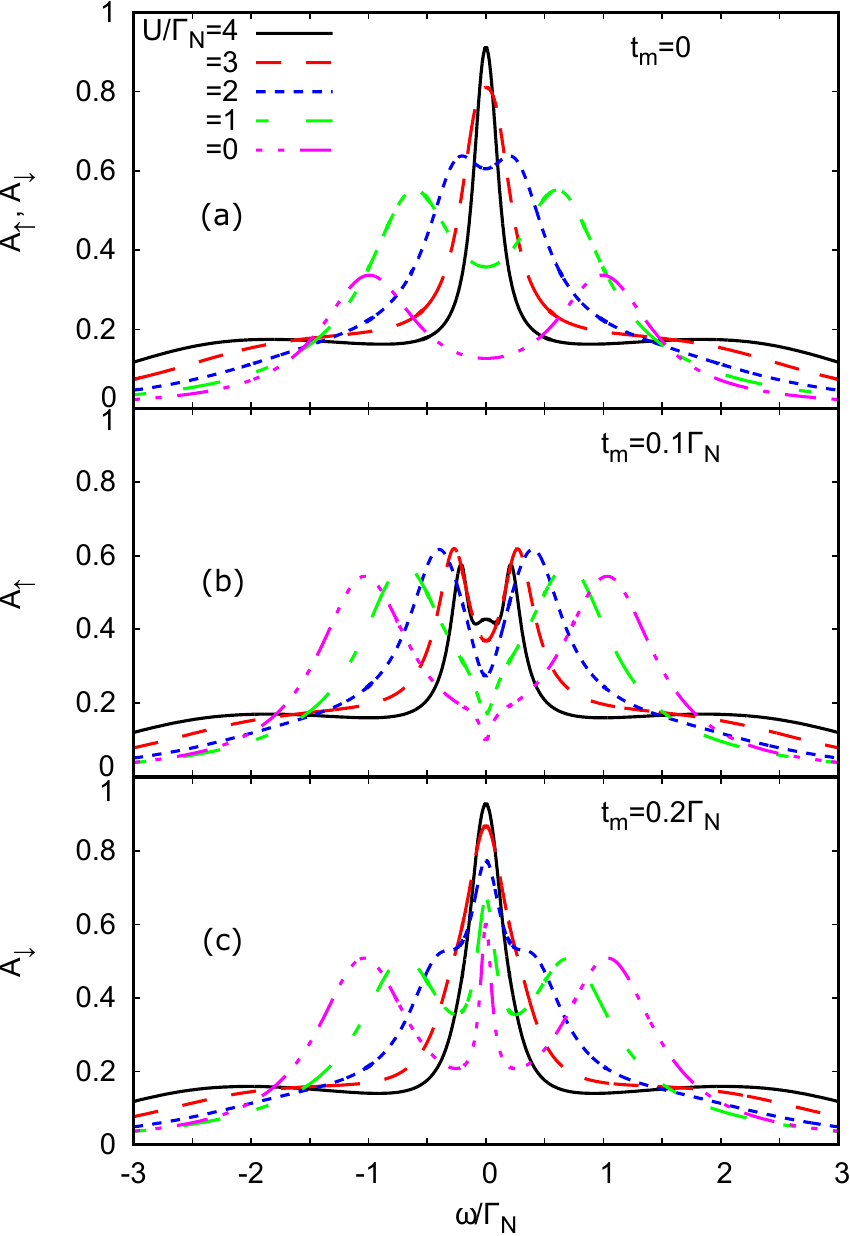}
\caption{The spin-resolved spectral function $A_{\sigma}(\omega)$ 
obtained by the SOPT method at zero temperature for the half-filled 
quantum dot ($\epsilon=-U/2$), using $t_{m}=0$ (upper panel) and
$t_{m}/\Gamma_{N}=0.2$ (middle/bottom panels).}
\label{spin_SOPT}
\end{figure}

Let us inspect the spectral function of the correlated quantum dot 
for each spin separately. Figure~\ref{spin_SOPT} shows the results  obtained at 
zero temperature by  perturbative treatment of the Coulomb potential. 
The top panel refers to N-QD-S junction in absence of the Majorana  mode. 
In the weak interaction $U$ regime it is characterized by two Andreev peaks. 
When approaching $U\approx \Gamma_{S}$ these quasiparticle peaks merge, 
signaling the quantum phase transition (formally for $\Gamma_{N}\neq 0$ 
it becomes a continuous crossover). In the strongly correlated limit 
($U>\Gamma_{S}$), we observe development of the subgap Kondo peak at 
$\omega=0$ whose broadening gradually shrinks upon increasing the ratio 
$U/\Gamma_{S}$ which has been  explained in Refs 
\cite{Domanski-2016,Yamada-2011}. 

In the presence of the side-attached nanowire we notice, that the 
Majorana mode has completely different influence on each spin channel
(panels b and c in Fig.\ \ref{spin_SOPT}). In some analogy to the 
non-interacting case the spectral function $A_{\uparrow}(\omega)$
reveals a depletion of the electronic states near $\omega \sim 0$.
We assign it to destructive interference caused by the side-attached 
Majorana mode \cite{Baranski-2017}. The other spectral function 
$A_{\downarrow}(\omega)$ shows an opposite effect. Indirect coupling 
of spin $\downarrow$ electrons with the Majorana mode contributes
more states near zero energy, the Kondo peak seems thus
to be magnified.

\begin{figure}
\centering
\includegraphics[width=\linewidth]{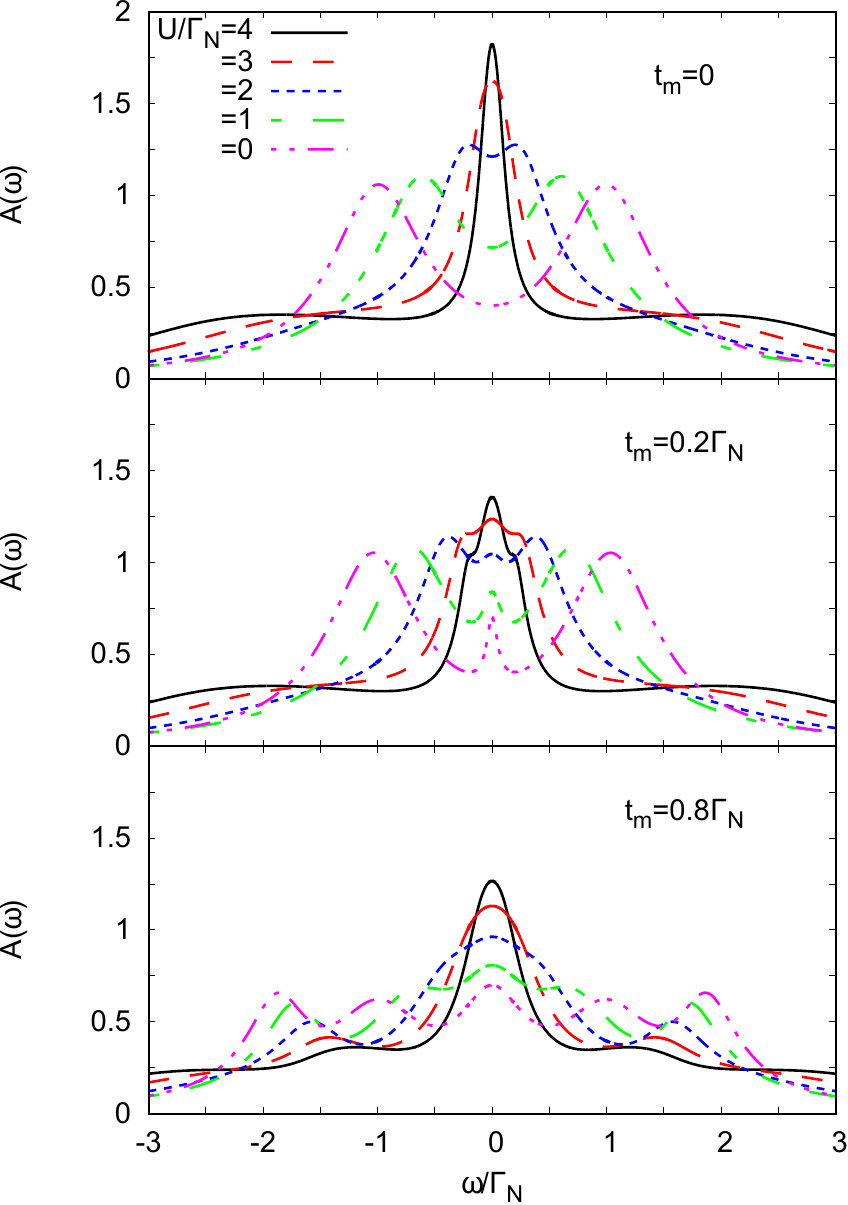}
\caption{The total normalized spectral function $A(\omega)=\sum_{\sigma}
A_{\sigma}(\omega)$ of the half-filled quantum dot ($\varepsilon
=-U/2$) obtained at $T=0$ for $\Gamma_S=2\Gamma_N$ and several values of 
the Coulomb potential $U$ (as indicated), assuming  (a) $t_m=0$, 
(b) $t_m=0.2$, and (c) $t_m=0.8$. All energies are expressed in
units of $\Gamma_{N}$.}
\label{Kondo_spectrum_2D}
\end{figure}

The subgap QD spectrum can be probed by the Andreev scattering which would 
engage both the spin components, we hence display (Fig. \ref{Kondo_spectrum_2D}) 
evolution of the total spectral function $A(\omega)=\sum_{\sigma}A_{\sigma}
(\omega)$ for various couplings $t_{m}$ and Coulomb potential $U$ (as indicated). 
Already in the weakly correlated limit the initial Kondo peak is substantially 
suppressed. Upon increasing the Coulomb potential $U$ the QD spectrum develops 
two separated Andreev quasiparticle states (broadened by $\Gamma_{N}$), coexisting  
with the Majorana feature. This behavior is reminiscent 
of the exact solution in the superconducting atomic limit $\Gamma_{N}=0$  
(see Figs.~\ref{janek} and \ref{both_spins}). On the other hand, for 
stronger coupling $t_{m}=0.8\Gamma_{N}$ the spectral function acquires 
qualitatively different `molecular' structure, in which the leaking 
Majorana mode is strongly mixed with the initial QD quasiparticle states. 
Under such circumstances we can hardly discriminate signatures 
of the Kondo state from Majorana quasiparticle.

\subsection{NRG results}
\label{sec:NRG}

For reliable analysis of the correlations, the induced electron 
pairing and the leaking Majorana quasiparticle, we have also performed 
calculations, based on the numerical renormalization 
group (NRG) algorithm \cite{Wilson}. Our purpose was to study 
the low energy spectrum of the following effective model
\begin{eqnarray}
H &=&   \sum_{\sigma} \epsilon d^{\dagger}_{\sigma} d_{\sigma}
+U n_{\downarrow}n_{\uparrow} - \frac{\Gamma_S}{2}(d_{\uparrow} 
d_{\downarrow}+d^{\dagger}_{\downarrow} d^{\dagger}_{\uparrow}) 
\nonumber \\
&+& t_m (d^{\dagger}_{\uparrow} - d_{\uparrow}) ( f + f^{\dagger} ) 
+ \epsilon_m \left( f^{\dagger} f 
- \frac{1}{2} \right) \nonumber \\
&+& H_{N} + H_{N - QD} .
\label{HAnd_eff}
\end{eqnarray}
\begin{figure}[t]
\centering
\includegraphics[width=0.9\linewidth]{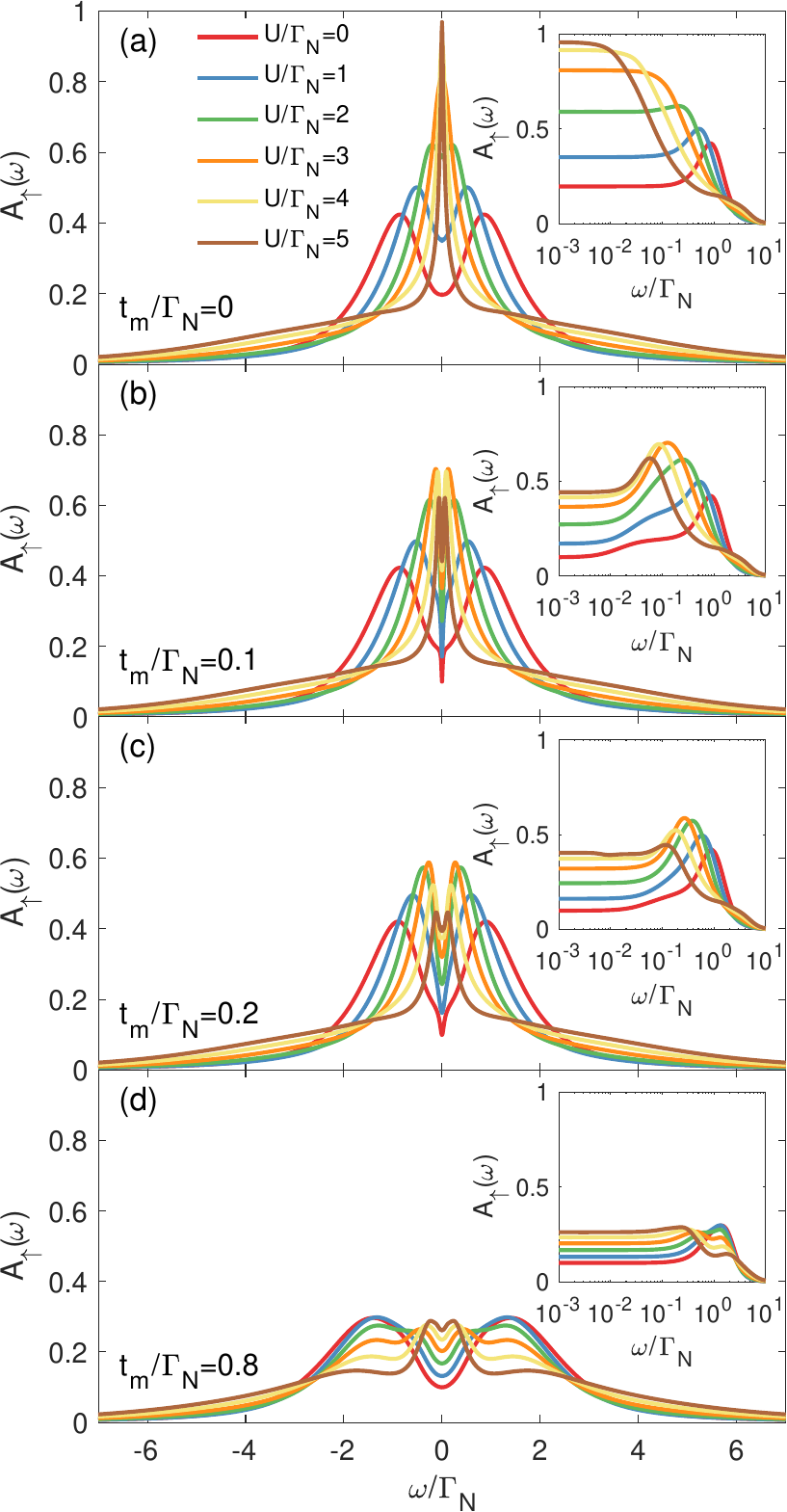}
\caption{The normalized spectral function $A_{\uparrow}(\omega)$
for spin $\uparrow$ obtained from NRG calculations for $\Gamma_{S}=2\Gamma_{N}$, various 
ratios of $U/\Gamma_{N}$ (displayed in the legend of the upper panel) 
and several values of the coupling $t_{m}$, as indicated.
The other parameters are $\epsilon=-U/2$, $\epsilon_m=0$ and 
$\Gamma_N = D/50$, with $D$ the band halfwidth.}
\label{up_NRG}
\end{figure} 
This Hamiltonian (\ref{HAnd_eff}) corresponds to the single-channel model, 
allowing for a good quality computational analysis. We have performed 
the NRG calculations using the Budapest Flexible DM-NRG code \cite{fnrg} 
for constructing the zero-temperature density matrix of the system 
and calculating the corresponding spin-resolved spectral functions
for arbitrary model parameters.
Because the coupling to Majorana zero-energy mode and superconducting pairing correlations
break the spin and charge symmetries,
only the charge parity symmetry of the total Hamiltonian was used.
In calculations we kept at least $1024$ states per iteration
and imposed the discretization parameter $\Lambda=2$.
Our results were averaged over $N_z=4$ interleaved discretizations \cite{Z},
using the logarithmic Gaussian broadening to obtain the smooth spectral functions.
We have assumed the flat density of states of the normal
lead with a cutoff $D \gg U$, with $D$ being the band halfwidth.

\begin{figure}
\centering
\includegraphics[width=0.9\linewidth]{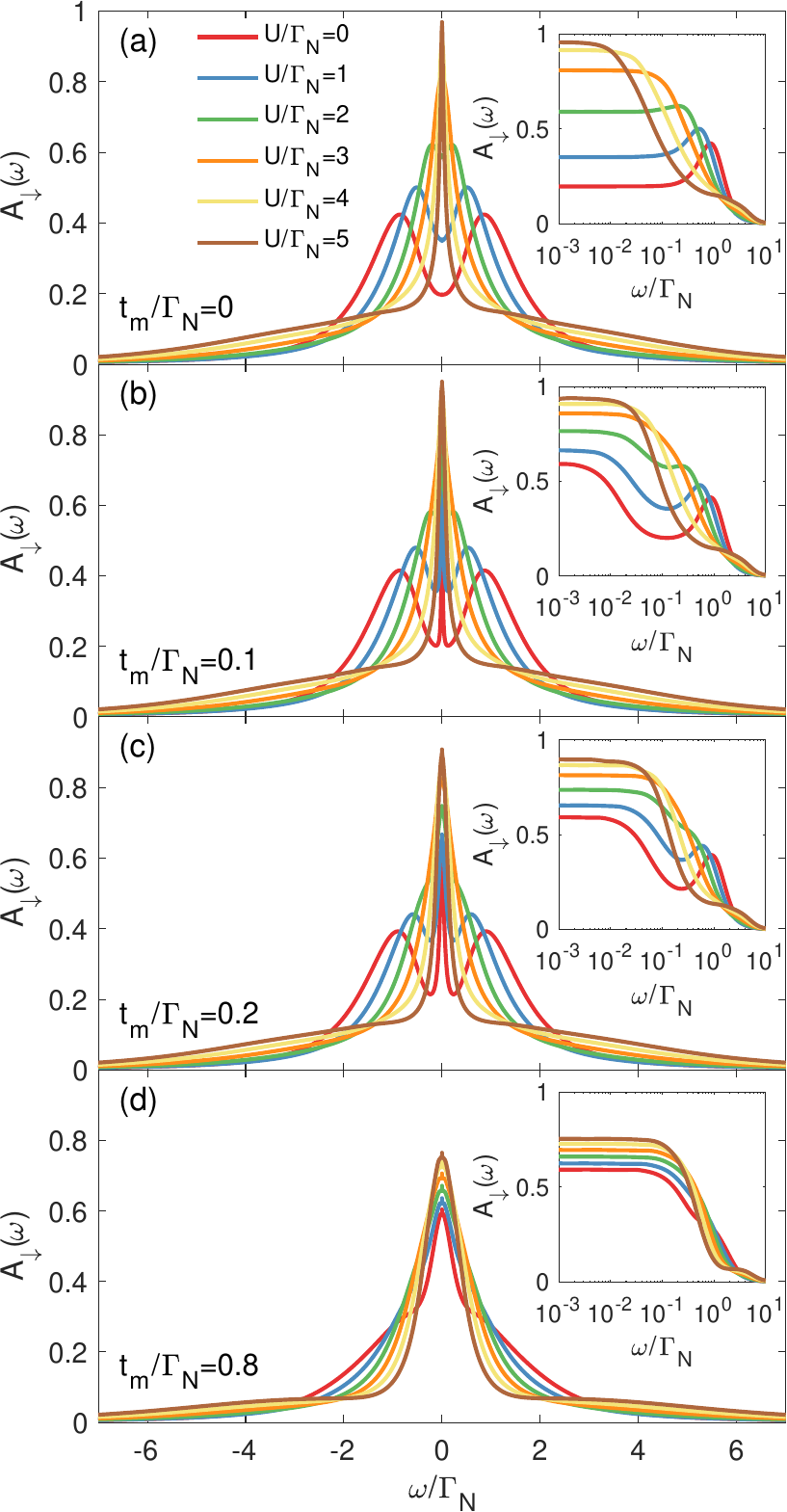}
\caption{The spectral function $A_{\downarrow}(\omega)$ obtained
by the NRG calculations for the same set of parameters as in Fig.~\ref{up_NRG}.}
\label{down_NRG}
\end{figure} 

Figures~\ref{up_NRG} and \ref{down_NRG} present the spectral functions 
obtained by NRG  for spin $\uparrow$ and $\downarrow$, 
respectively. Since we are interested in what happens to the Kondo 
state due to the side-attached Majorana mode, we additionally display 
the low energy spectrum in the logarithmic scale (in the insets).
Quasiparticle states of $\uparrow$ electron (directly coupled to 
the Majorana mode) are strongly suppressed near $\omega \sim 0$. 
In the weak Majorana-dot coupling regime (b \& c panels) we assign such 
effect to the destructive quantum interference \cite{Baranski-2017}. 
For stronger coupling $t_{m}=0.8\Gamma_{N}$, the QD states are 
hybridized with the Majorana mode, reducing the zero-energy
spectral function and developing the new (molecular) quasiparticles.
This is particularly evident when inspecting the inset
of Fig.~\ref{up_NRG}(d).

The spin $\downarrow$ sector (Fig.~\ref{down_NRG}) reveals an opposite 
tendency. In this case, the Majorana mode indirectly affects the states 
predominantly in a vicinity of $\omega\sim 0$. In the weak coupling 
limit the Kondo effect (existing for $U\geq \Gamma_{S}$) seems to be robust, 
but its shape slightly broadens (see the insets of panels b \& c). 
In the molecular regime (panel d) the electronic states 
cumulate near the zero energy, forming a single structureless peak. 
We interpret it as an indirect leakage of the Majorana quasiparticle 
driven by the on-dot pairing.    

Our numerical results obtained by the unbiased NRG calculations qualitatively 
agree with the selfconsistent perturbative treatment. For the Majorana mode 
weakly coupled to the QD, both methods show its detrimental influence on 
the subgap Kondo effect of $\uparrow$ spin and less severe (almost neutral) 
effect on $\downarrow$ spin sector. In the latter case the Kondo peak seems 
to be robust (it merely broadens). On the other hand, for the QD strongly 
coupled to the nanowire we find influence of the leaking Majorana mode
on both spin sectors, where it redistributes the overall quasiparticle
spectra. Under such circumstances the Kondo state is no longer evident.

\subsection{Andreev conductance}

\begin{figure}[b]
\centering
\includegraphics[width=0.9\linewidth]{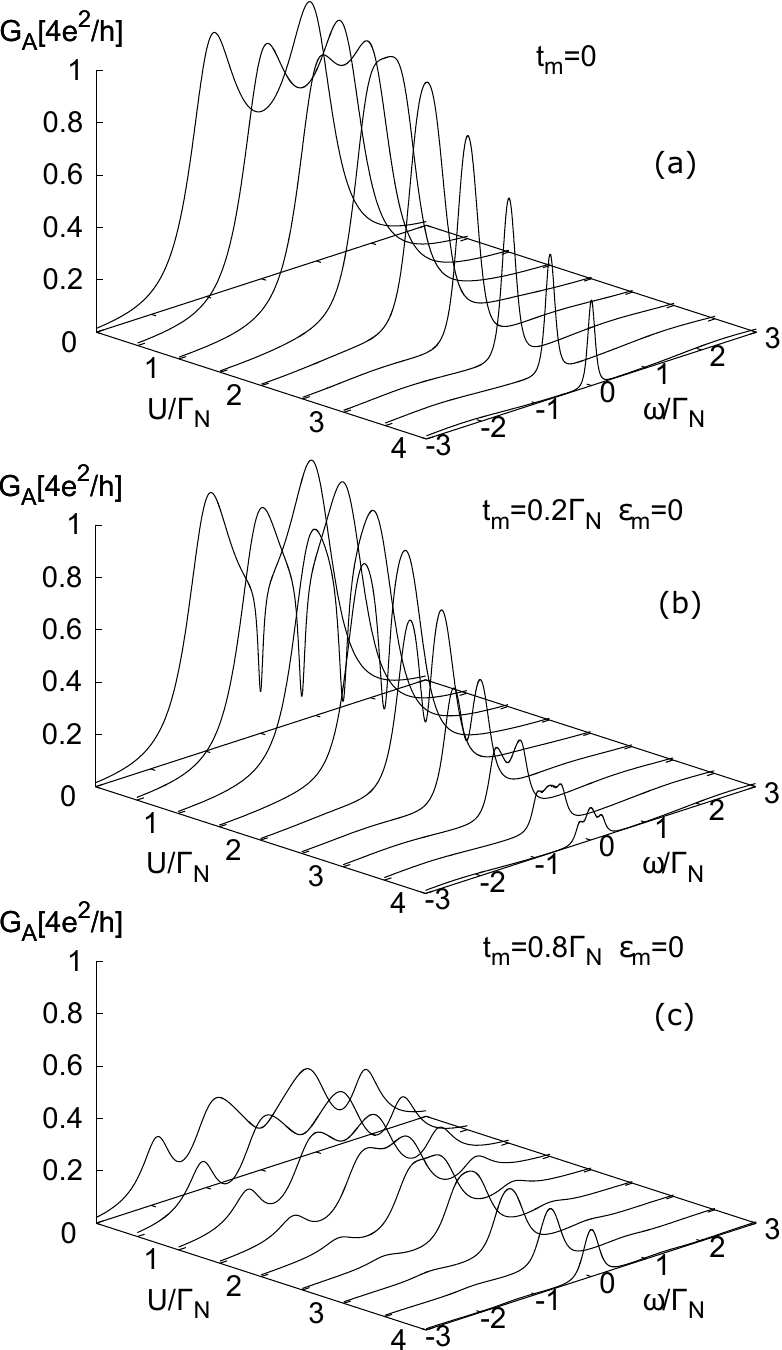}
\caption{The differential subgap conductance $G_{A}(V)$  as a function 
of the applied voltage $V$ and the Coulomb potential $U$ obtained at 
$T=0$ for the half-filled QD, using $\Gamma_{S}=2\Gamma_{N}$.}
\label{cond_vs_voltage}
\end{figure}

Verification of the above mentioned effects could be possible by measuring 
the Andreev current, cf. (\ref{I_A}).  Figure
\ref{cond_vs_voltage} shows the variation of the differential Andreev 
conductance $G_{A}(V)$ with respect to the Coulomb potential $U$  for 
the weak (b) and strong (c) coupling $t_{m}$ limits, respectively. Due 
to technical limitations (related with numerical precision for computation 
of the real part of the anomalous Green's function) we show here the results obtained 
within the perturbative scheme. In the weakly correlated case these plots 
resemble the results presented in Fig.\ \ref{Tra1} for the uncorrelated QD. 
Qualitative changes appear at stronger $U$, especially on the doublet 
side $U\geq \Gamma_S$.

In the absence of the Majorana mode ($t_{m}=0$) the differential conductance 
is characterized by two peaks at bias $V$, coinciding with 
 energies of the Andreev states. The additional zero-bias enhancement is 
due to the subgap Kondo effect, appearing in the doublet region 
(i.e.\ for $U\geq \Gamma_{S}$). Within the generalized Schrieffer-Wolff 
approach adopted for the N-QD-S setup we have previously estimated 
\cite{Domanski-2016}, that the effective Kondo temperature of half-filled 
QD scales as $\ln T_{K} \propto 1/ \left[ 1- \left( \frac{\Gamma_{S}}{U} 
\right)^{2}\right]$. In particular, it yields enhancement of $T_{K}$ 
with respect to $\Gamma_{S}$ upon approaching the doublet-singlet 
transition. This unique behavior is valid for arbitrary $\Delta$, 
as has been found by the NRG studies \cite{Zitko-2015a}. In the limit 
of $U \gg \Gamma_{S}$, the Andreev tunneling is strongly suppressed, 
because the off-diagonal Green's function (characterizing efficiency 
of the induced on-dot pairing) rapidly diminishes. We illustrate 
these effects in the upper panel of Fig.\ \ref{cond_vs_voltage}. 

\begin{figure}[b]
\centering
\includegraphics[width=0.95\linewidth]{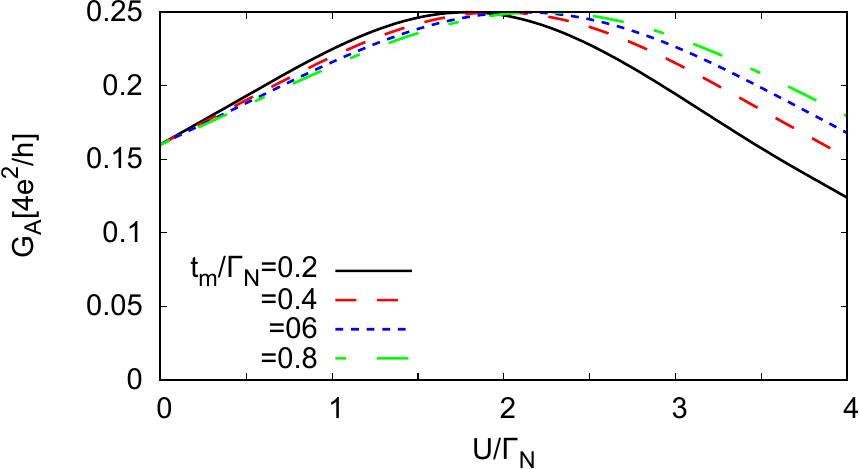}
\caption{The zero-bias Andreev conductance $G_{A}(V\!=\!0)$ versus
the Coulomb potential $U$ (in units of $\Gamma_{N}$) obtained 
perturbatively for $\epsilon=-U/2$, $\Gamma_S=2\Gamma_N$,  $\epsilon_m=0$.}
\label{Lin_cond_vs_voltage}
\end{figure}

Leakage of the Majorana mode on the QD remarkably affects the mentioned 
behavior. In the weak coupling limit (middle panel of Fig. \ref{cond_vs_voltage})
its influence merely shows up near the zero-bias conductance. For 
$\Gamma_S \sim U$, we observe a superposition of the leaking Majorana 
feature (whose width depends on $t_{m}$) with leftovers of the Kondo 
peak, surviving in spin $\downarrow$ channel . For the  strong coupling 
$t_m$ (bottom panel of Fig. \ref{cond_vs_voltage}), the differential 
conductance $G_A(V)$ develops the 'molecular' structure, 
characterized by four peaks. We interpret them as the bonding and 
anti-bonding mutations of the initial Andreev quasiparticles caused 
by  hybridization with the Majorana mode. Upon increasing the 
Coulomb potential the internal peaks gradually merge into 
a single central one, whereas the external peaks loose their spectral
weights.

We notice some universal feature, originating from the Majorana
mode in the linear conductance with respect to the Coulomb 
potential (Fig. \ref{Lin_cond_vs_voltage}). Its optimal value 
$e^2/h$ is 4 times smaller than the one obtained for the N-QD-S 
system without the side-attached Majorana mode \cite{Tanaka-2007,
Gorski-2016}. Maximum of the zero bias conductance $G_A(V\! =\! 0)$ 
occurs at $U\approx\Gamma_S$, which corresponds to a crossover 
between the  spinless  and spinfull configurations \cite{Tanaka-2007}. 
For $t_{m}\neq 0$, this maximum slightly shifts towards $U>\Gamma_S$, 
suggesting that the leaking Majorana mode does affect (though weakly) 
the singlet-doublet phase transition.

\section{Summary and discussion}
\label{sec.sum}

We have analyzed the spin-resolved spectroscopic features 
of the quantum dot side-coupled to the topologically 
superconducting nanowire, hosting the Majorana quasiparticles. 
Focusing on an STM-type geometry we have investigated its subgap 
electronic spectrum which can be probed by the Andreev tunneling,   
that simultaneously involves the particle and hole degrees of freedom.

In the uncorrelated case ($U=0$) the presence of the Majorana quasiparticle 
induces either the zero-energy peak or dip in the QD spectral density, 
depending on the spin (Fig.\ \ref{spectrum_free}). These effects originate 
from the constructive or destructive quantum interference \cite{Schuray-2017}. 
Our study predicts, that the differential Andreev conductance would be 
predominantly affected by a destructive interference, leading to the 
zero-bias dip preserved from the weak to strong hybridization $t_m$ regimes.  

We have also inspected the correlated quantum dot ($U \neq 0$) case, addressing
interplay between the on-dot pairing with the Kondo effect in the presence of 
the leaking Majorana quasiparticle. The Coulomb interaction along with the 
proximity induced electron pairing lead to the quantum phase transition 
from the spinless to spinfull configurations \cite{Bauer-2007,Rodero-2011}. 
When QD is additionally coupled to the normal electrode ($\Gamma_{N}\neq 0)$ 
such transition changes into a crossover and simultaneously the effective 
spin exchange (between QD and itinerant electrons) can induce the subgap 
Kondo effect \cite{Zitko-2015a,Domanski-2016}. By the selfconsistent treatment 
of the Coulomb potential $U$ and using the unbiased NRG calculations we 
have found that the side-attached Majorana mode would have spin-selective 
influence of this subgap Kondo effect. For spin $\uparrow$  (directly 
coupled to the Majorana mode) it would be detrimental, whereas for 
spin $\downarrow$ we predict an opposite tendency.

This spin-dependent influence of the Majorana mode on the subgap Kondo 
effect shows up in the low-energy spectrum of the QD. Such property can be 
added to the previously reported examples of: spin-selective Andreev 
processes \cite{He-2014}, spin-resolved current correlations \cite{Haim-2015}, 
or non-local spin blocking effect \cite{Ren-2017}, which are unique
consequences of the Majorana quasiparticles. Signatures of the 
spin-selective influence on the Kondo effect might be, however, 
difficult to detect in the Andreev tunneling because both spins are 
mixed (via particle-hole degrees of freedom). Nevertheless, we predict 
some universal features. For instance, the linear conductance shall 
be reduced to 25 \% of its perfect value typical for N-QD-S junctions 
\cite{Rodero-2011}. This is in contrast to what has been predicted 
for N-QD-N junctions, where the single particle conductance is reduced 
only to 75 \% of the unitary value \cite{Lee-2013}. Yet in both cases 
the underlying mechanism is related to the very same fractional character 
of the Majorana quasiparticles.

Our study of the Kondo state vs Majorana mode relationship is different 
from the previous considerations of the topological Kondo effect that 
could be realized in the correlated nanowires \cite{Beri-2012,Cheng-2014,
Logan-2014,Beek-2016,Tsvelik-2016,Beri-2017}. We hope that this analysis 
of the proximized quantum dot hybridized with the Majorana nanowire 
(Fig.\ \ref{schematics}) would be experimentally feasible in STM measurements  
and could verify nontrivial interplay between the on-dot pairing,  
the Kondo effect, and the exotic Majorana quasiparticles.   

\section*{Acknowledgments}

This work is supported by the National Science Centre in Poland 
via projects Nos. DEC-2014/13/B/ST3/04451 (TD) and 
DEC-2013/10/E/ST3/00213 (IW), and the
Faculty of Mathematics and Natural Sciences of the University
of Rzesz\'ow through the Project No. WMP/GD-06/2017 (GG).

\appendix

\section{Short wire case}
\label{sec.app}

\begin{figure}[t]
\centering
\includegraphics[width=0.9\linewidth]{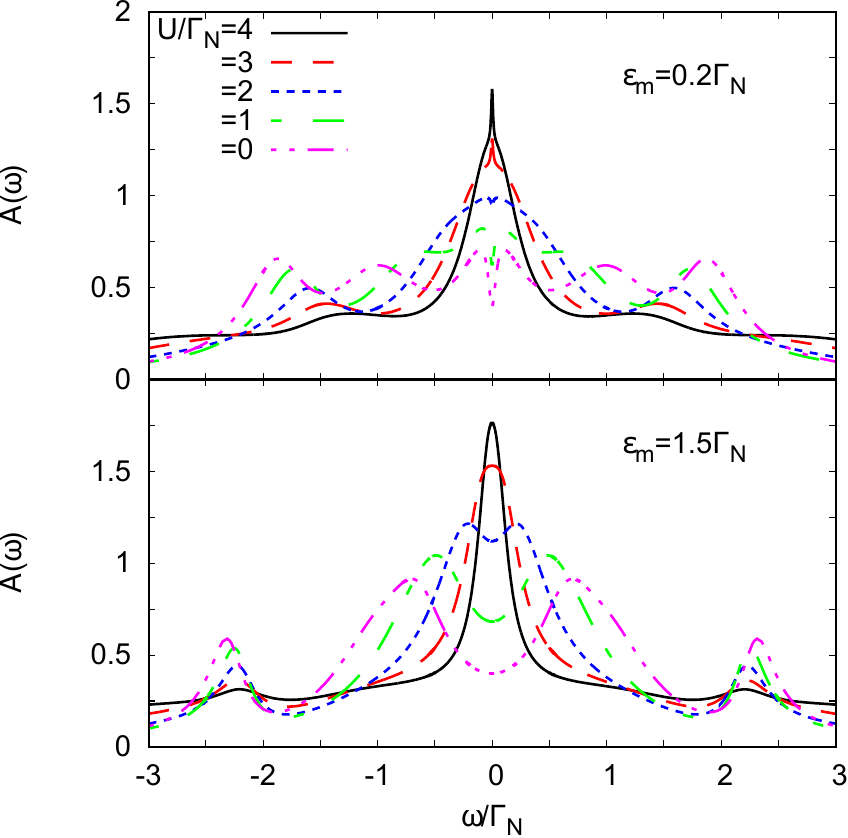}
\caption{The total spectral function $A(\omega)$ of the correlated dot
obtained for different values of $U$, using $\Gamma_S=2\Gamma_N$, $\epsilon=-U/2$, 
$t_m=0.8\Gamma_{N}$, $\epsilon_m=0.2\Gamma_{N}$ (a) and $\epsilon_m=1.5\Gamma_{N}$ (b).}
\label{Kondo_spectrum_kem}
\end{figure}

Main part of this paper is devoted to the infinitely long nanowire,
where the Majorana modes do not overlap ($\epsilon_m=0$) with each 
other. The Kondo state and the Majorana quasiparticle are there
manifested at $\omega=0$.
In finite nanowires the Majorana modes  partly overlap 
($\epsilon_m\neq0$). For
$\Gamma_{N}=0$ and $U=0$ we  obtain six quasiparticle states: 
two Andreev bound states ($\omega_=\pm \sqrt{\epsilon^2+
(\Gamma_S/2)^2}$) and four other ones mixed with the Majorana 
modes $\omega_=\pm (X/2 \pm 1/2\sqrt{X^2-4\epsilon_m^2(\epsilon^2
+\Gamma_S^2/4)})^{0.5}$, where $X=\epsilon^2+\epsilon_m^2+\Gamma_S^2/4+4t_m^2$. 
For $U\neq 0$ and  $\Gamma_{N}\neq 0$ these 
quasiparticle states appear away from the Fermi energy, so
they are less influential on the Kondo effect. Figure
\ref{Kondo_spectrum_kem} presents the total spectral function obtained 
for two values of $\epsilon_m$ in the Kondo regime. For small  
$\epsilon_m=0.2\Gamma_{N}$ (Fig. \ref{Kondo_spectrum_kem}a) and in the weak 
interaction limit, we observe six subgap quasiparticle peaks. 
With increasing $U$, two of them merge into the Kondo resonance peak 
(for $U\geq \Gamma_{S}$) with a tiny superstructure at $\omega=0$. 
For large overlap $\epsilon_m=1.5\Gamma_{N}$ 
(Fig. \ref{Kondo_spectrum_kem}b), we obtain the spectrum 
reminiscent of the N-QD-S system (Fig. \ref{Kondo_spectrum_2D}a) 
with only a redistribution of the spectral weight at higher energies.

\begin{figure}[b]
\centering
\includegraphics[width=0.9\linewidth]{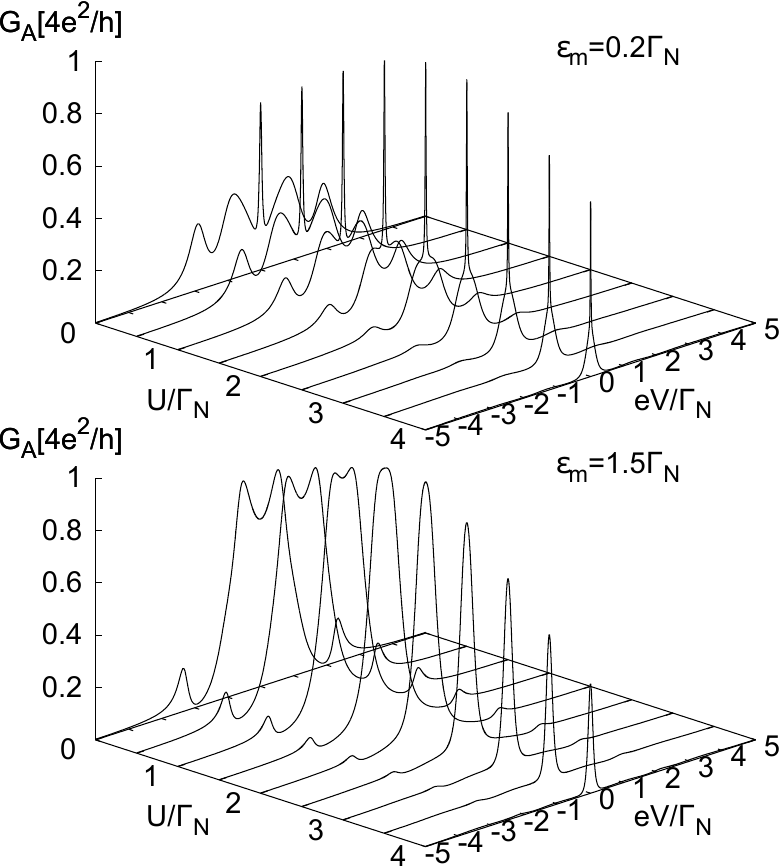}
\caption{The differential Andreev conductance $G_{A}$ as a function of the applied voltage $eV$ 
for $\Gamma_S=2\Gamma_N$, $\epsilon=-U/2$, $\epsilon_m=0.2\Gamma_N$ (a) and $\epsilon_m=1.5\Gamma_N$ (b).}
\label{Cond_vs_voltage_em}
\end{figure}

Finite overlap $\epsilon_m$ would show in the differential Andreev conductance. 
Figure \ref{Cond_vs_voltage_em} shows $G_{A}(V)$ as a function of the applied voltage
$V$ for the weakly ($\epsilon_m=0.2\Gamma_{N}$) and strongly ($\epsilon_m=1.5\Gamma_{N}$) 
overlapping Majorana modes. In the first case we observe the well pronounced zero-bias peak 
of a narrow width, dependent on $\epsilon_m$. Optimal value of the zero-bias peak 
approaches $4e^2/h$ for $U\approx \Gamma_S$. For  $U<\Gamma_S$, we observe two 
Andreev and two other Majorana-Andreev hybrids, appearing in the differential 
conductance. With increasing  $U$, the Andreev peaks 
evolve into the central peak, appearing at $V=0$. On the other hand, for the 
short nanowire ($\epsilon_m=1.5\Gamma_{N}$) and at small voltages, the differential 
conductance $G_A(eV)$ is similar as for  N-QD-S system \cite{Domanski-2016}. 

\begin{figure}[t]
\centering
\includegraphics[width=0.9\linewidth]{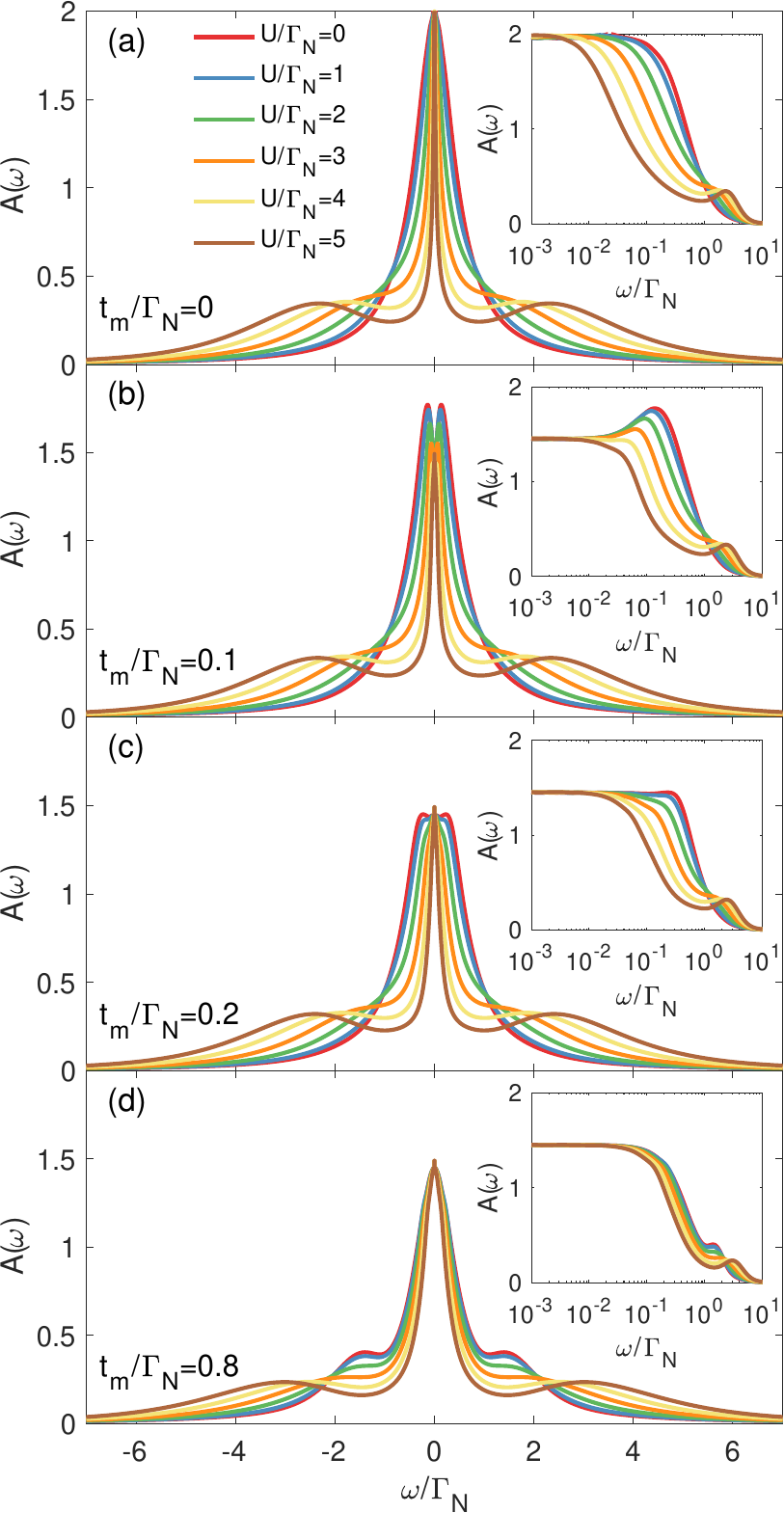}
\caption{The total spectral function of the half-filled quantum dot coupled
only to the normal lead ($\Gamma_{S}=0$). Results are obtained by NRG for 
various Coulomb potentials (indicated in the legend) and several couplings 
to the Majorana mode $t_{m}/\Gamma_{N}=0$, $0.1$, $0.2$, and $0.8$, assuming 
$\epsilon_{M}=0$. The other parameters are the same as in Fig.~\ref{up_NRG}.}
\label{NRG_dataN}
\end{figure} 

\section{Comparison with the normal QD}
\label{comparison_to_normal}

It might be instructive to give a comparison of the correlated quantum dot
spectrum (discussed in Sec.\ \ref{sec.Kondo}) with the one, when on-dot 
pairing is absent. We present here such results for the half-filled QD 
obtained by the nonperturbative NRG calculations.

\begin{figure}[t]
\centering
\includegraphics[width=0.9\linewidth]{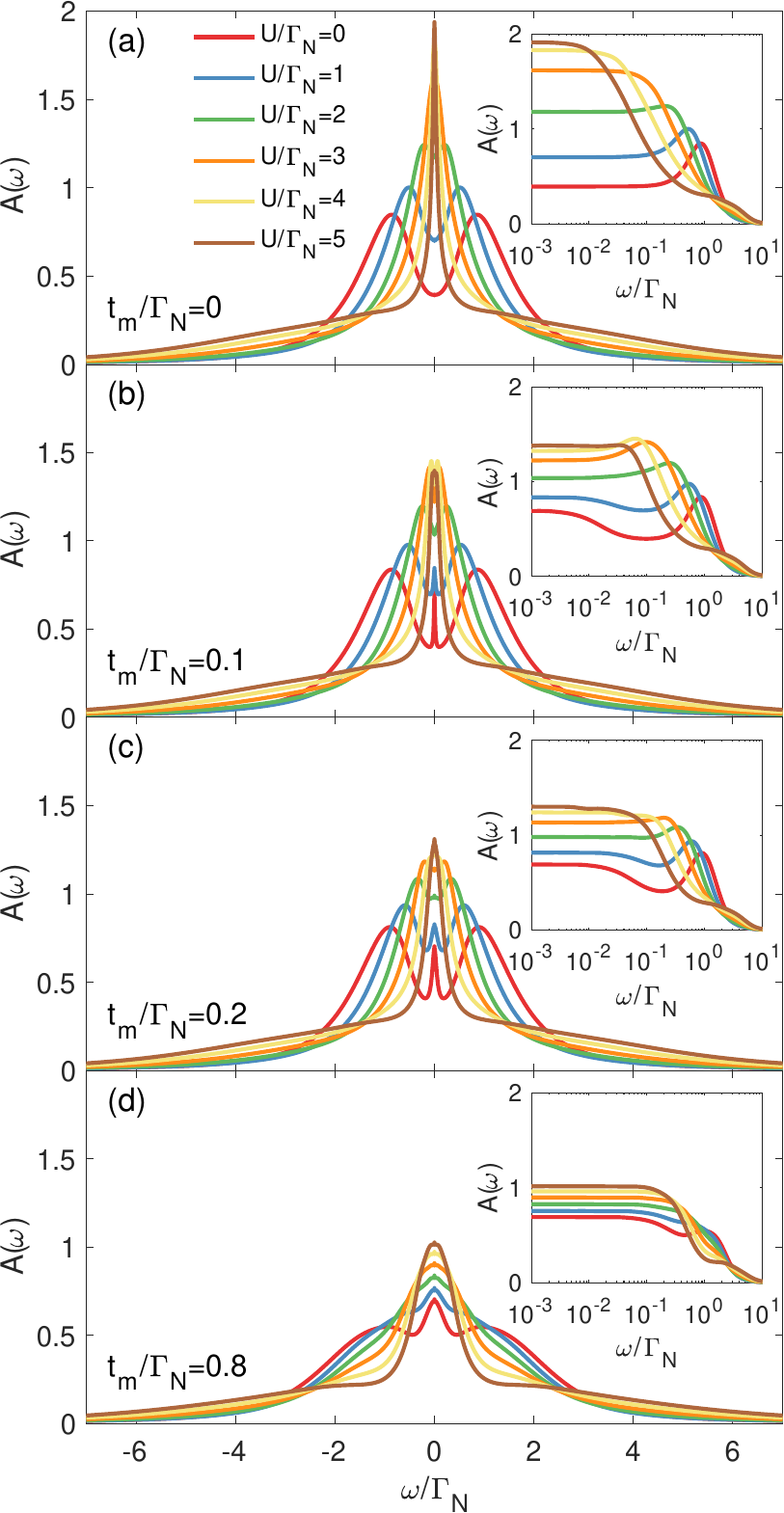}
\caption{The same as in figure \ref{NRG_dataN}, but for $\Gamma_{S}=2\Gamma_{N}$.}
\label{NRG_dataS}
\end{figure} 

Figure \ref{NRG_dataN} displays the total spectrum of the `normal' QD,
corresponding to the situation $\Gamma_{S}=0$. For $t_m=0$, we recognize 
the widely known structure with two quasiparticle peaks at $\epsilon$ 
and $\epsilon+U$ and the Kondo peak at the Fermi level (whose broadening 
depends on $U$ and $\Gamma_{N}$). The side-attached Majorana mode 
leads to noticeable changes mainly near the Fermi level, by suppressing 
the Kondo peak for the spin $\uparrow$ electrons, whereas the other spin 
is completely unaffected by the Majorana. For strong couplings $t_m$  
we observe development of the `molecular' structure, in which the quasiparticle
states are no longer present at $\epsilon$ and $\epsilon+U$.

This can be contrasted with our N-QD-S setup, taking into account the proximity 
induced on-dot pairing $\Gamma_{S}\neq 0$. For each spin sector we have discussed
already the corresponding spectra in Sec.\ \ref{sec:NRG}. Let us briefly 
comment on the total spectral function presented in  Fig.\ \ref{NRG_dataS}.
For $t_{m}=0$ (top panel in Fig.\ \ref{NRG_dataS}), we recognize the spinless 
to spinfull quantum phase transition/crossover  at $\Gamma_S \sim U$ discussed 
previously in Refs \cite{Zitko-2015a,Domanski-2016}, where the Kondo effect is 
observable on the spinfull side (i.e. for $\Gamma_{S} \leq U$). Leakage of
the Majorana mode (for $t_{m}\neq 0$) suppresses this subgap Kondo peak, as 
can be clearly visible in the logarithmic scale (see the insets).
On the other hand, on the BCS-type  side (corresponding to
$\Gamma_{S}>U$) we practically see the separated Andreev peaks
coexisting with a tiny feature at $\omega=0$ due to the leaking Majorana mode.
Such comparison  of Figs \ref{NRG_dataN} and \ref{NRG_dataS} 
emphasizes a qualitative influence of the proximity-induced 
pairing on the Kondo-Majorana interplay.

\section{Influence of magnetic field}
\label{sec:magn_field}

In realistic situations the Majorana end-modes emerge in the proximized nanowire
only above some critical magnetic field  on the order of $1$ Tesla 
\cite{Mourik-12}. The external magnetic field can have strong influence on 
the quantum dot (coupled to the Majorana modes) and on the resulting transport 
properties. Let us recall, that in experimental setup of the Copenhagen group 
\cite{Deng-2017, Deng-2017b} the finite-energy (Andreev) and the zero-energy 
(Majorana) quasiparticle states leak from the nanowire to the `normal' quantum 
dot. Actually, upon increasing the magnetic field there has been observed 
a coalescence of one pair of the Andreev bound states into the zero-energy 
Majorana mode. The quantum dot played a role of the spectrometer. 

Here we consider a different N-QD-SC setup, in which the quantum dot develops 
its initial Andreev states owing to the proximity with the superconducting 
substrate. The side-attached nanowire contributes the Majorana mode, which
modifies the QD spectrum. Obviously, the magnetic field would strongly
affect the subgap spectrum of such `proximitized' QD. Within our model 
described by the low-energy Hamiltonian (\ref{HAnd},\ref{Majorana_part}) 
we can consider the magnetic field through the spin-dependent energy 
levels $\epsilon_{\downarrow}=\epsilon+V_Z$ and $\epsilon_{\uparrow}=
\epsilon-V_Z$ with the Zeeman splitting $2V_Z=g\mu_BB$ \cite{DasSarma-2017,
Chevallier-2013,Rainis-2013}. 

\begin{figure}
\centering 
\includegraphics[width=0.9\columnwidth]{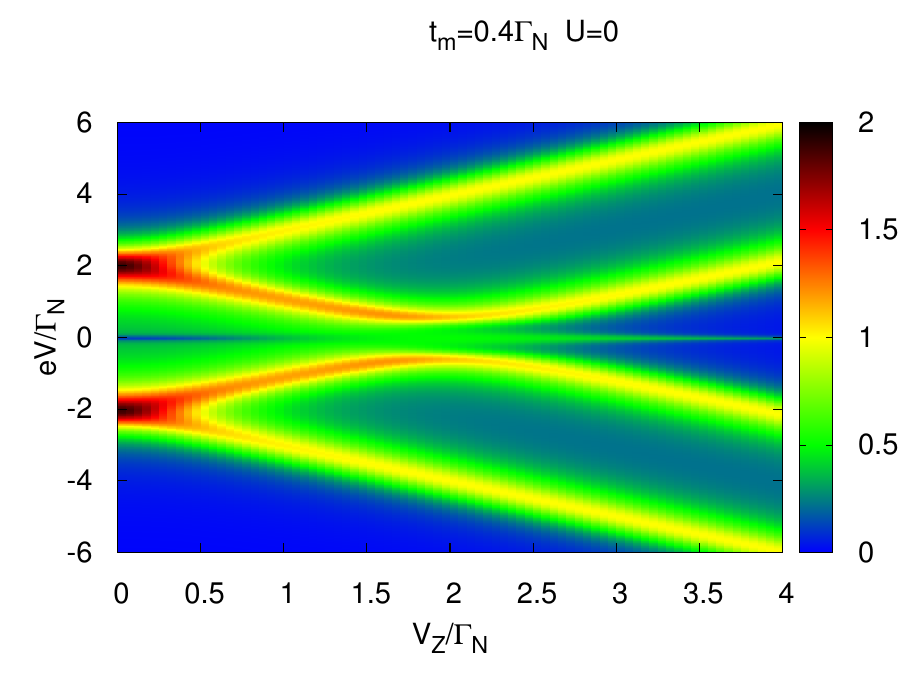}
\includegraphics[width=0.9\columnwidth]{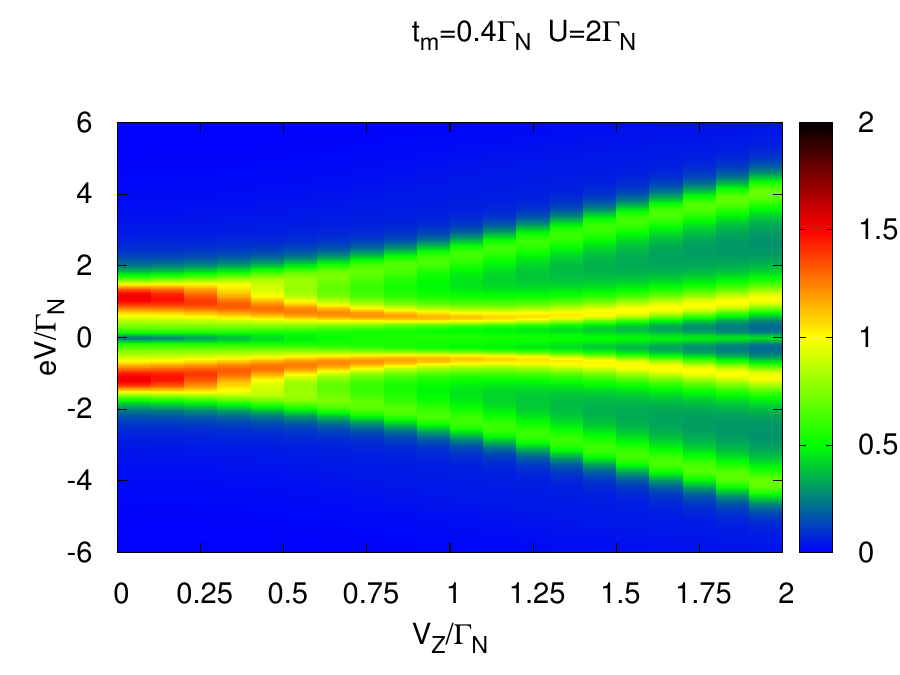}
\caption{\label{fig:figGVU05}
The differential Andreev conductance $G_{A}(V)$ [in units of $\frac{2e^{2}}{h}$] 
as function of voltage $V$ and Zeeman splitting $V_Z$ obtained at zero temperature, 
using $\Gamma_S=4\Gamma_N$, $U=0$ (top panel) and $U=2\Gamma_N$ (bottom panel).}
\end{figure}

To present this influence of magnetic field we focus on the strongly
asymmetric coupling $\Gamma_S=4\Gamma_N$, when the bound states are 
well pronounced (their broadening is then rather narrow). For brevity
we show in Fig. \ref{fig:figGVU05} the differential Andreev conductance 
as a function of Zeeman splitting energy $V_Z$ obtained for a moderate 
coupling $t_{m}=0.4\Gamma_{N}$ with Majorana mode. For the uncorrelated 
quantum dot (see the upper panel), we observe at $V_{Z}=0$ two  maxima 
(corresponding to the Andreev states energies) where the conductance 
is $G_A=4e^2/h$. Upon increasing the magnetic field they split and the 
internal branches start to approach, but they never cross  each other. 
At some critical magnetic field the internal branches are pushed aside, 
and simultaneously there emerges the  zero-bias conductance peak (ZBCP) 
signaling the leaking Majorana mode. Similar tendency (for the `normal'
quantum dots) has been reported recently by several groups 
\cite{Deng-2017b,Klinovaja-2018}. We have checked that in our setup 
this ZBCP achieves the optimal value near $V_Z\approx 
\Gamma_S/2$ and for stronger magnetic fields this zero-bias conductance
partly diminishes.

For the correlated case (bottom panel), the optimal differential conductance 
(which is $G_A<4e^2/h$) occurs at the renormalized Andreev energies. 
Again, the magnetic field has quite similar influence on $G_{A}(V)$. 
The emerging ZBCP shows at $2V_Z+U\approx \Gamma_S$. We thus conclude 
that, in our N-QD-SC system with the side-attached Majorana wire, 
the ZBCP can be induced above some critical magnetic field and
its value depends on the Coulomb potential $U$. Such appearance 
of the Majorana quasiparticles from the coalescing Andreev states 
is more exotic than in `normal' quantum dots coupled to the Majorana
modes. Influence of the magnetic field on the subgap Kondo effect
shall be discussed elsewhere, because there are a few characteristic
energy-scales controlling this effect.

\section{Majorana coupled to both spins}
\label{sec:polarization}

In realistic situations the spin-orbit coupling along with the Zeeman 
effect break a spin-rotational symmetry of the system. Formally, spin 
is hence no longer a good quantum number. Physically it means, that 
such nanowire with the strong spin-orbit interactions brought in contact 
with superconductor develops the intersite pairing of equal but `tilted'  
spins. Nevertheless one can project such pairing onto $\uparrow$ and 
$\downarrow$ components. In each of these sectors the intersite-pairing 
is characterized by different amplitudes, depending mainly on the magnetic 
field. In practice, the Majorana quasiparticles appear simultaneously in 
both spin channels, but with different spectral weights. This 
polarization of the Majorana modes has been recently considered by one 
of us \cite{MaskaDomanski-2017} and its experimental evidence has been 
indeed reported by A. Yazdani and coworkers from STM measurements 
of {\em Fe} atoms nanochain  using the ferromagnetic tip  
\cite{Yazdani-2017}.

Since magnetic polarization of the Majorana modes is in fact relevant, 
we have considered the coupling of both QD spins to Majorana modes
\begin{eqnarray}
H_{MQD} &=& t_{m\uparrow} (d^{\dagger}_{\uparrow} - d_{\uparrow}) 
( f + f^{\dagger} ) +t_{m\downarrow} (d^{\dagger}_{\downarrow} - d_{\downarrow}) 
( f + f^{\dagger} ) 
\nonumber \\
&+& \epsilon_m f^{\dagger} f + \frac{\epsilon_m}{2}  .
\end{eqnarray}
%
For the uncorrelated quantum dot the Green's function takes the following form
\begin{widetext}
\begin{eqnarray} 
{\cal{G}}^{-1}(\omega) =
\left( \begin{array}{cccccc}  
\omega-\epsilon_{\uparrow}+i\Gamma_N/2 &0&0& \Gamma_S/2 & -t_{m\uparrow} & -t_{m\uparrow}\\
0&\omega+\epsilon_{\uparrow}+i\Gamma_N/2&-\Gamma_S/2& 0 & t_{m\uparrow} & t_{m\uparrow}\\
0&-\Gamma_S/2&\omega-\epsilon_{\downarrow}+i\Gamma_N/2 &0 & -t_{m\downarrow} & -t_{m\downarrow}\\
\Gamma_S/2 &0&0&\omega+\epsilon_{\downarrow}+i\Gamma_N/2 & t_{m\downarrow} & t_{m\downarrow} \\
-t_{m\uparrow} & t_{m\uparrow} &-t_{m\downarrow} & t_{m\downarrow}& \omega-\epsilon_{m}& 0\\
-t_{m\uparrow} & t_{m\uparrow} &-t_{m\downarrow} & t_{m\downarrow}& 0    & \omega+\epsilon_{m} 
\end{array}\right),
\label{Gr66}
\end{eqnarray} 
\end{widetext}
For specific calculations, we have assumed imposed
\begin{eqnarray} 
t_{m\uparrow}&=&t_{m} \;\; p\nonumber \\
t_{m\downarrow}&=&t_{m} \;\; (1-p) \nonumber
\label{deftms}
\end{eqnarray}
where $p$ accounts for the asymmetric coupling of each spin component.
\begin{figure}[b]
\centering 
\includegraphics[width=0.9\columnwidth]{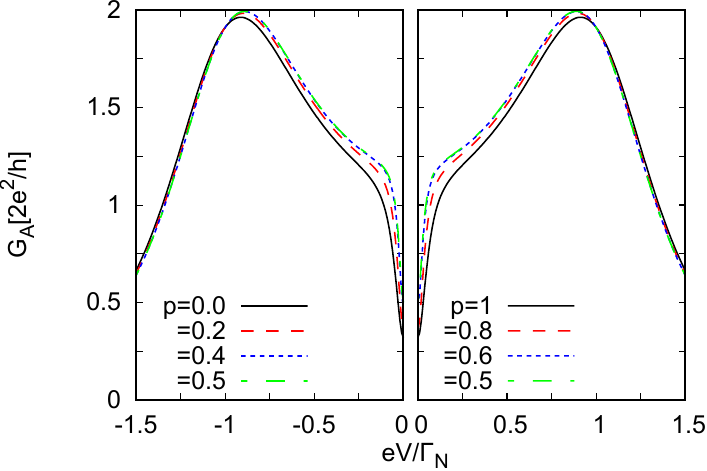}
\caption{\label{polarization}
The differential Andreev conductance $G_{A}(V)$ [in units of $\frac{2e^{2}}{h}$] 
as function of voltage $V$ obtained for the half-filled uncorrelated QD at zero 
temperature, using $\Gamma_S=2\Gamma_N$.}
\end{figure}
Fig.~\ref{polarization} shows the bias voltage dependence of the Andreev conductance 
$G_{A}(eV)$ obtained at zero temperature for several values of parameter $p$, as 
indicated. The differential conductance  is even with respect to 
$V=0$, so we present in the left h.s.\ panel results for $p<0.5$ and in 
the right h.s.\ panel for $p>0.5$, respectively. We notice that assymetry has 
rather negligible influence on the Andreev conductance (it slightly affects 
only a width of the Majorana interferrometric feature). This can be assigned 
to the fact that Andreev processes equally involve both spins, therefore 
$G_{A}(eV)$  symmetrizes the contributions of each sector. We thus conclude, 
that in our setup (Fig.\ \ref{schematics}) some influence of a polarized 
coupling to the Majorana modes would be rather residual.
More detailed anaylis of this problem (considering the correlations) 
is beyond a scope of this paper, thefore we shall discussed it elsewhere.

\section{Superconducting atomic limit}
\label{sec.QPT}

\begin{figure}
\centering
\includegraphics[width=0.63\linewidth]{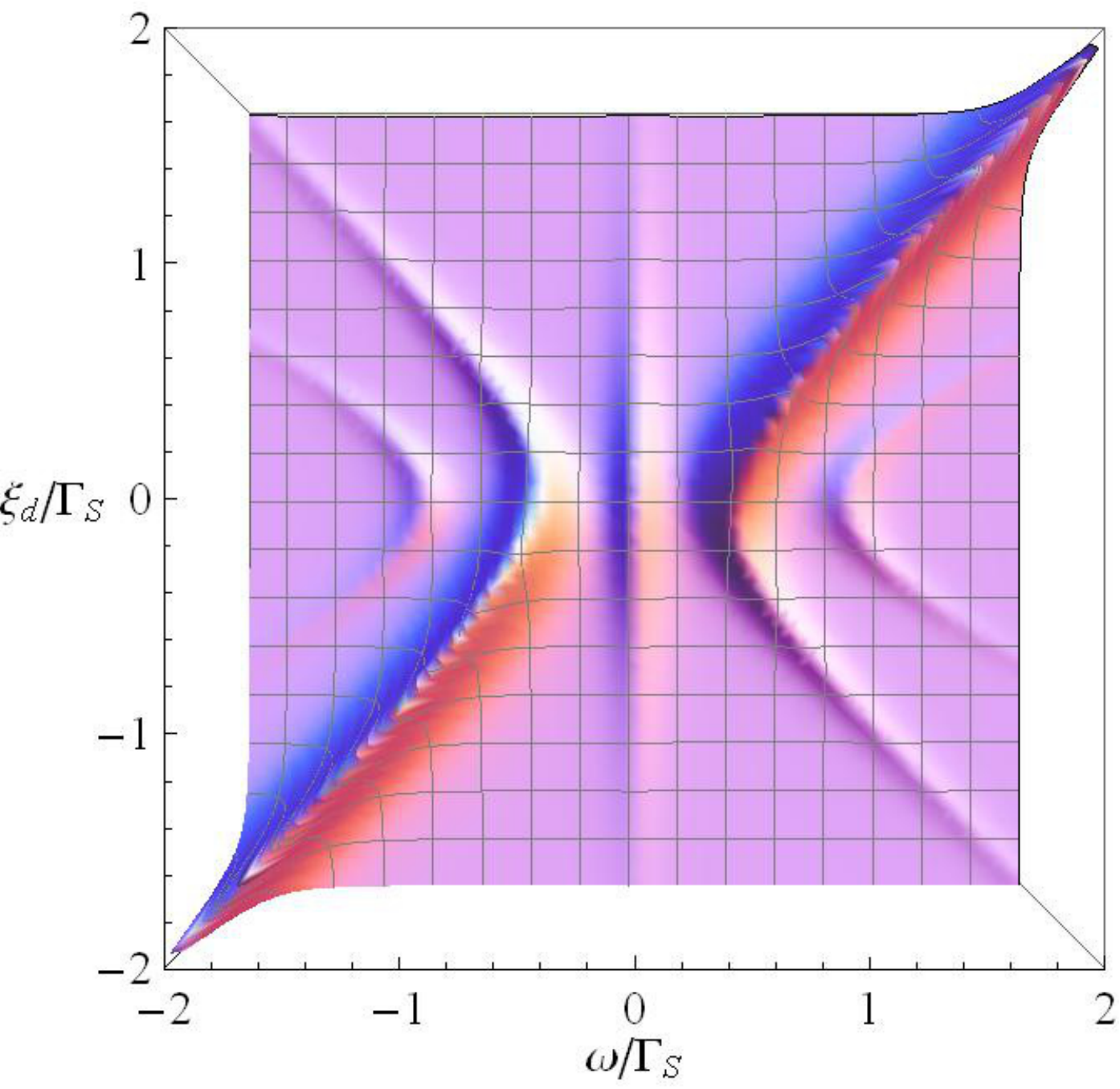}
\includegraphics[width=0.63\linewidth]{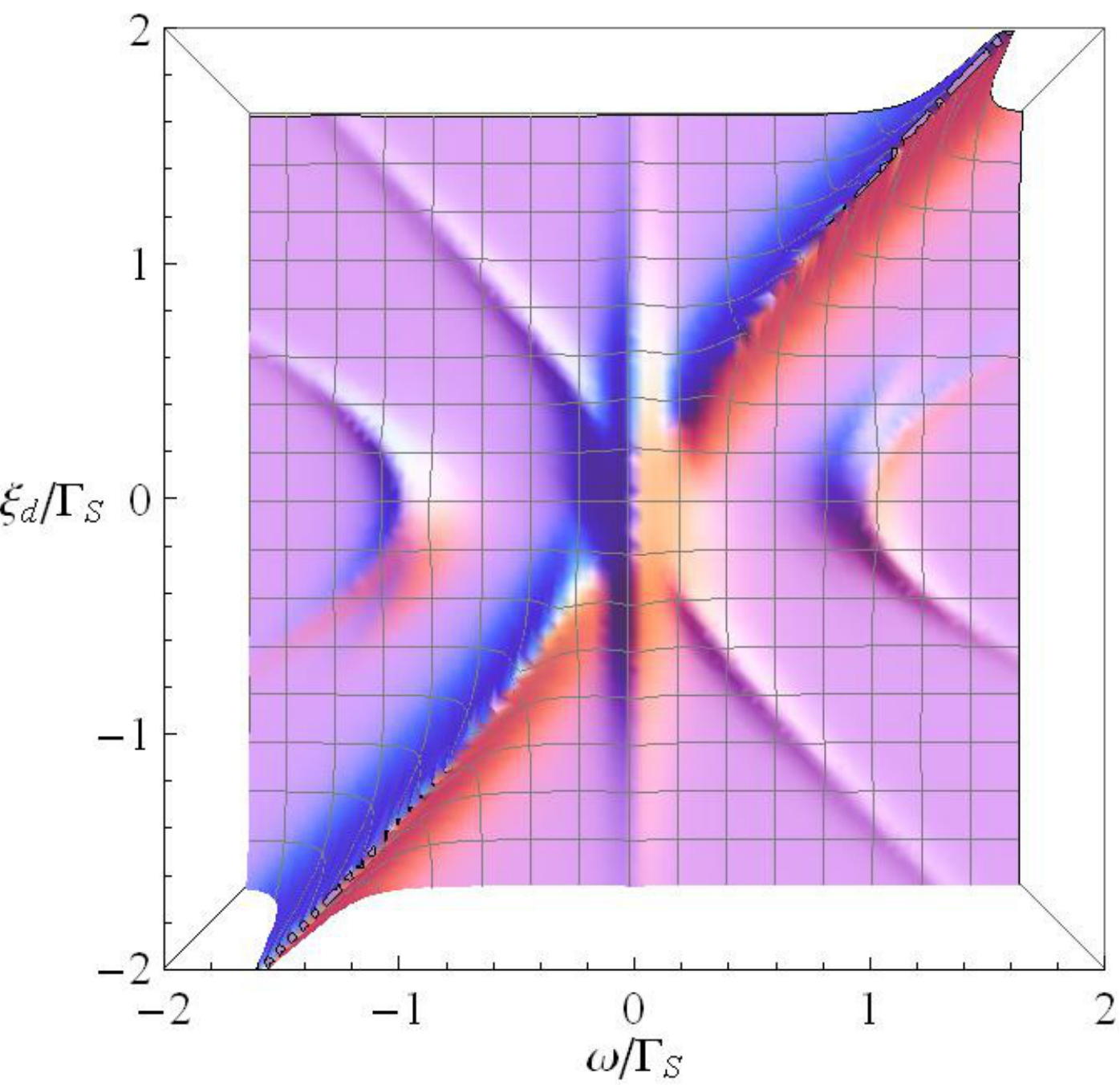}
\includegraphics[width=0.63\linewidth]{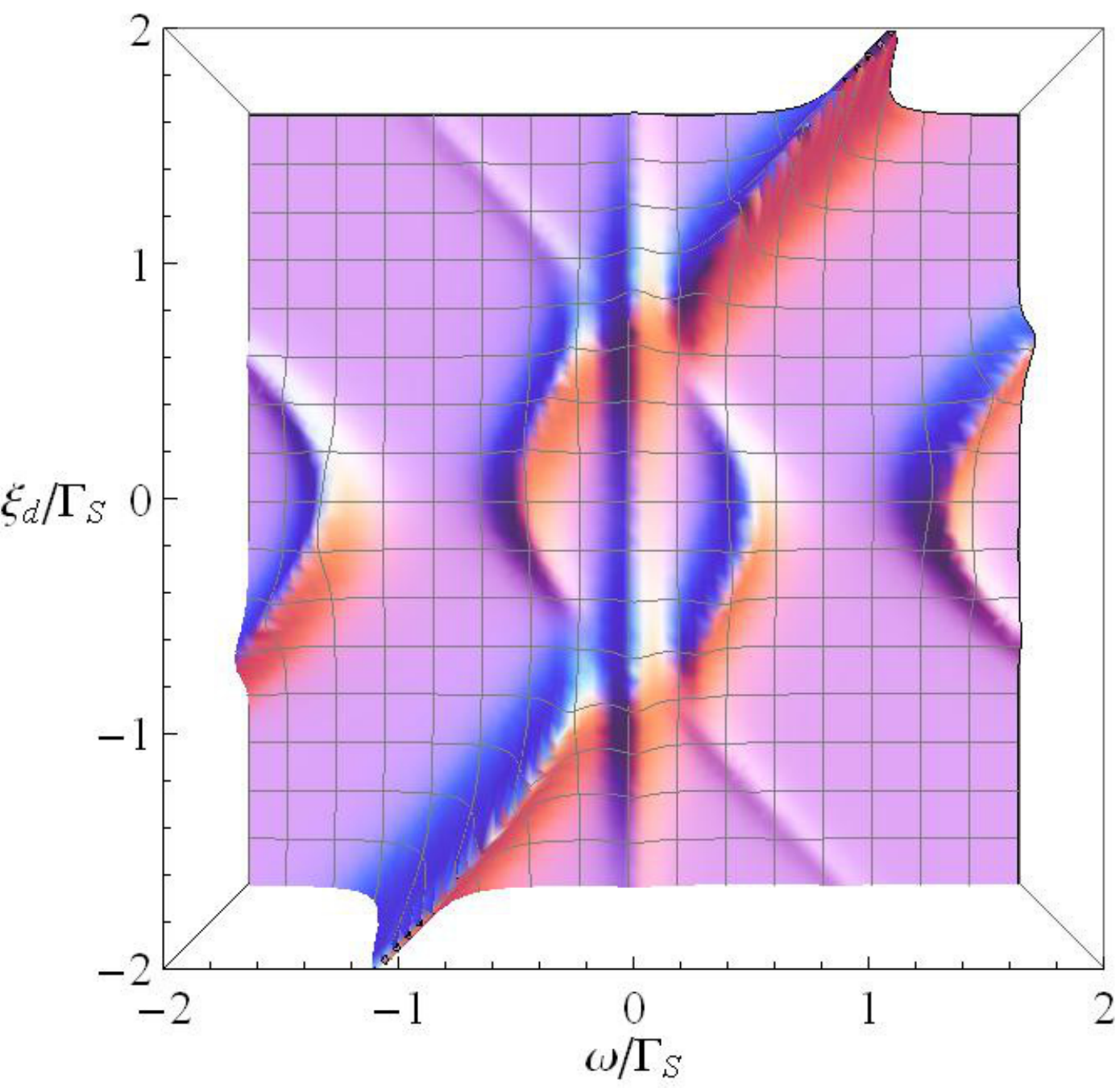}
\caption{The spectral function $A_{\uparrow}(\omega)$ of the correlated 
and proximitized QD obtained in the limit $\Gamma_{N}\rightarrow 0$, using 
$t_{m}/\Gamma_{S}=0.3$,  $U/\Gamma_{S}=0.05$ (top panel), $1$ (middle panel), 
and $2$ (bottom panel). Calculations have been done by numerical diagonalization 
of the Hamiltonian  (\ref{sc_at_lim}) subject to a small line broadening 
(that mimics finite life-time).}
\label{janek}
\end{figure}

To gain some insight into the singlet-doublet quantum phase transition 
(discussed in main part of this paper), let us first consider the limit 
$\Gamma_{S}\gg\Gamma_{N}$, assuming $\Gamma_{N}=0^{+}$. 
Influence of the Coulomb repulsion $U$ on the subgap Andreev states 
in such `superconducting atomic limit' has been first addressed by 
E.\ Vecino {\em et al} \cite{Vecino-2003}. In the absence of the Majorana
modes the effective bound states of such problem have been discussed 
in the review paper \cite{Rodero-2012}. For the present setup  we 
consider the effective Hamiltonian
\begin{eqnarray}
H^{\rm eff}_{QD} &\simeq & \sum_{\sigma} \epsilon d^{\dagger}_{\sigma} d_{\sigma}
+U n_{\downarrow}n_{\uparrow} - \frac{\Gamma_S}{2}(d_{\uparrow} 
d_{\downarrow}+d^{\dagger}_{\downarrow} d^{\dagger}_{\uparrow}) 
\nonumber \\
&+& t_m (d^{\dagger}_{\uparrow} - d_{\uparrow}) ( f + f^{\dagger} ) 
+ \epsilon_m \left( f^{\dagger} f 
+ \frac{1}{2} \right) .
\label{sc_at_lim}
\end{eqnarray}
Following Ref. \cite{Bauer-2007}
let us introduce the shorthand notations for the induced on-dot pairing 
$\Delta_{d} \equiv \frac{\Gamma_{S}}{2}$  and for the shifted QD level 
$\xi_{d} \equiv \epsilon + \frac{U}{2}$, respectively.

In the absence of the Majorana mode ($t_{m}=0$), the true eigenstates are
represented by the doublet configurations $\left| \uparrow\right>$ and 
$\left| \downarrow \right>$ (corresponding to spin $S=\frac{1}{2}$) 
and the BCS-like singlet states ($S=0$) 
\begin{eqnarray}
\left| \Psi_{-} \right> & = & u_{d} \left| 0 \right> - v_{d} \left|
\uparrow \downarrow \right> ,\\
\left| \Psi_{+} \right> & = & v_{d} \left| 0 \right> + u_{d} \left|
\uparrow \downarrow \right> .
\end{eqnarray}
with eigenenergies 
$E_{\mp}=\xi_{d} \mp \sqrt{\xi_{d}^{2}+\Delta_{d}^{2}}$
and BCS coefficients 
$u_{d}^{2} = \frac{1}{2} \left[ 1 + \frac{\xi_{d}}
{E_{d}} \right] =1 -v_{d}^{2}$,
where $E_{d}=\sqrt{\xi_{d}^{2}+\Delta_{d}^{2}}$.
In the presence of the side-coupled Majorana mode(s), we have to extend 
the Hilbert space by additional fermion state $f$ (that can be either 
empty or occupied). We have numerically diagonalized the Hamiltonian 
(\ref{sc_at_lim}) in the representation $\left| n_{d\uparrow} \right> 
\otimes \left| n_{d\downarrow} \right> \otimes \left| n_{f} \right>$,
determining the eigenvalues and eigenvectors. Using the Lehmann 
representation we have computed the spectral functions
$A_{\sigma}(\omega)$.

\begin{figure*}
\centering
\includegraphics[width=0.95\linewidth]{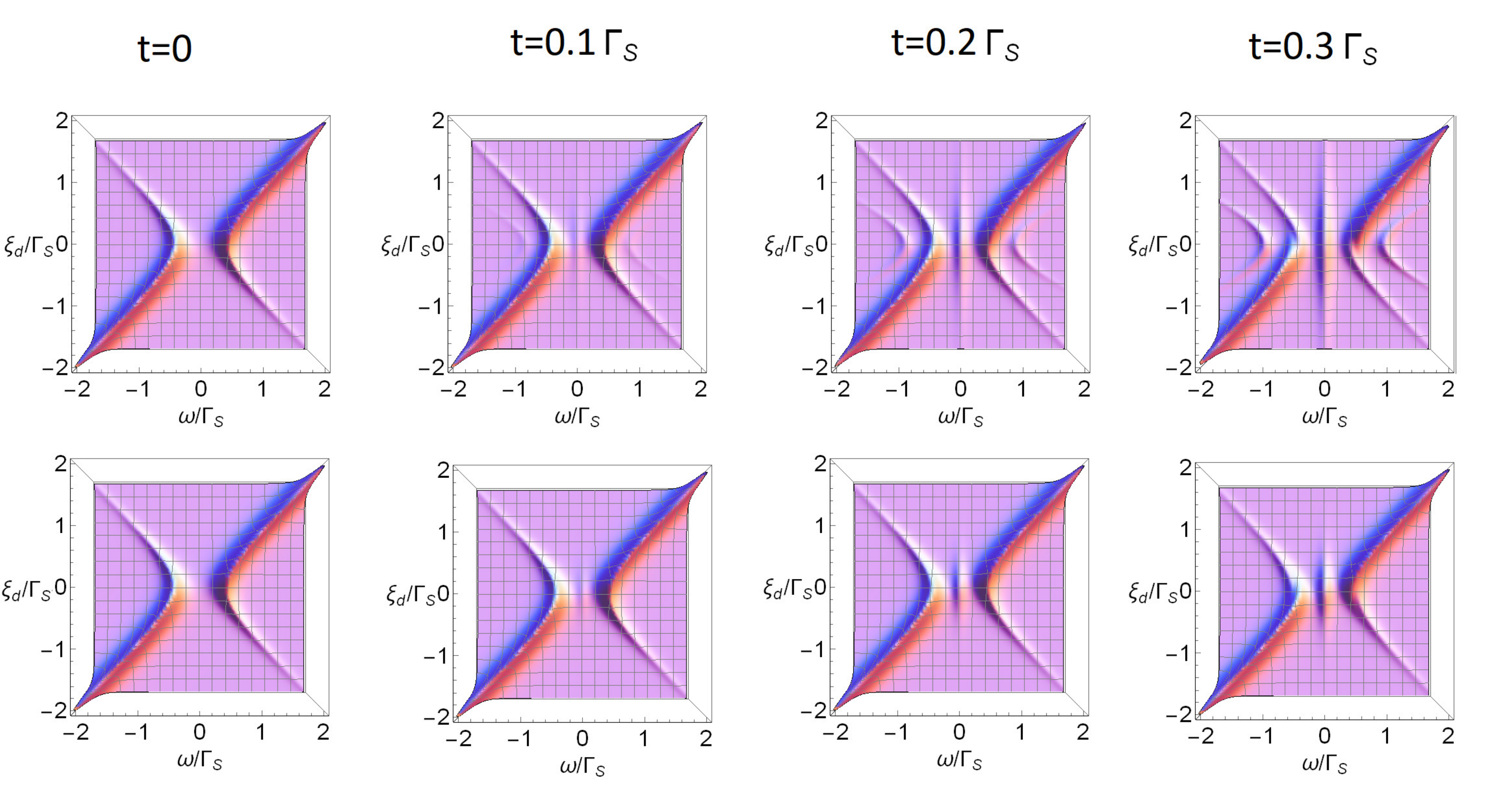}
\caption{Comparison of the spectral functions $A_{\sigma}(\omega)$
for $\uparrow$ (upper row) and $\downarrow$ electrons (bottom row) 
in the superconducting atomic limit obtained for $U/\Gamma_{S}=0.05$,
$\epsilon_{m}=0$ and several couplings $t_{m}$ (as indicated).
}
\label{both_spins}
\end{figure*}

Figure \ref{janek} shows the QD spectrum for three representative values 
of the Coulomb potential. In the weak interaction limit (top panel)
the spectrum exhibits the non-crossing Andreev quasiparticle branches 
coexisting with the zero-energy mode. For $U=\Gamma_{S}$, we observe 
a tendency towards avoided crossing of the Andreev bound states at 
$\xi_{d} \sim 0$ ($\epsilon \sim -\frac{U}{2}$).  In the strong 
correlation case $U=2\Gamma_{S}$ (bottom panel), there appear two such 
(avoided crossing) points aside from the half-filling \cite{Bauer-2007}. 
In all cases there exist the zero-energy quasiparticle state, although
its spectral weight is largest nearby these avoided-crossing points,
corresponding to the quantum phase transition from the  (spinfull) 
doublet to the (spinless) BCS-type configurations. This avoidance 
instead of true crossing is a hallmark of the states' hybridization
with the Majorana mode \cite{Prada-2017,Deng-2017b,Klinovaja-2018}.

Finally let us check, if the leaking Majorana mode can affect 
the opposite $\downarrow$ spin of the proximized correlated quantum 
dot. We compare the spectra of both spins in Fig.\ \ref{both_spins}. 
To be specific, we focus on the weak correlation limit ($\Gamma_{S}\gg U$)
when the pairing effects are most efficient. For $t_{m}=0$, both spectra 
are obviously identical and they are characterized by two gapped Andreev 
quasiparticle branches. With increasing $t_{m}$, we observe signatures of 
the leaking Majorana mode in both spins, with dominance in $\uparrow$ 
sector. Tiny feature of the zero-energy mode shows up in the spin 
$\downarrow$ sector only near $\xi_{d}\sim 0$, where the on-dot pairing 
between both spins is the most efficient. Another difference between 
the spin-resolved spectra is observed at large couplings $t_{m}$, 
where a number of the Andreev branches doubles for $\uparrow$ sector 
(due to the bonding/antibonding states formed in the `molecular' 
limit \cite{Baranski-2017}) whereas such effect is absent for 
$\downarrow$ electrons.

\bibliography{myBib}

\begin{thebibliography}{83}%
\makeatletter
\providecommand \@ifxundefined [1]{%
 \@ifx{#1\undefined}
}%
\providecommand \@ifnum [1]{%
 \ifnum #1\expandafter \@firstoftwo
 \else \expandafter \@secondoftwo
 \fi
}%
\providecommand \@ifx [1]{%
 \ifx #1\expandafter \@firstoftwo
 \else \expandafter \@secondoftwo
 \fi
}%
\providecommand \natexlab [1]{#1}%
\providecommand \enquote  [1]{``#1''}%
\providecommand \bibnamefont  [1]{#1}%
\providecommand \bibfnamefont [1]{#1}%
\providecommand \citenamefont [1]{#1}%
\providecommand \href@noop [0]{\@secondoftwo}%
\providecommand \href [0]{\begingroup \@sanitize@url \@href}%
\providecommand \@href[1]{\@@startlink{#1}\@@href}%
\providecommand \@@href[1]{\endgroup#1\@@endlink}%
\providecommand \@sanitize@url [0]{\catcode `\\12\catcode `\$12\catcode
  `\&12\catcode `\#12\catcode `\^12\catcode `\_12\catcode `\%12\relax}%
\providecommand \@@startlink[1]{}%
\providecommand \@@endlink[0]{}%
\providecommand \url  [0]{\begingroup\@sanitize@url \@url }%
\providecommand \@url [1]{\endgroup\@href {#1}{\urlprefix }}%
\providecommand \urlprefix  [0]{URL }%
\providecommand \Eprint [0]{\href }%
\providecommand \doibase [0]{http://dx.doi.org/}%
\providecommand \selectlanguage [0]{\@gobble}%
\providecommand \bibinfo  [0]{\@secondoftwo}%
\providecommand \bibfield  [0]{\@secondoftwo}%
\providecommand \translation [1]{[#1]}%
\providecommand \BibitemOpen [0]{}%
\providecommand \bibitemStop [0]{}%
\providecommand \bibitemNoStop [0]{.\EOS\space}%
\providecommand \EOS [0]{\spacefactor3000\relax}%
\providecommand \BibitemShut  [1]{\csname bibitem#1\endcsname}%
\let\auto@bib@innerbib\@empty
\bibitem [{\citenamefont {Alicea}(2012)}]{Alicea-12}%
  \BibitemOpen
  \bibfield  {author} {\bibinfo {author} {\bibfnamefont {J.}~\bibnamefont
  {Alicea}},\ }\bibfield  {title} {\enquote {\bibinfo {title} {New directions
  in the pursuit of {Majorana} fermions in solid state systems},}\ }\href
  {http://stacks.iop.org/0034-4885/75/i=7/a=076501} {\bibfield  {journal}
  {\bibinfo  {journal} {Rep. Prog. Phys.}\ }\textbf {\bibinfo {volume} {75}},\
  \bibinfo {pages} {076501} (\bibinfo {year} {2012})}\BibitemShut {NoStop}%
\bibitem [{\citenamefont {Leijnse}\ and\ \citenamefont
  {Flensberg}(2012)}]{Flensberg-12}%
  \BibitemOpen
  \bibfield  {author} {\bibinfo {author} {\bibfnamefont {M.}~\bibnamefont
  {Leijnse}}\ and\ \bibinfo {author} {\bibfnamefont {K.}~\bibnamefont
  {Flensberg}},\ }\bibfield  {title} {\enquote {\bibinfo {title} {Introduction
  to topological superconductivity and {Majorana} fermions},}\ }\href
  {http://stacks.iop.org/0268-1242/27/i=12/a=124003} {\bibfield  {journal}
  {\bibinfo  {journal} {Semicond. Sci. Technol.}\ }\textbf {\bibinfo {volume}
  {27}},\ \bibinfo {pages} {124003} (\bibinfo {year} {2012})}\BibitemShut
  {NoStop}%
\bibitem [{\citenamefont {Stanescu}\ and\ \citenamefont
  {Tewari}(2013)}]{Stanescu-13}%
  \BibitemOpen
  \bibfield  {author} {\bibinfo {author} {\bibfnamefont {T.~D.}\ \bibnamefont
  {Stanescu}}\ and\ \bibinfo {author} {\bibfnamefont {S.}~\bibnamefont
  {Tewari}},\ }\bibfield  {title} {\enquote {\bibinfo {title} {Majorana
  fermions in semiconductor nanowires: fundamentals, modeling, and
  experiment},}\ }\href {http://stacks.iop.org/0953-8984/25/i=23/a=233201}
  {\bibfield  {journal} {\bibinfo  {journal} {J. Phys.: Condens. Matter}\
  }\textbf {\bibinfo {volume} {25}},\ \bibinfo {pages} {233201} (\bibinfo
  {year} {2013})}\BibitemShut {NoStop}%
\bibitem [{\citenamefont {Beenakker}(2013)}]{Beenakker-13}%
  \BibitemOpen
  \bibfield  {author} {\bibinfo {author} {\bibfnamefont {C.~W.~J.}\
  \bibnamefont {Beenakker}},\ }\bibfield  {title} {\enquote {\bibinfo {title}
  {Search for {Majorana} fermions in superconductors},}\ }\href {\doibase
  10.1146/annurev-conmatphys-030212-184337} {\bibfield  {journal} {\bibinfo
  {journal} {Annu. Rev. Condens. Matt. Phys.}\ }\textbf {\bibinfo {volume}
  {4}},\ \bibinfo {pages} {113} (\bibinfo {year} {2013})}\BibitemShut {NoStop}%
\bibitem [{\citenamefont {Elliott}\ and\ \citenamefont
  {Franz}(2015)}]{Franz-15}%
  \BibitemOpen
  \bibfield  {author} {\bibinfo {author} {\bibfnamefont {S.~R.}\ \bibnamefont
  {Elliott}}\ and\ \bibinfo {author} {\bibfnamefont {M.}~\bibnamefont
  {Franz}},\ }\bibfield  {title} {\enquote {\bibinfo {title} {Colloquium:
  {Majorana} fermions in nuclear, particle, and solid-state physics},}\ }\href
  {\doibase 10.1103/RevModPhys.87.137} {\bibfield  {journal} {\bibinfo
  {journal} {Rev. Mod. Phys.}\ }\textbf {\bibinfo {volume} {87}},\ \bibinfo
  {pages} {137} (\bibinfo {year} {2015})}\BibitemShut {NoStop}%
\bibitem [{\citenamefont {Aguado}(2017)}]{Aguado-2017}%
  \BibitemOpen
  \bibfield  {author} {\bibinfo {author} {\bibfnamefont {R.}~\bibnamefont
  {Aguado}},\ }\bibfield  {title} {\enquote {\bibinfo {title} {Majorana
  quasiparticles in condensed matter},}\ }\href {\doibase
  10.1393/ncr/i2017-10141-9} {\bibfield  {journal} {\bibinfo  {journal} {Riv.
  Nuovo Cimento}\ }\textbf {\bibinfo {volume} {040}},\ \bibinfo {pages} {523}
  (\bibinfo {year} {2017})}\BibitemShut {NoStop}%
\bibitem [{\citenamefont {Lutchyn}\ \emph {et~al.}(2017)\citenamefont
  {Lutchyn}, \citenamefont {Bakkers}, \citenamefont {Kouwenhoven},
  \citenamefont {Krogstrup}, \citenamefont {Marcus},\ and\ \citenamefont
  {Oreg}}]{Lutchyn-2017}%
  \BibitemOpen
  \bibfield  {author} {\bibinfo {author} {\bibfnamefont {R.M.}\ \bibnamefont
  {Lutchyn}}, \bibinfo {author} {\bibfnamefont {E.P.A.M.}\ \bibnamefont
  {Bakkers}}, \bibinfo {author} {\bibfnamefont {L.P.}\ \bibnamefont
  {Kouwenhoven}}, \bibinfo {author} {\bibfnamefont {P.}~\bibnamefont
  {Krogstrup}}, \bibinfo {author} {\bibfnamefont {C.M.}\ \bibnamefont
  {Marcus}}, \ and\ \bibinfo {author} {\bibfnamefont {Y.}~\bibnamefont
  {Oreg}},\ }\href@noop {} {\enquote {\bibinfo {title} {Realizing {M}ajorana
  zero modes in superconductor-semiconductor heterostructures},}\ } (\bibinfo
  {year} {2017}),\ \Eprint {http://arxiv.org/abs/arXiv:1707.04899}
  {arXiv:1707.04899} \BibitemShut {NoStop}%
\bibitem [{\citenamefont {Volovik}(1999)}]{Volovik-1999}%
  \BibitemOpen
  \bibfield  {author} {\bibinfo {author} {\bibfnamefont {G.~E.}\ \bibnamefont
  {Volovik}},\ }\bibfield  {title} {\enquote {\bibinfo {title} {Fermion zero
  modes on vortices in chiral superconductors},}\ }\href {\doibase
  10.1134/1.568223} {\bibfield  {journal} {\bibinfo  {journal} {JETP Lett.}\
  }\textbf {\bibinfo {volume} {70}},\ \bibinfo {pages} {609} (\bibinfo {year}
  {1999})}\BibitemShut {NoStop}%
\bibitem [{\citenamefont {Read}\ and\ \citenamefont {Green}(2000)}]{Read-2000}%
  \BibitemOpen
  \bibfield  {author} {\bibinfo {author} {\bibfnamefont {N.}~\bibnamefont
  {Read}}\ and\ \bibinfo {author} {\bibfnamefont {D.}~\bibnamefont {Green}},\
  }\bibfield  {title} {\enquote {\bibinfo {title} {Paired states of fermions in
  two dimensions with breaking of parity and time-reversal symmetries and the
  fractional quantum {Hall} effect},}\ }\href {\doibase
  10.1103/PhysRevB.61.10267} {\bibfield  {journal} {\bibinfo  {journal} {Phys.
  Rev. B}\ }\textbf {\bibinfo {volume} {61}},\ \bibinfo {pages} {10267}
  (\bibinfo {year} {2000})}\BibitemShut {NoStop}%
\bibitem [{\citenamefont {Kitaev}(2001)}]{Kitaev-2001}%
  \BibitemOpen
  \bibfield  {author} {\bibinfo {author} {\bibfnamefont {A.~Y.}\ \bibnamefont
  {Kitaev}},\ }\bibfield  {title} {\enquote {\bibinfo {title} {Unpaired
  {Majorana} fermions in quantum wires},}\ }\href
  {http://stacks.iop.org/1063-7869/44/i=10S/a=S29} {\bibfield  {journal}
  {\bibinfo  {journal} {Phys. Usp.}\ }\textbf {\bibinfo {volume} {44}},\
  \bibinfo {pages} {131} (\bibinfo {year} {2001})}\BibitemShut {NoStop}%
\bibitem [{\citenamefont {Liu}\ \emph {et~al.}(2016)\citenamefont {Liu},
  \citenamefont {Li}, \citenamefont {Deng}, \citenamefont {Liu},\ and\
  \citenamefont {Das~Sarma}}]{DasSarma-2016}%
  \BibitemOpen
  \bibfield  {author} {\bibinfo {author} {\bibfnamefont {X.}~\bibnamefont
  {Liu}}, \bibinfo {author} {\bibfnamefont {X.}~\bibnamefont {Li}}, \bibinfo
  {author} {\bibfnamefont {D.-L.}\ \bibnamefont {Deng}}, \bibinfo {author}
  {\bibfnamefont {X.-J.}\ \bibnamefont {Liu}}, \ and\ \bibinfo {author}
  {\bibfnamefont {S.}~\bibnamefont {Das~Sarma}},\ }\bibfield  {title} {\enquote
  {\bibinfo {title} {Majorana spintronics},}\ }\href {\doibase
  10.1103/PhysRevB.94.014511} {\bibfield  {journal} {\bibinfo  {journal} {Phys.
  Rev. B}\ }\textbf {\bibinfo {volume} {94}},\ \bibinfo {pages} {014511}
  (\bibinfo {year} {2016})}\BibitemShut {NoStop}%
\bibitem [{\citenamefont {Tewari}\ \emph {et~al.}(2007)\citenamefont {Tewari},
  \citenamefont {Das~Sarma}, \citenamefont {Nayak}, \citenamefont {Zhang},\
  and\ \citenamefont {Zoller}}]{Tewari-2007}%
  \BibitemOpen
  \bibfield  {author} {\bibinfo {author} {\bibfnamefont {S.}~\bibnamefont
  {Tewari}}, \bibinfo {author} {\bibfnamefont {S.}~\bibnamefont {Das~Sarma}},
  \bibinfo {author} {\bibfnamefont {C.}~\bibnamefont {Nayak}}, \bibinfo
  {author} {\bibfnamefont {C.}~\bibnamefont {Zhang}}, \ and\ \bibinfo {author}
  {\bibfnamefont {P.}~\bibnamefont {Zoller}},\ }\bibfield  {title} {\enquote
  {\bibinfo {title} {Quantum computation using vortices and {Majorana} zero
  modes of a ${p}_{x}+i{p}_{y}$ superfluid of fermionic cold atoms},}\ }\href
  {\doibase 10.1103/PhysRevLett.98.010506} {\bibfield  {journal} {\bibinfo
  {journal} {Phys. Rev. Lett.}\ }\textbf {\bibinfo {volume} {98}},\ \bibinfo
  {pages} {010506} (\bibinfo {year} {2007})}\BibitemShut {NoStop}%
\bibitem [{\citenamefont {Fu}\ and\ \citenamefont {Kane}(2008)}]{Fu-2008}%
  \BibitemOpen
  \bibfield  {author} {\bibinfo {author} {\bibfnamefont {L.}~\bibnamefont
  {Fu}}\ and\ \bibinfo {author} {\bibfnamefont {C.~L.}\ \bibnamefont {Kane}},\
  }\bibfield  {title} {\enquote {\bibinfo {title} {Superconducting proximity
  effect and {Majorana} fermions at the surface of a topological insulator},}\
  }\href {\doibase 10.1103/PhysRevLett.100.096407} {\bibfield  {journal}
  {\bibinfo  {journal} {Phys. Rev. Lett.}\ }\textbf {\bibinfo {volume} {100}},\
  \bibinfo {pages} {096407} (\bibinfo {year} {2008})}\BibitemShut {NoStop}%
\bibitem [{\citenamefont {Nilsson}\ \emph {et~al.}(2008)\citenamefont
  {Nilsson}, \citenamefont {Akhmerov},\ and\ \citenamefont
  {Beenakker}}]{Nilsson-2008}%
  \BibitemOpen
  \bibfield  {author} {\bibinfo {author} {\bibfnamefont {J.}~\bibnamefont
  {Nilsson}}, \bibinfo {author} {\bibfnamefont {A.~R.}\ \bibnamefont
  {Akhmerov}}, \ and\ \bibinfo {author} {\bibfnamefont {C.~W.~J.}\ \bibnamefont
  {Beenakker}},\ }\bibfield  {title} {\enquote {\bibinfo {title} {Splitting of
  a {Cooper} pair by a pair of {Majorana} bound states},}\ }\href {\doibase
  10.1103/PhysRevLett.101.120403} {\bibfield  {journal} {\bibinfo  {journal}
  {Phys. Rev. Lett.}\ }\textbf {\bibinfo {volume} {101}},\ \bibinfo {pages}
  {120403} (\bibinfo {year} {2008})}\BibitemShut {NoStop}%
\bibitem [{\citenamefont {Sato}\ and\ \citenamefont
  {Fujimoto}(2009)}]{Sato-2009}%
  \BibitemOpen
  \bibfield  {author} {\bibinfo {author} {\bibfnamefont {M.}~\bibnamefont
  {Sato}}\ and\ \bibinfo {author} {\bibfnamefont {S.}~\bibnamefont
  {Fujimoto}},\ }\bibfield  {title} {\enquote {\bibinfo {title} {Topological
  phases of noncentrosymmetric superconductors: Edge states, {Majorana}
  fermions, and non-{Abelian} statistics},}\ }\href {\doibase
  10.1103/PhysRevB.79.094504} {\bibfield  {journal} {\bibinfo  {journal} {Phys.
  Rev. B}\ }\textbf {\bibinfo {volume} {79}},\ \bibinfo {pages} {094504}
  (\bibinfo {year} {2009})}\BibitemShut {NoStop}%
\bibitem [{\citenamefont {Wimmer}\ \emph {et~al.}(2010)\citenamefont {Wimmer},
  \citenamefont {Akhmerov}, \citenamefont {Medvedyeva}, \citenamefont
  {Tworzyd\l{}o},\ and\ \citenamefont {Beenakker}}]{Tworzydlo-2010}%
  \BibitemOpen
  \bibfield  {author} {\bibinfo {author} {\bibfnamefont {M.}~\bibnamefont
  {Wimmer}}, \bibinfo {author} {\bibfnamefont {A.~R.}\ \bibnamefont
  {Akhmerov}}, \bibinfo {author} {\bibfnamefont {M.~V.}\ \bibnamefont
  {Medvedyeva}}, \bibinfo {author} {\bibfnamefont {J.}~\bibnamefont
  {Tworzyd\l{}o}}, \ and\ \bibinfo {author} {\bibfnamefont {C.~W.~J.}\
  \bibnamefont {Beenakker}},\ }\bibfield  {title} {\enquote {\bibinfo {title}
  {Majorana bound states without vortices in topological superconductors with
  electrostatic defects},}\ }\href {\doibase 10.1103/PhysRevLett.105.046803}
  {\bibfield  {journal} {\bibinfo  {journal} {Phys. Rev. Lett.}\ }\textbf
  {\bibinfo {volume} {105}},\ \bibinfo {pages} {046803} (\bibinfo {year}
  {2010})}\BibitemShut {NoStop}%
\bibitem [{\citenamefont {Sau}\ \emph {et~al.}(2010)\citenamefont {Sau},
  \citenamefont {Lutchyn}, \citenamefont {Tewari},\ and\ \citenamefont
  {Das~Sarma}}]{Sau-2010}%
  \BibitemOpen
  \bibfield  {author} {\bibinfo {author} {\bibfnamefont {J.~D.}\ \bibnamefont
  {Sau}}, \bibinfo {author} {\bibfnamefont {R.~M.}\ \bibnamefont {Lutchyn}},
  \bibinfo {author} {\bibfnamefont {S.}~\bibnamefont {Tewari}}, \ and\ \bibinfo
  {author} {\bibfnamefont {S.}~\bibnamefont {Das~Sarma}},\ }\bibfield  {title}
  {\enquote {\bibinfo {title} {Generic new platform for topological quantum
  computation using semiconductor heterostructures},}\ }\href {\doibase
  10.1103/PhysRevLett.104.040502} {\bibfield  {journal} {\bibinfo  {journal}
  {Phys. Rev. Lett.}\ }\textbf {\bibinfo {volume} {104}},\ \bibinfo {pages}
  {040502} (\bibinfo {year} {2010})}\BibitemShut {NoStop}%
\bibitem [{\citenamefont {Oreg}\ \emph {et~al.}(2010)\citenamefont {Oreg},
  \citenamefont {Refael},\ and\ \citenamefont {von Oppen}}]{Oreg-2010}%
  \BibitemOpen
  \bibfield  {author} {\bibinfo {author} {\bibfnamefont {Y.}~\bibnamefont
  {Oreg}}, \bibinfo {author} {\bibfnamefont {G.}~\bibnamefont {Refael}}, \ and\
  \bibinfo {author} {\bibfnamefont {F.}~\bibnamefont {von Oppen}},\ }\bibfield
  {title} {\enquote {\bibinfo {title} {Helical liquids and {Majorana} bound
  states in quantum wires},}\ }\href {\doibase 10.1103/PhysRevLett.105.177002}
  {\bibfield  {journal} {\bibinfo  {journal} {Phys. Rev. Lett.}\ }\textbf
  {\bibinfo {volume} {105}},\ \bibinfo {pages} {177002} (\bibinfo {year}
  {2010})}\BibitemShut {NoStop}%
\bibitem [{\citenamefont {Lutchyn}\ \emph {et~al.}(2010)\citenamefont
  {Lutchyn}, \citenamefont {Sau},\ and\ \citenamefont
  {Das~Sarma}}]{Lutchyn-2010}%
  \BibitemOpen
  \bibfield  {author} {\bibinfo {author} {\bibfnamefont {R.~M.}\ \bibnamefont
  {Lutchyn}}, \bibinfo {author} {\bibfnamefont {J.~D.}\ \bibnamefont {Sau}}, \
  and\ \bibinfo {author} {\bibfnamefont {S.}~\bibnamefont {Das~Sarma}},\
  }\bibfield  {title} {\enquote {\bibinfo {title} {Majorana fermions and a
  topological phase transition in semiconductor-superconductor
  heterostructures},}\ }\href {\doibase 10.1103/PhysRevLett.105.077001}
  {\bibfield  {journal} {\bibinfo  {journal} {Phys. Rev. Lett.}\ }\textbf
  {\bibinfo {volume} {105}},\ \bibinfo {pages} {077001} (\bibinfo {year}
  {2010})}\BibitemShut {NoStop}%
\bibitem [{\citenamefont {Choy}\ \emph {et~al.}(2011)\citenamefont {Choy},
  \citenamefont {Edge}, \citenamefont {Akhmerov},\ and\ \citenamefont
  {Beenakker}}]{Choy-2011}%
  \BibitemOpen
  \bibfield  {author} {\bibinfo {author} {\bibfnamefont {T.-P.}\ \bibnamefont
  {Choy}}, \bibinfo {author} {\bibfnamefont {J.~M.}\ \bibnamefont {Edge}},
  \bibinfo {author} {\bibfnamefont {A.~R.}\ \bibnamefont {Akhmerov}}, \ and\
  \bibinfo {author} {\bibfnamefont {C.~W.~J.}\ \bibnamefont {Beenakker}},\
  }\bibfield  {title} {\enquote {\bibinfo {title} {Majorana fermions emerging
  from magnetic nanoparticles on a superconductor without spin-orbit
  coupling},}\ }\href {\doibase 10.1103/PhysRevB.84.195442} {\bibfield
  {journal} {\bibinfo  {journal} {Phys. Rev. B}\ }\textbf {\bibinfo {volume}
  {84}},\ \bibinfo {pages} {195442} (\bibinfo {year} {2011})}\BibitemShut
  {NoStop}%
\bibitem [{\citenamefont {Jiang}\ \emph {et~al.}(2011)\citenamefont {Jiang},
  \citenamefont {Kitagawa}, \citenamefont {Alicea}, \citenamefont {Akhmerov},
  \citenamefont {Pekker}, \citenamefont {Refael}, \citenamefont {Cirac},
  \citenamefont {Demler}, \citenamefont {Lukin},\ and\ \citenamefont
  {Zoller}}]{ultracold}%
  \BibitemOpen
  \bibfield  {author} {\bibinfo {author} {\bibfnamefont {L.}~\bibnamefont
  {Jiang}}, \bibinfo {author} {\bibfnamefont {T.}~\bibnamefont {Kitagawa}},
  \bibinfo {author} {\bibfnamefont {J.}~\bibnamefont {Alicea}}, \bibinfo
  {author} {\bibfnamefont {A.~R.}\ \bibnamefont {Akhmerov}}, \bibinfo {author}
  {\bibfnamefont {D.}~\bibnamefont {Pekker}}, \bibinfo {author} {\bibfnamefont
  {G.}~\bibnamefont {Refael}}, \bibinfo {author} {\bibfnamefont {J.~I.}\
  \bibnamefont {Cirac}}, \bibinfo {author} {\bibfnamefont {E.}~\bibnamefont
  {Demler}}, \bibinfo {author} {\bibfnamefont {M.~D.}\ \bibnamefont {Lukin}}, \
  and\ \bibinfo {author} {\bibfnamefont {P.}~\bibnamefont {Zoller}},\
  }\bibfield  {title} {\enquote {\bibinfo {title} {Majorana fermions in
  equilibrium and in driven cold-atom quantum wires},}\ }\href {\doibase
  10.1103/PhysRevLett.106.220402} {\bibfield  {journal} {\bibinfo  {journal}
  {Phys. Rev. Lett.}\ }\textbf {\bibinfo {volume} {106}},\ \bibinfo {pages}
  {220402} (\bibinfo {year} {2011})}\BibitemShut {NoStop}%
\bibitem [{\citenamefont {San-Jose}\ \emph {et~al.}(2012)\citenamefont
  {San-Jose}, \citenamefont {Prada},\ and\ \citenamefont
  {Aguado}}]{Aguado-2012}%
  \BibitemOpen
  \bibfield  {author} {\bibinfo {author} {\bibfnamefont {P.}~\bibnamefont
  {San-Jose}}, \bibinfo {author} {\bibfnamefont {E.}~\bibnamefont {Prada}}, \
  and\ \bibinfo {author} {\bibfnamefont {R.}~\bibnamefont {Aguado}},\
  }\bibfield  {title} {\enquote {\bibinfo {title} {ac {Josephson} effect in
  finite-length nanowire junctions with {Majorana} modes},}\ }\href {\doibase
  10.1103/PhysRevLett.108.257001} {\bibfield  {journal} {\bibinfo  {journal}
  {Phys. Rev. Lett.}\ }\textbf {\bibinfo {volume} {108}},\ \bibinfo {pages}
  {257001} (\bibinfo {year} {2012})}\BibitemShut {NoStop}%
\bibitem [{\citenamefont {Mourik}\ \emph {et~al.}(2012)\citenamefont {Mourik},
  \citenamefont {Zuo}, \citenamefont {Frolov}, \citenamefont {Plissard},
  \citenamefont {Bakkers},\ and\ \citenamefont {Kouwenhoven}}]{Mourik-12}%
  \BibitemOpen
  \bibfield  {author} {\bibinfo {author} {\bibfnamefont {V.}~\bibnamefont
  {Mourik}}, \bibinfo {author} {\bibfnamefont {K.}~\bibnamefont {Zuo}},
  \bibinfo {author} {\bibfnamefont {S.~M.}\ \bibnamefont {Frolov}}, \bibinfo
  {author} {\bibfnamefont {S.~R.}\ \bibnamefont {Plissard}}, \bibinfo {author}
  {\bibfnamefont {E.~P. A.~M.}\ \bibnamefont {Bakkers}}, \ and\ \bibinfo
  {author} {\bibfnamefont {L.~P.}\ \bibnamefont {Kouwenhoven}},\ }\bibfield
  {title} {\enquote {\bibinfo {title} {Signatures of {Majorana} fermions in
  hybrid superconductor-semiconductor nanowire devices},}\ }\href {\doibase
  10.1126/science.1222360} {\bibfield  {journal} {\bibinfo  {journal}
  {Science}\ }\textbf {\bibinfo {volume} {336}},\ \bibinfo {pages} {1003}
  (\bibinfo {year} {2012})}\BibitemShut {NoStop}%
\bibitem [{\citenamefont {et~al.}(2016)}]{Kouwenhoven-2016}%
  \BibitemOpen
  \bibfield  {author} {\bibinfo {author} {\bibfnamefont {H.~Zhang}\
  \bibnamefont {et~al.}},\ }\href@noop {} {\enquote {\bibinfo {title}
  {Ballistic {Majorana} nanowire devices},}\ } (\bibinfo {year} {2016}),\
  \Eprint {http://arxiv.org/abs/arXiv:1603.04069} {arXiv:1603.04069}
  \BibitemShut {NoStop}%
\bibitem [{\citenamefont {Nadj-Perge}\ \emph {et~al.}(2014)\citenamefont
  {Nadj-Perge}, \citenamefont {Drozdov}, \citenamefont {Li}, \citenamefont
  {Chen}, \citenamefont {Jeon}, \citenamefont {Seo}, \citenamefont {MacDonald},
  \citenamefont {Bernevig},\ and\ \citenamefont {Yazdani}}]{Yazdani-14}%
  \BibitemOpen
  \bibfield  {author} {\bibinfo {author} {\bibfnamefont {S.}~\bibnamefont
  {Nadj-Perge}}, \bibinfo {author} {\bibfnamefont {I.~K.}\ \bibnamefont
  {Drozdov}}, \bibinfo {author} {\bibfnamefont {J.}~\bibnamefont {Li}},
  \bibinfo {author} {\bibfnamefont {H.}~\bibnamefont {Chen}}, \bibinfo {author}
  {\bibfnamefont {S.}~\bibnamefont {Jeon}}, \bibinfo {author} {\bibfnamefont
  {J.}~\bibnamefont {Seo}}, \bibinfo {author} {\bibfnamefont {A.~H.}\
  \bibnamefont {MacDonald}}, \bibinfo {author} {\bibfnamefont {B.~A.}\
  \bibnamefont {Bernevig}}, \ and\ \bibinfo {author} {\bibfnamefont
  {A.}~\bibnamefont {Yazdani}},\ }\bibfield  {title} {\enquote {\bibinfo
  {title} {Observation of {Majorana} fermions in ferromagnetic atomic chains on
  a superconductor},}\ }\href {\doibase 10.1126/science.1259327} {\bibfield
  {journal} {\bibinfo  {journal} {Science}\ }\textbf {\bibinfo {volume}
  {346}},\ \bibinfo {pages} {602} (\bibinfo {year} {2014})}\BibitemShut
  {NoStop}%
\bibitem [{\citenamefont {Pawlak}\ \emph {et~al.}(2016)\citenamefont {Pawlak},
  \citenamefont {Kisiel}, \citenamefont {Klinovaja}, \citenamefont {Maier},
  \citenamefont {Kawai}, \citenamefont {Glatzel}, \citenamefont {Loss},\ and\
  \citenamefont {Meyer}}]{Kisiel-15}%
  \BibitemOpen
  \bibfield  {author} {\bibinfo {author} {\bibfnamefont {R.}~\bibnamefont
  {Pawlak}}, \bibinfo {author} {\bibfnamefont {M.}~\bibnamefont {Kisiel}},
  \bibinfo {author} {\bibfnamefont {J.}~\bibnamefont {Klinovaja}}, \bibinfo
  {author} {\bibfnamefont {T.}~\bibnamefont {Maier}}, \bibinfo {author}
  {\bibfnamefont {S.}~\bibnamefont {Kawai}}, \bibinfo {author} {\bibfnamefont
  {T.}~\bibnamefont {Glatzel}}, \bibinfo {author} {\bibfnamefont
  {D.}~\bibnamefont {Loss}}, \ and\ \bibinfo {author} {\bibfnamefont
  {E.}~\bibnamefont {Meyer}},\ }\bibfield  {title} {\enquote {\bibinfo {title}
  {Probing atomic structure and {Majorana} wave-functions in mono-atomic
  {Fe}-chains on superconducting {Pb}-surface},}\ }\href {\doibase
  10.1038/npjqi.2016.35} {\bibfield  {journal} {\bibinfo  {journal} {npj
  Quantum Info}\ }\textbf {\bibinfo {volume} {2}},\ \bibinfo {pages} {16035}
  (\bibinfo {year} {2016})}\BibitemShut {NoStop}%
\bibitem [{\citenamefont {Ruby}\ \emph {et~al.}(2015)\citenamefont {Ruby},
  \citenamefont {Pientka}, \citenamefont {Peng}, \citenamefont {von Oppen},
  \citenamefont {Heinrich},\ and\ \citenamefont {Franke}}]{Franke-15}%
  \BibitemOpen
  \bibfield  {author} {\bibinfo {author} {\bibfnamefont {M.}~\bibnamefont
  {Ruby}}, \bibinfo {author} {\bibfnamefont {F.}~\bibnamefont {Pientka}},
  \bibinfo {author} {\bibfnamefont {Y.}~\bibnamefont {Peng}}, \bibinfo {author}
  {\bibfnamefont {F.}~\bibnamefont {von Oppen}}, \bibinfo {author}
  {\bibfnamefont {B.~W.}\ \bibnamefont {Heinrich}}, \ and\ \bibinfo {author}
  {\bibfnamefont {K.~J.}\ \bibnamefont {Franke}},\ }\bibfield  {title}
  {\enquote {\bibinfo {title} {End states and subgap structure in
  proximity-coupled chains of magnetic adatoms},}\ }\href {\doibase
  10.1103/PhysRevLett.115.197204} {\bibfield  {journal} {\bibinfo  {journal}
  {Phys. Rev. Lett.}\ }\textbf {\bibinfo {volume} {115}},\ \bibinfo {pages}
  {197204} (\bibinfo {year} {2015})}\BibitemShut {NoStop}%
\bibitem [{\citenamefont {Jeon}\ \emph {et~al.}(2017)\citenamefont {Jeon},
  \citenamefont {Xie}, \citenamefont {Li}, \citenamefont {Wang}, \citenamefont
  {Bernevig},\ and\ \citenamefont {Yazdani}}]{Yazdani-2017}%
  \BibitemOpen
  \bibfield  {author} {\bibinfo {author} {\bibfnamefont {S.}~\bibnamefont
  {Jeon}}, \bibinfo {author} {\bibfnamefont {Y.}~\bibnamefont {Xie}}, \bibinfo
  {author} {\bibfnamefont {J.}~\bibnamefont {Li}}, \bibinfo {author}
  {\bibfnamefont {Z.}~\bibnamefont {Wang}}, \bibinfo {author} {\bibfnamefont
  {B.~A.}\ \bibnamefont {Bernevig}}, \ and\ \bibinfo {author} {\bibfnamefont
  {A.}~\bibnamefont {Yazdani}},\ }\bibfield  {title} {\enquote {\bibinfo
  {title} {Distinguishing a {Majorana} zero mode using spin-resolved
  measurements},}\ }\href {\doibase 10.1126/science.aan3670} {\bibfield
  {journal} {\bibinfo  {journal} {Science}\ }\textbf {\bibinfo {volume}
  {358}},\ \bibinfo {pages} {772} (\bibinfo {year} {2017})}\BibitemShut
  {NoStop}%
\bibitem [{\citenamefont {Deng}\ \emph {et~al.}(2016)\citenamefont {Deng},
  \citenamefont {Vaitiek\.{e}nas}, \citenamefont {Hansen}, \citenamefont
  {Danon}, \citenamefont {Leijnse}, \citenamefont {Flensberg}, \citenamefont
  {Nyg{\r a}rd}, \citenamefont {Krogstrup},\ and\ \citenamefont
  {Marcus}}]{Deng-2017}%
  \BibitemOpen
  \bibfield  {author} {\bibinfo {author} {\bibfnamefont {M.~T.}\ \bibnamefont
  {Deng}}, \bibinfo {author} {\bibfnamefont {S.}~\bibnamefont
  {Vaitiek\.{e}nas}}, \bibinfo {author} {\bibfnamefont {E.~B.}\ \bibnamefont
  {Hansen}}, \bibinfo {author} {\bibfnamefont {J.}~\bibnamefont {Danon}},
  \bibinfo {author} {\bibfnamefont {M.}~\bibnamefont {Leijnse}}, \bibinfo
  {author} {\bibfnamefont {K.}~\bibnamefont {Flensberg}}, \bibinfo {author}
  {\bibfnamefont {J.}~\bibnamefont {Nyg{\r a}rd}}, \bibinfo {author}
  {\bibfnamefont {P.}~\bibnamefont {Krogstrup}}, \ and\ \bibinfo {author}
  {\bibfnamefont {C.~M.}\ \bibnamefont {Marcus}},\ }\bibfield  {title}
  {\enquote {\bibinfo {title} {Majorana bound state in a coupled quantum-dot
  hybrid-nanowire system},}\ }\href {\doibase 10.1126/science.aaf3961}
  {\bibfield  {journal} {\bibinfo  {journal} {Science}\ }\textbf {\bibinfo
  {volume} {354}},\ \bibinfo {pages} {1557} (\bibinfo {year}
  {2016})}\BibitemShut {NoStop}%
\bibitem [{\citenamefont {Vernek}\ \emph {et~al.}(2014)\citenamefont {Vernek},
  \citenamefont {Penteado}, \citenamefont {Seridonio},\ and\ \citenamefont
  {Egues}}]{Vernek-2014}%
  \BibitemOpen
  \bibfield  {author} {\bibinfo {author} {\bibfnamefont {E.}~\bibnamefont
  {Vernek}}, \bibinfo {author} {\bibfnamefont {P.~H.}\ \bibnamefont
  {Penteado}}, \bibinfo {author} {\bibfnamefont {A.~C.}\ \bibnamefont
  {Seridonio}}, \ and\ \bibinfo {author} {\bibfnamefont {J.~C.}\ \bibnamefont
  {Egues}},\ }\bibfield  {title} {\enquote {\bibinfo {title} {Subtle leakage of
  a {Majorana} mode into a quantum dot},}\ }\href {\doibase
  10.1103/PhysRevB.89.165314} {\bibfield  {journal} {\bibinfo  {journal} {Phys.
  Rev. B}\ }\textbf {\bibinfo {volume} {89}},\ \bibinfo {pages} {165314}
  (\bibinfo {year} {2014})}\BibitemShut {NoStop}%
\bibitem [{\citenamefont {Liu}\ \emph {et~al.}(2017)\citenamefont {Liu},
  \citenamefont {Sau}, \citenamefont {Stanescu},\ and\ \citenamefont
  {Das~Sarma}}]{DasSarma-2017}%
  \BibitemOpen
  \bibfield  {author} {\bibinfo {author} {\bibfnamefont {C.-X.}\ \bibnamefont
  {Liu}}, \bibinfo {author} {\bibfnamefont {J.~D.}\ \bibnamefont {Sau}},
  \bibinfo {author} {\bibfnamefont {T.~D.}\ \bibnamefont {Stanescu}}, \ and\
  \bibinfo {author} {\bibfnamefont {S.}~\bibnamefont {Das~Sarma}},\ }\bibfield
  {title} {\enquote {\bibinfo {title} {Andreev bound states versus {Majorana}
  bound states in quantum dot-nanowire-superconductor hybrid structures:
  Trivial versus topological zero-bias conductance peaks},}\ }\href {\doibase
  10.1103/PhysRevB.96.075161} {\bibfield  {journal} {\bibinfo  {journal} {Phys.
  Rev. B}\ }\textbf {\bibinfo {volume} {96}},\ \bibinfo {pages} {075161}
  (\bibinfo {year} {2017})}\BibitemShut {NoStop}%
\bibitem [{\citenamefont {Hoffman}\ \emph {et~al.}(2017)\citenamefont
  {Hoffman}, \citenamefont {Chevallier}, \citenamefont {Loss},\ and\
  \citenamefont {Klinovaja}}]{Klinovaja-2017}%
  \BibitemOpen
  \bibfield  {author} {\bibinfo {author} {\bibfnamefont {S.}~\bibnamefont
  {Hoffman}}, \bibinfo {author} {\bibfnamefont {D.}~\bibnamefont {Chevallier}},
  \bibinfo {author} {\bibfnamefont {D.}~\bibnamefont {Loss}}, \ and\ \bibinfo
  {author} {\bibfnamefont {J.}~\bibnamefont {Klinovaja}},\ }\bibfield  {title}
  {\enquote {\bibinfo {title} {Spin-dependent coupling between quantum dots and
  topological quantum wires},}\ }\href {\doibase 10.1103/PhysRevB.96.045440}
  {\bibfield  {journal} {\bibinfo  {journal} {Phys. Rev. B}\ }\textbf {\bibinfo
  {volume} {96}},\ \bibinfo {pages} {045440} (\bibinfo {year}
  {2017})}\BibitemShut {NoStop}%
\bibitem [{\citenamefont {Ptok}\ \emph {et~al.}(2017)\citenamefont {Ptok},
  \citenamefont {Kobia\l{}ka},\ and\ \citenamefont {Doma\ifmmode~\acute{n}\else
  \'{n}\fi{}ski}}]{Ptok-2017}%
  \BibitemOpen
  \bibfield  {author} {\bibinfo {author} {\bibfnamefont {Andrzej}\ \bibnamefont
  {Ptok}}, \bibinfo {author} {\bibfnamefont {Aksel}\ \bibnamefont
  {Kobia\l{}ka}}, \ and\ \bibinfo {author} {\bibfnamefont {Tadeusz}\
  \bibnamefont {Doma\ifmmode~\acute{n}\else \'{n}\fi{}ski}},\ }\bibfield
  {title} {\enquote {\bibinfo {title} {Controlling the bound states in a
  quantum-dot hybrid nanowire},}\ }\href {\doibase 10.1103/PhysRevB.96.195430}
  {\bibfield  {journal} {\bibinfo  {journal} {Phys. Rev. B}\ }\textbf {\bibinfo
  {volume} {96}},\ \bibinfo {pages} {195430} (\bibinfo {year}
  {2017})}\BibitemShut {NoStop}%
\bibitem [{\citenamefont {Prada}\ \emph {et~al.}(2017)\citenamefont {Prada},
  \citenamefont {Aguado},\ and\ \citenamefont {San-Jose}}]{Prada-2017}%
  \BibitemOpen
  \bibfield  {author} {\bibinfo {author} {\bibfnamefont {E.}~\bibnamefont
  {Prada}}, \bibinfo {author} {\bibfnamefont {R.}~\bibnamefont {Aguado}}, \
  and\ \bibinfo {author} {\bibfnamefont {P.}~\bibnamefont {San-Jose}},\
  }\bibfield  {title} {\enquote {\bibinfo {title} {Measuring {Majorana}
  nonlocality and spin structure with a quantum dot},}\ }\href {\doibase
  10.1103/PhysRevB.96.085418} {\bibfield  {journal} {\bibinfo  {journal} {Phys.
  Rev. B}\ }\textbf {\bibinfo {volume} {96}},\ \bibinfo {pages} {085418}
  (\bibinfo {year} {2017})}\BibitemShut {NoStop}%
\bibitem [{\citenamefont {Bara\'nski}\ \emph {et~al.}(2017)\citenamefont
  {Bara\'nski}, \citenamefont {Kobia\l{}ka},\ and\ \citenamefont
  {Doma\'nski}}]{Baranski-2017}%
  \BibitemOpen
  \bibfield  {author} {\bibinfo {author} {\bibfnamefont {J.}~\bibnamefont
  {Bara\'nski}}, \bibinfo {author} {\bibfnamefont {A.}~\bibnamefont
  {Kobia\l{}ka}}, \ and\ \bibinfo {author} {\bibfnamefont {T.}~\bibnamefont
  {Doma\'nski}},\ }\bibfield  {title} {\enquote {\bibinfo {title}
  {Spin-sensitive interference due to {Majorana} state on the interface between
  normal and superconducting leads},}\ }\href
  {http://stacks.iop.org/0953-8984/29/i=7/a=075603} {\bibfield  {journal}
  {\bibinfo  {journal} {J. Phys.: Condens. Matter}\ }\textbf {\bibinfo {volume}
  {29}},\ \bibinfo {pages} {075603} (\bibinfo {year} {2017})}\BibitemShut
  {NoStop}%
\bibitem [{\citenamefont {Chirla}\ and\ \citenamefont
  {Moca}(2016)}]{Chirla-2016}%
  \BibitemOpen
  \bibfield  {author} {\bibinfo {author} {\bibfnamefont {R.}~\bibnamefont
  {Chirla}}\ and\ \bibinfo {author} {\bibfnamefont {C.~P.}\ \bibnamefont
  {Moca}},\ }\bibfield  {title} {\enquote {\bibinfo {title} {Fingerprints of
  {Majorana} fermions in spin-resolved subgap spectroscopy},}\ }\href {\doibase
  10.1103/PhysRevB.94.045405} {\bibfield  {journal} {\bibinfo  {journal} {Phys.
  Rev. B}\ }\textbf {\bibinfo {volume} {94}},\ \bibinfo {pages} {045405}
  (\bibinfo {year} {2016})}\BibitemShut {NoStop}%
\bibitem [{\citenamefont {\ifmmode~\check{Z}\else \v{Z}\fi{}itko}\ \emph
  {et~al.}(2015)\citenamefont {\ifmmode~\check{Z}\else \v{Z}\fi{}itko},
  \citenamefont {Lim}, \citenamefont {L\'opez},\ and\ \citenamefont
  {Aguado}}]{Zitko-2015a}%
  \BibitemOpen
  \bibfield  {author} {\bibinfo {author} {\bibfnamefont {R.}~\bibnamefont
  {\ifmmode~\check{Z}\else \v{Z}\fi{}itko}}, \bibinfo {author} {\bibfnamefont
  {J.~Soo}\ \bibnamefont {Lim}}, \bibinfo {author} {\bibfnamefont
  {R.}~\bibnamefont {L\'opez}}, \ and\ \bibinfo {author} {\bibfnamefont
  {R.}~\bibnamefont {Aguado}},\ }\bibfield  {title} {\enquote {\bibinfo {title}
  {Shiba states and zero-bias anomalies in the hybrid normal-superconductor
  {Anderson} model},}\ }\href {\doibase 10.1103/PhysRevB.91.045441} {\bibfield
  {journal} {\bibinfo  {journal} {Phys. Rev. B}\ }\textbf {\bibinfo {volume}
  {91}},\ \bibinfo {pages} {045441} (\bibinfo {year} {2015})}\BibitemShut
  {NoStop}%
\bibitem [{\citenamefont {Doma\'{n}ski}\ \emph {et~al.}(2016)\citenamefont
  {Doma\'{n}ski}, \citenamefont {Weymann}, \citenamefont {Bara\'{n}ska},\ and\
  \citenamefont {G\'orski}}]{Domanski-2016}%
  \BibitemOpen
  \bibfield  {author} {\bibinfo {author} {\bibfnamefont {T.}~\bibnamefont
  {Doma\'{n}ski}}, \bibinfo {author} {\bibfnamefont {I.}~\bibnamefont
  {Weymann}}, \bibinfo {author} {\bibfnamefont {M.}~\bibnamefont
  {Bara\'{n}ska}}, \ and\ \bibinfo {author} {\bibfnamefont {G.}~\bibnamefont
  {G\'orski}},\ }\bibfield  {title} {\enquote {\bibinfo {title} {Constructive
  influence of the induced electron pairing on the {K}ondo state},}\ }\href
  {\doibase 10.1038/srep23336} {\bibfield  {journal} {\bibinfo  {journal} {Sci.
  Rep.}\ }\textbf {\bibinfo {volume} {6}},\ \bibinfo {pages} {23336} (\bibinfo
  {year} {2016})}\BibitemShut {NoStop}%
\bibitem [{\citenamefont {Lee}\ \emph {et~al.}(2017)\citenamefont {Lee},
  \citenamefont {Jiang}, \citenamefont {\v{Z}itko}, \citenamefont {Aguado},
  \citenamefont {Lieber},\ and\ \citenamefont {De~Franceschi}}]{Lee-2017}%
  \BibitemOpen
  \bibfield  {author} {\bibinfo {author} {\bibfnamefont {E.~J.~H.}\
  \bibnamefont {Lee}}, \bibinfo {author} {\bibfnamefont {X.}~\bibnamefont
  {Jiang}}, \bibinfo {author} {\bibfnamefont {R.}~\bibnamefont {\v{Z}itko}},
  \bibinfo {author} {\bibfnamefont {R.}~\bibnamefont {Aguado}}, \bibinfo
  {author} {\bibfnamefont {C.~M.}\ \bibnamefont {Lieber}}, \ and\ \bibinfo
  {author} {\bibfnamefont {S.}~\bibnamefont {De~Franceschi}},\ }\bibfield
  {title} {\enquote {\bibinfo {title} {Scaling of subgap excitations in a
  superconductor-semiconductor nanowire quantum dot},}\ }\href {\doibase
  10.1103/PhysRevB.95.180502} {\bibfield  {journal} {\bibinfo  {journal} {Phys.
  Rev. B}\ }\textbf {\bibinfo {volume} {95}},\ \bibinfo {pages} {180502}
  (\bibinfo {year} {2017})}\BibitemShut {NoStop}%
\bibitem [{\citenamefont {Liu}\ and\ \citenamefont
  {Baranger}(2011)}]{Baranger-2011}%
  \BibitemOpen
  \bibfield  {author} {\bibinfo {author} {\bibfnamefont {D.~E.}\ \bibnamefont
  {Liu}}\ and\ \bibinfo {author} {\bibfnamefont {H.~U.}\ \bibnamefont
  {Baranger}},\ }\bibfield  {title} {\enquote {\bibinfo {title} {Detecting a
  {Majorana}-fermion zero mode using a quantum dot},}\ }\href {\doibase
  10.1103/PhysRevB.84.201308} {\bibfield  {journal} {\bibinfo  {journal} {Phys.
  Rev. B}\ }\textbf {\bibinfo {volume} {84}},\ \bibinfo {pages} {201308}
  (\bibinfo {year} {2011})}\BibitemShut {NoStop}%
\bibitem [{\citenamefont {L\'opez}\ \emph {et~al.}(2014)\citenamefont
  {L\'opez}, \citenamefont {Lee}, \citenamefont {Serra},\ and\ \citenamefont
  {Lim}}]{Lopez-2014}%
  \BibitemOpen
  \bibfield  {author} {\bibinfo {author} {\bibfnamefont {R.}~\bibnamefont
  {L\'opez}}, \bibinfo {author} {\bibfnamefont {M.}~\bibnamefont {Lee}},
  \bibinfo {author} {\bibfnamefont {L.}~\bibnamefont {Serra}}, \ and\ \bibinfo
  {author} {\bibfnamefont {J.~S.}\ \bibnamefont {Lim}},\ }\bibfield  {title}
  {\enquote {\bibinfo {title} {Thermoelectrical detection of {Majorana}
  states},}\ }\href {\doibase 10.1103/PhysRevB.89.205418} {\bibfield  {journal}
  {\bibinfo  {journal} {Phys. Rev. B}\ }\textbf {\bibinfo {volume} {89}},\
  \bibinfo {pages} {205418} (\bibinfo {year} {2014})}\BibitemShut {NoStop}%
\bibitem [{\citenamefont {Lee}\ \emph {et~al.}(2013)\citenamefont {Lee},
  \citenamefont {Lim},\ and\ \citenamefont {L\'opez}}]{Lee-2013}%
  \BibitemOpen
  \bibfield  {author} {\bibinfo {author} {\bibfnamefont {M.}~\bibnamefont
  {Lee}}, \bibinfo {author} {\bibfnamefont {J.~S.}\ \bibnamefont {Lim}}, \ and\
  \bibinfo {author} {\bibfnamefont {R.}~\bibnamefont {L\'opez}},\ }\bibfield
  {title} {\enquote {\bibinfo {title} {Kondo effect in a quantum dot
  side-coupled to a topological superconductor},}\ }\href {\doibase
  10.1103/PhysRevB.87.241402} {\bibfield  {journal} {\bibinfo  {journal} {Phys.
  Rev. B}\ }\textbf {\bibinfo {volume} {87}},\ \bibinfo {pages} {241402}
  (\bibinfo {year} {2013})}\BibitemShut {NoStop}%
\bibitem [{\citenamefont {Cao}\ \emph {et~al.}(2012)\citenamefont {Cao},
  \citenamefont {Wang}, \citenamefont {Xiong}, \citenamefont {Gong},\ and\
  \citenamefont {Li}}]{Cao-2012}%
  \BibitemOpen
  \bibfield  {author} {\bibinfo {author} {\bibfnamefont {Y.}~\bibnamefont
  {Cao}}, \bibinfo {author} {\bibfnamefont {P.}~\bibnamefont {Wang}}, \bibinfo
  {author} {\bibfnamefont {G.}~\bibnamefont {Xiong}}, \bibinfo {author}
  {\bibfnamefont {M.}~\bibnamefont {Gong}}, \ and\ \bibinfo {author}
  {\bibfnamefont {X.-Q.}\ \bibnamefont {Li}},\ }\bibfield  {title} {\enquote
  {\bibinfo {title} {Probing the existence and dynamics of {Majorana} fermion
  via transport through a quantum dot},}\ }\href {\doibase
  10.1103/PhysRevB.86.115311} {\bibfield  {journal} {\bibinfo  {journal} {Phys.
  Rev. B}\ }\textbf {\bibinfo {volume} {86}},\ \bibinfo {pages} {115311}
  (\bibinfo {year} {2012})}\BibitemShut {NoStop}%
\bibitem [{\citenamefont {Gong}\ \emph {et~al.}(2014)\citenamefont {Gong},
  \citenamefont {Zhang}, \citenamefont {Li}, \citenamefont {Yi},\ and\
  \citenamefont {Zheng}}]{Gong-2014}%
  \BibitemOpen
  \bibfield  {author} {\bibinfo {author} {\bibfnamefont {W.-J.}\ \bibnamefont
  {Gong}}, \bibinfo {author} {\bibfnamefont {S.-F.}\ \bibnamefont {Zhang}},
  \bibinfo {author} {\bibfnamefont {Z.-C.}\ \bibnamefont {Li}}, \bibinfo
  {author} {\bibfnamefont {G.}~\bibnamefont {Yi}}, \ and\ \bibinfo {author}
  {\bibfnamefont {Y.-S.}\ \bibnamefont {Zheng}},\ }\bibfield  {title} {\enquote
  {\bibinfo {title} {Detection of a {Majorana} fermion zero mode by a
  {T}-shaped quantum-dot structure},}\ }\href {\doibase
  10.1103/PhysRevB.89.245413} {\bibfield  {journal} {\bibinfo  {journal} {Phys.
  Rev. B}\ }\textbf {\bibinfo {volume} {89}},\ \bibinfo {pages} {245413}
  (\bibinfo {year} {2014})}\BibitemShut {NoStop}%
\bibitem [{\citenamefont {Liu}\ \emph {et~al.}(2015)\citenamefont {Liu},
  \citenamefont {Cheng},\ and\ \citenamefont {Lutchyn}}]{Lutchyn-2015}%
  \BibitemOpen
  \bibfield  {author} {\bibinfo {author} {\bibfnamefont {D.~E.}\ \bibnamefont
  {Liu}}, \bibinfo {author} {\bibfnamefont {M.}~\bibnamefont {Cheng}}, \ and\
  \bibinfo {author} {\bibfnamefont {R.~M.}\ \bibnamefont {Lutchyn}},\
  }\bibfield  {title} {\enquote {\bibinfo {title} {Probing {Majorana} physics
  in quantum-dot shot-noise experiments},}\ }\href {\doibase
  10.1103/PhysRevB.91.081405} {\bibfield  {journal} {\bibinfo  {journal} {Phys.
  Rev. B}\ }\textbf {\bibinfo {volume} {91}},\ \bibinfo {pages} {081405}
  (\bibinfo {year} {2015})}\BibitemShut {NoStop}%
\bibitem [{\citenamefont {Stefa\'nski}(2015)}]{Stefanski-2015}%
  \BibitemOpen
  \bibfield  {author} {\bibinfo {author} {\bibfnamefont {P.}~\bibnamefont
  {Stefa\'nski}},\ }\bibfield  {title} {\enquote {\bibinfo {title} {Signatures
  of {Majorana} states in electron transport through a quantum dot coupled to
  topological wire},}\ }\href {\doibase 10.12693/APhysPolA.127.198} {\bibfield
  {journal} {\bibinfo  {journal} {Acta Phys. Pol. A}\ }\textbf {\bibinfo
  {volume} {127}},\ \bibinfo {pages} {198} (\bibinfo {year}
  {2015})}\BibitemShut {NoStop}%
\bibitem [{\citenamefont {Li}\ \emph {et~al.}(2015)\citenamefont {Li},
  \citenamefont {Lam},\ and\ \citenamefont {You}}]{Li-2015}%
  \BibitemOpen
  \bibfield  {author} {\bibinfo {author} {\bibfnamefont {Z.-Z.}\ \bibnamefont
  {Li}}, \bibinfo {author} {\bibfnamefont {C.-H.}\ \bibnamefont {Lam}}, \ and\
  \bibinfo {author} {\bibfnamefont {J.~Q.}\ \bibnamefont {You}},\ }\bibfield
  {title} {\enquote {\bibinfo {title} {Probing {Majorana} bound states via
  counting statistics of a single electron transistor},}\ }\href {\doibase
  10.1038/srep11416} {\bibfield  {journal} {\bibinfo  {journal} {Sci. Rep.}\
  }\textbf {\bibinfo {volume} {5}},\ \bibinfo {pages} {11416} (\bibinfo {year}
  {2015})}\BibitemShut {NoStop}%
\bibitem [{\citenamefont {Weymann}(2017)}]{Weymann-2017}%
  \BibitemOpen
  \bibfield  {author} {\bibinfo {author} {\bibfnamefont {I.}~\bibnamefont
  {Weymann}},\ }\bibfield  {title} {\enquote {\bibinfo {title} {Spin {Seebeck}
  effect in quantum dot side-coupled to topological superconductor},}\ }\href
  {http://stacks.iop.org/0953-8984/29/i=9/a=095301} {\bibfield  {journal}
  {\bibinfo  {journal} {J. Phys.: Condens. Matter}\ }\textbf {\bibinfo {volume}
  {29}},\ \bibinfo {pages} {095301} (\bibinfo {year} {2017})}\BibitemShut
  {NoStop}%
\bibitem [{\citenamefont {Weymann}\ and\ \citenamefont
  {W\'ojcik}(2017)}]{Wojcik-2017}%
  \BibitemOpen
  \bibfield  {author} {\bibinfo {author} {\bibfnamefont {I.}~\bibnamefont
  {Weymann}}\ and\ \bibinfo {author} {\bibfnamefont {K.~P.}\ \bibnamefont
  {W\'ojcik}},\ }\bibfield  {title} {\enquote {\bibinfo {title} {Transport
  properties of a hybrid {Majorana} wire-quantum dot system with ferromagnetic
  contacts},}\ }\href {\doibase 10.1103/PhysRevB.95.155427} {\bibfield
  {journal} {\bibinfo  {journal} {Phys. Rev. B}\ }\textbf {\bibinfo {volume}
  {95}},\ \bibinfo {pages} {155427} (\bibinfo {year} {2017})}\BibitemShut
  {NoStop}%
\bibitem [{\citenamefont {Wang}\ \emph {et~al.}(2016)\citenamefont {Wang},
  \citenamefont {Li}, \citenamefont {Wang},\ and\ \citenamefont
  {Liu}}]{Wang-2016}%
  \BibitemOpen
  \bibfield  {author} {\bibinfo {author} {\bibfnamefont {S.-X.}\ \bibnamefont
  {Wang}}, \bibinfo {author} {\bibfnamefont {Y.-X.}\ \bibnamefont {Li}},
  \bibinfo {author} {\bibfnamefont {N.}~\bibnamefont {Wang}}, \ and\ \bibinfo
  {author} {\bibfnamefont {J.-J.}\ \bibnamefont {Liu}},\ }\bibfield  {title}
  {\enquote {\bibinfo {title} {Andreev reflection in a {T}-shaped double
  quantum-dot with coupled {Majorana} bound states},}\ }\href {\doibase
  10.7498/aps.65.137302} {\bibfield  {journal} {\bibinfo  {journal} {Acta Phys.
  Sin.}\ }\textbf {\bibinfo {volume} {65}},\ \bibinfo {pages} {137302}
  (\bibinfo {year} {2016})}\BibitemShut {NoStop}%
\bibitem [{\citenamefont {Balatsky}\ \emph {et~al.}(2006)\citenamefont
  {Balatsky}, \citenamefont {Vekhter},\ and\ \citenamefont
  {Zhu}}]{Balatsky-2006}%
  \BibitemOpen
  \bibfield  {author} {\bibinfo {author} {\bibfnamefont {A.~V.}\ \bibnamefont
  {Balatsky}}, \bibinfo {author} {\bibfnamefont {I.}~\bibnamefont {Vekhter}}, \
  and\ \bibinfo {author} {\bibfnamefont {J.-X.}\ \bibnamefont {Zhu}},\
  }\bibfield  {title} {\enquote {\bibinfo {title} {Impurity-induced states in
  conventional and unconventional superconductors},}\ }\href {\doibase
  10.1103/RevModPhys.78.373} {\bibfield  {journal} {\bibinfo  {journal} {Rev.
  Mod. Phys.}\ }\textbf {\bibinfo {volume} {78}},\ \bibinfo {pages} {373}
  (\bibinfo {year} {2006})}\BibitemShut {NoStop}%
\bibitem [{\citenamefont {Doma\'nski}(2010)}]{Domanski-2010}%
  \BibitemOpen
  \bibfield  {author} {\bibinfo {author} {\bibfnamefont {T.}~\bibnamefont
  {Doma\'nski}},\ }\bibfield  {title} {\enquote {\bibinfo {title}
  {Particle-hole mixing driven by the superconducting fluctuations},}\ }\href
  {\doibase 10.1140/epjb/e2010-00100-0} {\bibfield  {journal} {\bibinfo
  {journal} {Eur. Phys. J. B}\ }\textbf {\bibinfo {volume} {74}},\ \bibinfo
  {pages} {437} (\bibinfo {year} {2010})}\BibitemShut {NoStop}%
\bibitem [{\citenamefont {Golub}(2015)}]{Golub-2015}%
  \BibitemOpen
  \bibfield  {author} {\bibinfo {author} {\bibfnamefont {A.}~\bibnamefont
  {Golub}},\ }\bibfield  {title} {\enquote {\bibinfo {title} {Multiple
  {Andreev} reflections in $s$-wave superconductor-quantum dot-topological
  superconductor tunnel junctions and {Majorana} bound states},}\ }\href
  {\doibase 10.1103/PhysRevB.91.205105} {\bibfield  {journal} {\bibinfo
  {journal} {Phys. Rev. B}\ }\textbf {\bibinfo {volume} {91}},\ \bibinfo
  {pages} {205105} (\bibinfo {year} {2015})}\BibitemShut {NoStop}%
\bibitem [{\citenamefont {Ruiz-Tijerina}\ \emph {et~al.}(2015)\citenamefont
  {Ruiz-Tijerina}, \citenamefont {Vernek}, \citenamefont {Dias~da Silva},\ and\
  \citenamefont {Egues}}]{Vernek-2015}%
  \BibitemOpen
  \bibfield  {author} {\bibinfo {author} {\bibfnamefont {D.~A.}\ \bibnamefont
  {Ruiz-Tijerina}}, \bibinfo {author} {\bibfnamefont {E.}~\bibnamefont
  {Vernek}}, \bibinfo {author} {\bibfnamefont {L.~G. G.~V.}\ \bibnamefont
  {Dias~da Silva}}, \ and\ \bibinfo {author} {\bibfnamefont {J.~C.}\
  \bibnamefont {Egues}},\ }\bibfield  {title} {\enquote {\bibinfo {title}
  {Interaction effects on a {Majorana} zero mode leaking into a quantum dot},}\
  }\href {\doibase 10.1103/PhysRevB.91.115435} {\bibfield  {journal} {\bibinfo
  {journal} {Phys. Rev. B}\ }\textbf {\bibinfo {volume} {91}},\ \bibinfo
  {pages} {115435} (\bibinfo {year} {2015})}\BibitemShut {NoStop}%
\bibitem [{\citenamefont {Bauer}\ \emph {et~al.}(2007)\citenamefont {Bauer},
  \citenamefont {Oguri},\ and\ \citenamefont {Hewson}}]{Bauer-2007}%
  \BibitemOpen
  \bibfield  {author} {\bibinfo {author} {\bibfnamefont {J.}~\bibnamefont
  {Bauer}}, \bibinfo {author} {\bibfnamefont {A.}~\bibnamefont {Oguri}}, \ and\
  \bibinfo {author} {\bibfnamefont {A.~C.}\ \bibnamefont {Hewson}},\ }\bibfield
   {title} {\enquote {\bibinfo {title} {Spectral properties of locally
  correlated electrons in a {B}ardeen-–{C}ooper–-{Schrieffer}
  superconductor},}\ }\href {http://stacks.iop.org/0953-8984/19/i=48/a=486211}
  {\bibfield  {journal} {\bibinfo  {journal} {J. Phys.: Condens. Matter}\
  }\textbf {\bibinfo {volume} {19}},\ \bibinfo {pages} {486211} (\bibinfo
  {year} {2007})}\BibitemShut {NoStop}%
\bibitem [{\citenamefont {Yamada}\ \emph {et~al.}(2011)\citenamefont {Yamada},
  \citenamefont {Tanaka},\ and\ \citenamefont {Kawakami}}]{Yamada-2011}%
  \BibitemOpen
  \bibfield  {author} {\bibinfo {author} {\bibfnamefont {Y.}~\bibnamefont
  {Yamada}}, \bibinfo {author} {\bibfnamefont {Y.}~\bibnamefont {Tanaka}}, \
  and\ \bibinfo {author} {\bibfnamefont {N.}~\bibnamefont {Kawakami}},\
  }\bibfield  {title} {\enquote {\bibinfo {title} {Interplay of {Kondo} and
  superconducting correlations in the nonequilibrium {Andreev} transport
  through a quantum dot},}\ }\href {\doibase 10.1103/PhysRevB.84.075484}
  {\bibfield  {journal} {\bibinfo  {journal} {Phys. Rev. B}\ }\textbf {\bibinfo
  {volume} {84}},\ \bibinfo {pages} {075484} (\bibinfo {year}
  {2011})}\BibitemShut {NoStop}%
\bibitem [{\citenamefont {Mart\'in-Rodero}\ and\ \citenamefont
  {Levy~Yeyati}(2011)}]{Rodero-2011}%
  \BibitemOpen
  \bibfield  {author} {\bibinfo {author} {\bibfnamefont {A.}~\bibnamefont
  {Mart\'in-Rodero}}\ and\ \bibinfo {author} {\bibfnamefont {A.}~\bibnamefont
  {Levy~Yeyati}},\ }\bibfield  {title} {\enquote {\bibinfo {title} {Josephson
  and {Andreev} transport through quantum dots},}\ }\href {\doibase
  10.1080/00018732.2011.624266} {\bibfield  {journal} {\bibinfo  {journal}
  {Advances in Physics}\ }\textbf {\bibinfo {volume} {60}},\ \bibinfo {pages}
  {899} (\bibinfo {year} {2011})}\BibitemShut {NoStop}%
\bibitem [{\citenamefont {Bara\'nski}\ and\ \citenamefont
  {Doma\'nski}(2013)}]{Baranski-2013}%
  \BibitemOpen
  \bibfield  {author} {\bibinfo {author} {\bibfnamefont {J.}~\bibnamefont
  {Bara\'nski}}\ and\ \bibinfo {author} {\bibfnamefont {T.}~\bibnamefont
  {Doma\'nski}},\ }\bibfield  {title} {\enquote {\bibinfo {title} {In-gap
  states of a quantum dot coupled between a normal and a superconducting
  lead},}\ }\href {http://stacks.iop.org/0953-8984/25/i=43/a=435305} {\bibfield
   {journal} {\bibinfo  {journal} {J. Phys.: Condens. Matter}\ }\textbf
  {\bibinfo {volume} {25}},\ \bibinfo {pages} {435305} (\bibinfo {year}
  {2013})}\BibitemShut {NoStop}%
\bibitem [{\citenamefont {Andreev}(1964)}]{Andreev-1964}%
  \BibitemOpen
  \bibfield  {author} {\bibinfo {author} {\bibfnamefont {A.~F.}\ \bibnamefont
  {Andreev}},\ }\bibfield  {title} {\enquote {\bibinfo {title} {The thermal
  conductivity of the intermediate state in superconductors.}}\ }\href@noop {}
  {\bibfield  {journal} {\bibinfo  {journal} {J. Exp. Theor. Phys}\ }\textbf
  {\bibinfo {volume} {19}},\ \bibinfo {pages} {1228} (\bibinfo {year}
  {1964})}\BibitemShut {NoStop}%
\bibitem [{\citenamefont {Krawiec}\ and\ \citenamefont
  {Wysoki\'{n}ski}(2004)}]{Krawiec-2004}%
  \BibitemOpen
  \bibfield  {author} {\bibinfo {author} {\bibfnamefont {M.}~\bibnamefont
  {Krawiec}}\ and\ \bibinfo {author} {\bibfnamefont {K.I.}\ \bibnamefont
  {Wysoki\'{n}ski}},\ }\bibfield  {title} {\enquote {\bibinfo {title} {Electron
  transport through a strongly interacting quantum dot coupled to a normal
  metal and {BCS} superconductor},}\ }\href {\doibase
  10.1088/0953-2048/17/1/018} {\bibfield  {journal} {\bibinfo  {journal}
  {Supercond. Sci. Technol.}\ }\textbf {\bibinfo {volume} {17}},\ \bibinfo
  {pages} {103} (\bibinfo {year} {2004})}\BibitemShut {NoStop}%
\bibitem [{\citenamefont {Schuray}\ \emph {et~al.}(2017)\citenamefont
  {Schuray}, \citenamefont {Weithofer},\ and\ \citenamefont
  {Recher}}]{Schuray-2017}%
  \BibitemOpen
  \bibfield  {author} {\bibinfo {author} {\bibfnamefont {A.}~\bibnamefont
  {Schuray}}, \bibinfo {author} {\bibfnamefont {L.}~\bibnamefont {Weithofer}},
  \ and\ \bibinfo {author} {\bibfnamefont {P.}~\bibnamefont {Recher}},\
  }\bibfield  {title} {\enquote {\bibinfo {title} {Fano resonances in
  {M}ajorana bound states-quantum dot hybrid systems},}\ }\href {\doibase
  10.1103/PhysRevB.96.085417} {\bibfield  {journal} {\bibinfo  {journal} {Phys.
  Rev. B}\ }\textbf {\bibinfo {volume} {96}},\ \bibinfo {pages} {085417}
  (\bibinfo {year} {2017})}\BibitemShut {NoStop}%
\bibitem [{\citenamefont {Cheng}\ \emph {et~al.}(2014)\citenamefont {Cheng},
  \citenamefont {Becker}, \citenamefont {Bauer},\ and\ \citenamefont
  {Lutchyn}}]{Cheng-2014}%
  \BibitemOpen
  \bibfield  {author} {\bibinfo {author} {\bibfnamefont {M.}~\bibnamefont
  {Cheng}}, \bibinfo {author} {\bibfnamefont {M.}~\bibnamefont {Becker}},
  \bibinfo {author} {\bibfnamefont {B.}~\bibnamefont {Bauer}}, \ and\ \bibinfo
  {author} {\bibfnamefont {R.~M.}\ \bibnamefont {Lutchyn}},\ }\bibfield
  {title} {\enquote {\bibinfo {title} {Interplay between {Kondo} and {Majorana}
  interactions in quantum dots},}\ }\href {\doibase 10.1103/PhysRevX.4.031051}
  {\bibfield  {journal} {\bibinfo  {journal} {Phys. Rev. X}\ }\textbf {\bibinfo
  {volume} {4}},\ \bibinfo {pages} {031051} (\bibinfo {year}
  {2014})}\BibitemShut {NoStop}%
\bibitem [{\citenamefont {van Beek}\ and\ \citenamefont
  {Braunecker}(2016)}]{Beek-2016}%
  \BibitemOpen
  \bibfield  {author} {\bibinfo {author} {\bibfnamefont {I.~J.}\ \bibnamefont
  {van Beek}}\ and\ \bibinfo {author} {\bibfnamefont {B.}~\bibnamefont
  {Braunecker}},\ }\bibfield  {title} {\enquote {\bibinfo {title} {Non-{Kondo}
  many-body physics in a {Majorana}-based {Kondo} type system},}\ }\href
  {\doibase 10.1103/PhysRevB.94.115416} {\bibfield  {journal} {\bibinfo
  {journal} {Phys. Rev. B}\ }\textbf {\bibinfo {volume} {94}},\ \bibinfo
  {pages} {115416} (\bibinfo {year} {2016})}\BibitemShut {NoStop}%
\bibitem [{\citenamefont {Tanaka}\ \emph {et~al.}(2007)\citenamefont {Tanaka},
  \citenamefont {Kawakami},\ and\ \citenamefont {Oguri}}]{Tanaka-2007}%
  \BibitemOpen
  \bibfield  {author} {\bibinfo {author} {\bibfnamefont {Y.}~\bibnamefont
  {Tanaka}}, \bibinfo {author} {\bibfnamefont {N.}~\bibnamefont {Kawakami}}, \
  and\ \bibinfo {author} {\bibfnamefont {A.}~\bibnamefont {Oguri}},\ }\bibfield
   {title} {\enquote {\bibinfo {title} {Numerical renormalization group
  approach to a quantum dot coupled to normal and superconducting leads},}\
  }\href {\doibase 10.1143/JPSJ.76.074701} {\bibfield  {journal} {\bibinfo
  {journal} {J. Phys. Soc. Japan}\ }\textbf {\bibinfo {volume} {76}},\ \bibinfo
  {pages} {074701} (\bibinfo {year} {2007})}\BibitemShut {NoStop}%
\bibitem [{\citenamefont {Doma\'nski}\ \emph {et~al.}(2017)\citenamefont
  {Doma\'nski}, \citenamefont {\v{Z}onda}, \citenamefont {Pokorn\'y},
  \citenamefont {G\'orski}, \citenamefont {Jani\v{s}},\ and\ \citenamefont
  {Novotn\'y}}]{Domanski-2017}%
  \BibitemOpen
  \bibfield  {author} {\bibinfo {author} {\bibfnamefont {T.}~\bibnamefont
  {Doma\'nski}}, \bibinfo {author} {\bibfnamefont {M.}~\bibnamefont
  {\v{Z}onda}}, \bibinfo {author} {\bibfnamefont {V.}~\bibnamefont
  {Pokorn\'y}}, \bibinfo {author} {\bibfnamefont {G.}~\bibnamefont {G\'orski}},
  \bibinfo {author} {\bibfnamefont {V.}~\bibnamefont {Jani\v{s}}}, \ and\
  \bibinfo {author} {\bibfnamefont {T.}~\bibnamefont {Novotn\'y}},\ }\bibfield
  {title} {\enquote {\bibinfo {title} {Josephson-phase-controlled interplay
  between correlation effects and electron pairing in a three-terminal
  nanostructure},}\ }\href {\doibase 10.1103/PhysRevB.95.045104} {\bibfield
  {journal} {\bibinfo  {journal} {Phys. Rev. B}\ }\textbf {\bibinfo {volume}
  {95}},\ \bibinfo {pages} {045104} (\bibinfo {year} {2017})}\BibitemShut
  {NoStop}%
\bibitem [{\citenamefont {Vecino}\ \emph {et~al.}(2003)\citenamefont {Vecino},
  \citenamefont {Mart\'{\i}n-Rodero},\ and\ \citenamefont
  {Levy~Yeyati}}]{Vecino-2003}%
  \BibitemOpen
  \bibfield  {author} {\bibinfo {author} {\bibfnamefont {E.}~\bibnamefont
  {Vecino}}, \bibinfo {author} {\bibfnamefont {A.}~\bibnamefont
  {Mart\'{\i}n-Rodero}}, \ and\ \bibinfo {author} {\bibfnamefont
  {A.}~\bibnamefont {Levy~Yeyati}},\ }\bibfield  {title} {\enquote {\bibinfo
  {title} {Josephson current through a correlated quantum level: {Andreev}
  states and $\ensuremath{\pi}$ junction behavior},}\ }\href {\doibase
  10.1103/PhysRevB.68.035105} {\bibfield  {journal} {\bibinfo  {journal} {Phys.
  Rev. B}\ }\textbf {\bibinfo {volume} {68}},\ \bibinfo {pages} {035105}
  (\bibinfo {year} {2003})}\BibitemShut {NoStop}%
\bibitem [{\citenamefont {Wilson}(1975)}]{Wilson}%
  \BibitemOpen
  \bibfield  {author} {\bibinfo {author} {\bibfnamefont {K.~G.}\ \bibnamefont
  {Wilson}},\ }\bibfield  {title} {\enquote {\bibinfo {title} {The
  renormalization group: {C}ritical phenomena and the {K}ondo problem},}\
  }\href {\doibase 10.1103/RevModPhys.47.773} {\bibfield  {journal} {\bibinfo
  {journal} {Rev. Mod. Phys.}\ }\textbf {\bibinfo {volume} {47}},\ \bibinfo
  {pages} {773--840} (\bibinfo {year} {1975})}\BibitemShut {NoStop}%
\bibitem [{\citenamefont {Legeza}\ \emph {et~al.}(2008)\citenamefont {Legeza},
  \citenamefont {Moca}, \citenamefont {Toth}, \citenamefont {Weymann},\ and\
  \citenamefont {Zarand}}]{fnrg}%
  \BibitemOpen
  \bibfield  {author} {\bibinfo {author} {\bibfnamefont {O.}~\bibnamefont
  {Legeza}}, \bibinfo {author} {\bibfnamefont {C.~P.}\ \bibnamefont {Moca}},
  \bibinfo {author} {\bibfnamefont {A.~I.}\ \bibnamefont {Toth}}, \bibinfo
  {author} {\bibfnamefont {I.}~\bibnamefont {Weymann}}, \ and\ \bibinfo
  {author} {\bibfnamefont {G.}~\bibnamefont {Zarand}},\ }\href@noop {}
  {\enquote {\bibinfo {title} {Manual for the {F}lexible {DM-NRG} code},}\ }
  (\bibinfo {year} {2008}),\ \Eprint {http://arxiv.org/abs/arXiv:0809.3143}
  {arXiv:0809.3143} \BibitemShut {NoStop}%
\bibitem [{\citenamefont {Oliveira}\ and\ \citenamefont {Oliveira}(1994)}]{Z}%
  \BibitemOpen
  \bibfield  {author} {\bibinfo {author} {\bibfnamefont {W.~C.}\ \bibnamefont
  {Oliveira}}\ and\ \bibinfo {author} {\bibfnamefont {Luiz~N.}\ \bibnamefont
  {Oliveira}},\ }\bibfield  {title} {\enquote {\bibinfo {title} {Generalized
  numerical renormalization-group method to calculate the thermodynamical
  properties of impurities in metals},}\ }\href {\doibase
  10.1103/PhysRevB.49.11986} {\bibfield  {journal} {\bibinfo  {journal} {Phys.
  Rev. B}\ }\textbf {\bibinfo {volume} {49}},\ \bibinfo {pages} {11986--11994}
  (\bibinfo {year} {1994})}\BibitemShut {NoStop}%
\bibitem [{\citenamefont {G\'orski}(2016)}]{Gorski-2016}%
  \BibitemOpen
  \bibfield  {author} {\bibinfo {author} {\bibfnamefont {G.}~\bibnamefont
  {G\'orski}},\ }\bibfield  {title} {\enquote {\bibinfo {title} {Irreducible
  {Green} functions method applied to nanoscopic systems},}\ }\href {\doibase
  10.12693/APhysPolA.130.551} {\bibfield  {journal} {\bibinfo  {journal} {Acta
  Phys. Pol.}\ }\textbf {\bibinfo {volume} {130}},\ \bibinfo {pages} {551}
  (\bibinfo {year} {2016})}\BibitemShut {NoStop}%
\bibitem [{\citenamefont {He}\ \emph {et~al.}(2014)\citenamefont {He},
  \citenamefont {Ng}, \citenamefont {Lee},\ and\ \citenamefont
  {Law}}]{He-2014}%
  \BibitemOpen
  \bibfield  {author} {\bibinfo {author} {\bibfnamefont {J.~J.}\ \bibnamefont
  {He}}, \bibinfo {author} {\bibfnamefont {T.~K.}\ \bibnamefont {Ng}}, \bibinfo
  {author} {\bibfnamefont {P.~A.}\ \bibnamefont {Lee}}, \ and\ \bibinfo
  {author} {\bibfnamefont {K.~T.}\ \bibnamefont {Law}},\ }\bibfield  {title}
  {\enquote {\bibinfo {title} {Selective equal-spin {Andreev} reflections
  induced by {Majorana} fermions},}\ }\href {\doibase
  10.1103/PhysRevLett.112.037001} {\bibfield  {journal} {\bibinfo  {journal}
  {Phys. Rev. Lett.}\ }\textbf {\bibinfo {volume} {112}},\ \bibinfo {pages}
  {037001} (\bibinfo {year} {2014})}\BibitemShut {NoStop}%
\bibitem [{\citenamefont {Haim}\ \emph {et~al.}(2015)\citenamefont {Haim},
  \citenamefont {Berg}, \citenamefont {von Oppen},\ and\ \citenamefont
  {Oreg}}]{Haim-2015}%
  \BibitemOpen
  \bibfield  {author} {\bibinfo {author} {\bibfnamefont {A.}~\bibnamefont
  {Haim}}, \bibinfo {author} {\bibfnamefont {E.}~\bibnamefont {Berg}}, \bibinfo
  {author} {\bibfnamefont {F.}~\bibnamefont {von Oppen}}, \ and\ \bibinfo
  {author} {\bibfnamefont {Y.}~\bibnamefont {Oreg}},\ }\bibfield  {title}
  {\enquote {\bibinfo {title} {Signatures of {M}ajorana zero modes in
  spin-resolved current correlations},}\ }\href {\doibase
  10.1103/PhysRevLett.114.166406} {\bibfield  {journal} {\bibinfo  {journal}
  {Phys. Rev. Lett.}\ }\textbf {\bibinfo {volume} {114}},\ \bibinfo {pages}
  {166406} (\bibinfo {year} {2015})}\BibitemShut {NoStop}%
\bibitem [{\citenamefont {Ren}\ \emph {et~al.}(2017)\citenamefont {Ren},
  \citenamefont {Yang}, \citenamefont {Xiang}, \citenamefont {Wang},\ and\
  \citenamefont {Tian}}]{Ren-2017}%
  \BibitemOpen
  \bibfield  {author} {\bibinfo {author} {\bibfnamefont {C.}~\bibnamefont
  {Ren}}, \bibinfo {author} {\bibfnamefont {J}~\bibnamefont {Yang}}, \bibinfo
  {author} {\bibfnamefont {J.}~\bibnamefont {Xiang}}, \bibinfo {author}
  {\bibfnamefont {S.}~\bibnamefont {Wang}}, \ and\ \bibinfo {author}
  {\bibfnamefont {H.}~\bibnamefont {Tian}},\ }\bibfield  {title} {\enquote
  {\bibinfo {title} {Non-local spin blocking effect of zero-energy {M}ajorana
  fermions},}\ }\href {\doibase 10.7566/JPSJ.86.124715} {\bibfield  {journal}
  {\bibinfo  {journal} {Journal of the Physical Society of Japan}\ }\textbf
  {\bibinfo {volume} {86}},\ \bibinfo {pages} {124715} (\bibinfo {year}
  {2017})},\ \Eprint
  {http://arxiv.org/abs/https://doi.org/10.7566/JPSJ.86.124715}
  {https://doi.org/10.7566/JPSJ.86.124715} \BibitemShut {NoStop}%
\bibitem [{\citenamefont {B\'eri}\ and\ \citenamefont
  {Cooper}(2012)}]{Beri-2012}%
  \BibitemOpen
  \bibfield  {author} {\bibinfo {author} {\bibfnamefont {B.}~\bibnamefont
  {B\'eri}}\ and\ \bibinfo {author} {\bibfnamefont {N.~R.}\ \bibnamefont
  {Cooper}},\ }\bibfield  {title} {\enquote {\bibinfo {title} {Topological
  kondo effect with majorana fermions},}\ }\href {\doibase
  10.1103/PhysRevLett.109.156803} {\bibfield  {journal} {\bibinfo  {journal}
  {Phys. Rev. Lett.}\ }\textbf {\bibinfo {volume} {109}},\ \bibinfo {pages}
  {156803} (\bibinfo {year} {2012})}\BibitemShut {NoStop}%
\bibitem [{\citenamefont {Galpin}\ \emph {et~al.}(2014)\citenamefont {Galpin},
  \citenamefont {Mitchell}, \citenamefont {Temaismithi}, \citenamefont {Logan},
  \citenamefont {B\'eri},\ and\ \citenamefont {Cooper}}]{Logan-2014}%
  \BibitemOpen
  \bibfield  {author} {\bibinfo {author} {\bibfnamefont {M.~R.}\ \bibnamefont
  {Galpin}}, \bibinfo {author} {\bibfnamefont {A.~K.}\ \bibnamefont
  {Mitchell}}, \bibinfo {author} {\bibfnamefont {J.}~\bibnamefont
  {Temaismithi}}, \bibinfo {author} {\bibfnamefont {D.~E.}\ \bibnamefont
  {Logan}}, \bibinfo {author} {\bibfnamefont {B.}~\bibnamefont {B\'eri}}, \
  and\ \bibinfo {author} {\bibfnamefont {N.~R.}\ \bibnamefont {Cooper}},\
  }\bibfield  {title} {\enquote {\bibinfo {title} {Conductance fingerprint of
  majorana fermions in the topological kondo effect},}\ }\href {\doibase
  10.1103/PhysRevB.89.045143} {\bibfield  {journal} {\bibinfo  {journal} {Phys.
  Rev. B}\ }\textbf {\bibinfo {volume} {89}},\ \bibinfo {pages} {045143}
  (\bibinfo {year} {2014})}\BibitemShut {NoStop}%
\bibitem [{\citenamefont {Plugge}\ \emph {et~al.}(2016)\citenamefont {Plugge},
  \citenamefont {Zazunov}, \citenamefont {Eriksson}, \citenamefont {Tsvelik},\
  and\ \citenamefont {Egger}}]{Tsvelik-2016}%
  \BibitemOpen
  \bibfield  {author} {\bibinfo {author} {\bibfnamefont {S.}~\bibnamefont
  {Plugge}}, \bibinfo {author} {\bibfnamefont {A.}~\bibnamefont {Zazunov}},
  \bibinfo {author} {\bibfnamefont {E.}~\bibnamefont {Eriksson}}, \bibinfo
  {author} {\bibfnamefont {A.~M.}\ \bibnamefont {Tsvelik}}, \ and\ \bibinfo
  {author} {\bibfnamefont {R.}~\bibnamefont {Egger}},\ }\bibfield  {title}
  {\enquote {\bibinfo {title} {Kondo physics from quasiparticle poisoning in
  majorana devices},}\ }\href {\doibase 10.1103/PhysRevB.93.104524} {\bibfield
  {journal} {\bibinfo  {journal} {Phys. Rev. B}\ }\textbf {\bibinfo {volume}
  {93}},\ \bibinfo {pages} {104524} (\bibinfo {year} {2016})}\BibitemShut
  {NoStop}%
\bibitem [{\citenamefont {B\'eri}(2017)}]{Beri-2017}%
  \BibitemOpen
  \bibfield  {author} {\bibinfo {author} {\bibfnamefont {B.}~\bibnamefont
  {B\'eri}},\ }\bibfield  {title} {\enquote {\bibinfo {title} {Exact
  nonequilibrium transport in the topological kondo effect},}\ }\href {\doibase
  10.1103/PhysRevLett.119.027701} {\bibfield  {journal} {\bibinfo  {journal}
  {Phys. Rev. Lett.}\ }\textbf {\bibinfo {volume} {119}},\ \bibinfo {pages}
  {027701} (\bibinfo {year} {2017})}\BibitemShut {NoStop}%
\bibitem [{\citenamefont {Deng}\ \emph {et~al.}(2017)\citenamefont {Deng},
  \citenamefont {Vaitiek\.{e}nas}, \citenamefont {Prada}, \citenamefont
  {San-Jose}, \citenamefont {Nyg{\r a}rd}, \citenamefont {Krogstrup},
  \citenamefont {Aguado},\ and\ \citenamefont {Marcus}}]{Deng-2017b}%
  \BibitemOpen
  \bibfield  {author} {\bibinfo {author} {\bibfnamefont {M.~T.}\ \bibnamefont
  {Deng}}, \bibinfo {author} {\bibfnamefont {S.}~\bibnamefont
  {Vaitiek\.{e}nas}}, \bibinfo {author} {\bibfnamefont {E.}~\bibnamefont
  {Prada}}, \bibinfo {author} {\bibfnamefont {P.}~\bibnamefont {San-Jose}},
  \bibinfo {author} {\bibfnamefont {J.}~\bibnamefont {Nyg{\r a}rd}}, \bibinfo
  {author} {\bibfnamefont {P.}~\bibnamefont {Krogstrup}}, \bibinfo {author}
  {\bibfnamefont {R.}~\bibnamefont {Aguado}}, \ and\ \bibinfo {author}
  {\bibfnamefont {C.~M.}\ \bibnamefont {Marcus}},\ }\href@noop {} {\enquote
  {\bibinfo {title} {Majorana non-locality in hybrid nanowires},}\ } (\bibinfo
  {year} {2017}),\ \Eprint {http://arxiv.org/abs/arXiv:1712.03536}
  {arXiv:1712.03536} \BibitemShut {NoStop}%
\bibitem [{\citenamefont {Chevallier}\ \emph {et~al.}(2013)\citenamefont
  {Chevallier}, \citenamefont {Simon},\ and\ \citenamefont
  {Bena}}]{Chevallier-2013}%
  \BibitemOpen
  \bibfield  {author} {\bibinfo {author} {\bibfnamefont {D.}~\bibnamefont
  {Chevallier}}, \bibinfo {author} {\bibfnamefont {P.}~\bibnamefont {Simon}}, \
  and\ \bibinfo {author} {\bibfnamefont {C.}~\bibnamefont {Bena}},\ }\bibfield
  {title} {\enquote {\bibinfo {title} {From {A}ndreev bound states to
  {M}ajorana fermions in topological wires on superconducting substrates: A
  story of mutation},}\ }\href {\doibase 10.1103/PhysRevB.88.165401} {\bibfield
   {journal} {\bibinfo  {journal} {Phys. Rev. B}\ }\textbf {\bibinfo {volume}
  {88}},\ \bibinfo {pages} {165401} (\bibinfo {year} {2013})}\BibitemShut
  {NoStop}%
\bibitem [{\citenamefont {Rainis}\ \emph {et~al.}(2013)\citenamefont {Rainis},
  \citenamefont {Trifunovic}, \citenamefont {Klinovaja},\ and\ \citenamefont
  {Loss}}]{Rainis-2013}%
  \BibitemOpen
  \bibfield  {author} {\bibinfo {author} {\bibfnamefont {D.}~\bibnamefont
  {Rainis}}, \bibinfo {author} {\bibfnamefont {L.}~\bibnamefont {Trifunovic}},
  \bibinfo {author} {\bibfnamefont {J.}~\bibnamefont {Klinovaja}}, \ and\
  \bibinfo {author} {\bibfnamefont {D.}~\bibnamefont {Loss}},\ }\bibfield
  {title} {\enquote {\bibinfo {title} {Towards a realistic transport modeling
  in a superconducting nanowire with {Majorana} fermions},}\ }\href {\doibase
  10.1103/PhysRevB.87.024515} {\bibfield  {journal} {\bibinfo  {journal} {Phys.
  Rev. B}\ }\textbf {\bibinfo {volume} {87}},\ \bibinfo {pages} {024515}
  (\bibinfo {year} {2013})}\BibitemShut {NoStop}%
\bibitem [{\citenamefont {Chevallier}\ \emph {et~al.}(2018)\citenamefont
  {Chevallier}, \citenamefont {Szumniak}, \citenamefont {Hoffman},
  \citenamefont {Loss},\ and\ \citenamefont {Klinovaja}}]{Klinovaja-2018}%
  \BibitemOpen
  \bibfield  {author} {\bibinfo {author} {\bibfnamefont {D.}~\bibnamefont
  {Chevallier}}, \bibinfo {author} {\bibfnamefont {P.}~\bibnamefont
  {Szumniak}}, \bibinfo {author} {\bibfnamefont {S.}~\bibnamefont {Hoffman}},
  \bibinfo {author} {\bibfnamefont {D.}~\bibnamefont {Loss}}, \ and\ \bibinfo
  {author} {\bibfnamefont {J.}~\bibnamefont {Klinovaja}},\ }\bibfield  {title}
  {\enquote {\bibinfo {title} {Topological phase detection in {R}ashba
  nanowires with a quantum dot},}\ }\href {\doibase 10.1103/PhysRevB.97.045404}
  {\bibfield  {journal} {\bibinfo  {journal} {Phys. Rev. B}\ }\textbf {\bibinfo
  {volume} {97}},\ \bibinfo {pages} {045404} (\bibinfo {year}
  {2018})}\BibitemShut {NoStop}%
\bibitem [{\citenamefont {Ma\'ska}\ and\ \citenamefont
  {Doma\'nski}(2017)}]{MaskaDomanski-2017}%
  \BibitemOpen
  \bibfield  {author} {\bibinfo {author} {\bibfnamefont {M.~M.}\ \bibnamefont
  {Ma\'ska}}\ and\ \bibinfo {author} {\bibfnamefont {T.}~\bibnamefont
  {Doma\'nski}},\ }\bibfield  {title} {\enquote {\bibinfo {title} {Polarization
  of the {M}ajorana quasiparticles in the {R}ashba chain},}\ }\href {\doibase
  10.1038/s41598-017-16323-3} {\bibfield  {journal} {\bibinfo  {journal} {Sci.
  Rep.}\ }\textbf {\bibinfo {volume} {7}},\ \bibinfo {pages} {16193} (\bibinfo
  {year} {2017})}\BibitemShut {NoStop}%
\bibitem [{\citenamefont {Mart\'in-Rodero}\ and\ \citenamefont
  {Levy~Yeyati}(2012)}]{Rodero-2012}%
  \BibitemOpen
  \bibfield  {author} {\bibinfo {author} {\bibfnamefont {A.}~\bibnamefont
  {Mart\'in-Rodero}}\ and\ \bibinfo {author} {\bibfnamefont {A.}~\bibnamefont
  {Levy~Yeyati}},\ }\bibfield  {title} {\enquote {\bibinfo {title} {The
  {Andreev} states of a superconducting quantum dot: mean field versus exact
  numerical results},}\ }\href
  {http://stacks.iop.org/0953-8984/24/i=38/a=385303} {\bibfield  {journal}
  {\bibinfo  {journal} {J. Phys.: Condens. Matter}\ }\textbf {\bibinfo {volume}
  {24}},\ \bibinfo {pages} {385303} (\bibinfo {year} {2012})}\BibitemShut
  {NoStop}%
\end{thebibliography}%

\end{document}